%% file: access.tex
\documentclass{ieeeaccess}
\usepackage{array}
\usepackage[caption=false,font=normalsize,labelfont=sf,textfont=sf]{subfig}
\usepackage{stfloats}
\usepackage{url}
\usepackage{verbatim}
\usepackage{adjustbox}
\usepackage{amsmath,amssymb,amsfonts}
\usepackage{algorithmic}
\usepackage{graphicx}
\usepackage{graphicx}
\usepackage{caption}

\usepackage{textcomp}
\usepackage{soul}
\usepackage{booktabs} 
\usepackage{multirow}
\usepackage{balance}
\usepackage{booktabs}
\usepackage{tabularx}
\usepackage{makecell} 
\usepackage{ragged2e}

\usepackage[backend=biber,style=numeric-comp,sorting=none,maxbibnames=99]{biblatex}
\newcolumntype{L}[1]{>{\RaggedRight\arraybackslash}p{#1}}   
\newcolumntype{C}[1]{>{\centering\arraybackslash}p{#1}}    
\usepackage[hidelinks]{hyperref}
\begin{document}
\history{Date of publication xxxx 00, 0000, date of current version xxxx 00, 0000.}
\doi{10.1109/ACCESS.2017.DOI}

\title{scaleTRIM: Scalable TRuncation-Based Integer Approximate Multiplier with Linearization and Compensation}
\author{\uppercase{Ebrahim Farahmand}\authorrefmark{1,5}, Mohammad~Javad~Askarizadeh\authorrefmark{2}, Ali Mahani\authorrefmark{2,3},Behnam~Ghavami\authorrefmark{4}, Hassan Ghasemzadeh\authorrefmark{1,5},~\IEEEmembership{Senior Member,~IEEE}, Muhammad Abdullah Hanif\authorrefmark{6},~\IEEEmembership{Student Member,~IEEE},
Muhammad Shafique\authorrefmark{6},~\IEEEmembership{Senior Member,~IEEE}}
\address[1]{School of Computing and Augmented Intelligence, Arizona State University, Tempe, United States}
\address[2]{Department of Electrical Engineering, Shahid Bahonar University of Kerman, Iran}
\address[3]{York University, Toronto, ON, Canada}
\address[4]{Simon Fraser University,Burnaby, BC, Canada}
\address[5]{College of Health Solutions, Arizona State University, Phoenix, United States}
\address[6]{eBrain Lab, Division of Engineering, New York University Abu Dhabi, UAE}
\tfootnote{This research is partly supported by the ASPIRE AARE Grant (S1561) on ”Towards Extreme Energy Efficiency through Cross-Layer Approximate Computing".}

\corresp{Corresponding author: Ebrahim Farahmand (efarahma@asu.edu).}

\begin{abstract}
In this paper, we propose a scalable approximate multiplier design, scaleTRIM, that approximates the multiplication operation using fitted linear functions, also referred to as linearization. We show that multiplication operations can be completely replaced by low-cost addition and bit-wise shift operations by exploiting linearization. Moreover, our proposed design utilizes a lookup table (LUT)-based compensation unit as a novel error-reduction method. In essence, input operands are truncated to a reduced bit-width representation (i.e., $h$ bits) based on their leading-one positions. Then, a curve-fitting method is employed to map the product term to a linear function. Additionally, a piecewise constant error-correction term is used to reduce the approximation error. To compute the piecewise constant, we divide the function space into $M$ segments and average the errors within each segment. In particular, our multiplier supports various degrees of truncation and error compensation to offer a range of accuracy-efficiency trade-offs. The proposed multiplier improves the Mean Relative Error Distance (MRED) by about 15.2\% while satisfying the efficiency constraint and improves the Power Delay Product (PDP) by about 22.8\%  while satisfying the accuracy and efficiency constraints compared to different state-of-the-art approximate multipliers. From a usability perspective, our evaluation of the proposed design for image classification using Deep Neural Networks (DNNs) demonstrates that scaleTRIM offers a better accuracy-efficiency trade-off than state-of-the-art approximate multiplier designs.
\end{abstract}


\titlepgskip=-15pt

\maketitle

\section{Introduction}
\label{sec:introduction}
\PARstart{D}{ue} to the inherently error-tolerant behavior of a large number of application domains, such as digital signal processing (DSP)~\cite{6858039}, 
 biomedical informatics, 
data mining, and deep learning,~\cite{8342139}, 
Approximate Computing (AC) has emerged as a promising solution to improve energy/power, area, and performance efficiency of systems~\cite{venkataramani2015approximate,shafique2016cross,farahmand2021high}. A major part of energy consumption in these applications is associated with multiplication operations, which motivated researchers to investigate novel highly efficient approximate multiplier architectures~\cite{jiang2017review}. 
Research efforts focus on achieving efficient hardware designs at minimal accuracy loss. 
The accuracy can be further improved by using low-cost error compensation in successive computations~\cite{8493590}. 
\textit{As different applications require different accuracy targets, an approximate multiplier design with a wide range of trade-offs between accuracy and resource efficiency is highly desirable}~\cite{rehman2016architectural,venkataramani2015approximate}.

A multiplier is generally composed of three stages: (1) partial product generation, (2) partial product accumulation, and (3) addition of the remaining two rows using a fast adder.
To approximate a multiplier, approximations can be applied to each of the three aforementioned stages. 
In some works, such as~ \cite{ullah2021high,7434040,7372600,7517375}, approximations are applied in the first stage by reducing the number of partial products or reducing the complexity of partial product generation units. 
Other works, such as \cite{7817900} and \cite{fang2021approximate}, aim to improve energy and performance efficiency by using approximate compressors at different levels of the partial product accumulation stage. 
The type of adder used in the final accumulation stage also affects the overall energy consumption of a multiplier~\cite{8050437}. Thus, employing approximations in the final stage can also further reduce the energy consumption of a multiplier. 

\begin{figure}
    \centering
    \includegraphics[width=1\linewidth]{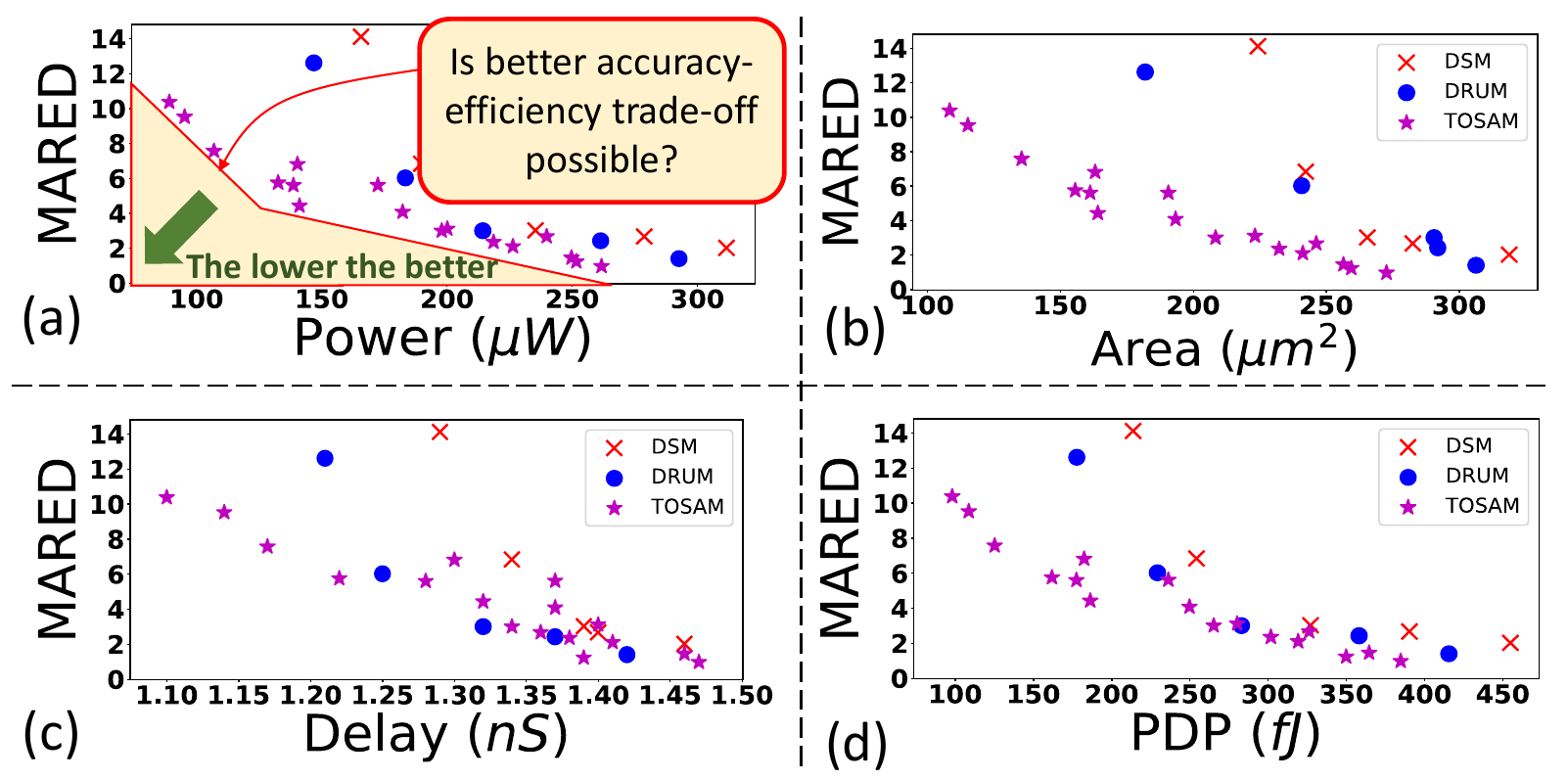}
    \caption{The complete design space of 8-bit TOSAM~\cite{8626488}, DSM~\cite{6858039} and DRUM~\cite{7372600} multipliers. MRED corresponds to mean average relative error distance, a standard metric used to quantify the accuracy (quality) of an approximate multiplier. (a) MRED vs. Power, (b) MRED vs. Area, (c) MRED vs. Delay, and (d) MRED vs. PDP.}
    \label{fig:initial_design_space}
    \vspace{-15pt}
\end{figure} 

\subsection{Motivational Analysis}
It is important to note that fixed approximate multiplier designs cannot accommodate a wide range of applications, due to their diverse resilience profiles and different accuracy requirements. 
To address this concern, several design-time configurable approximate multipliers have been proposed that enable an explicit accuracy-efficiency trade-off by adjusting the amount of operand information used during multiplication. These designs include DSM~\cite{6858039}, DRUM~\cite{7372600}, and TOSAM~\cite{8626488}. 
Although these designs offer various configurations, their efficiency in terms of area, power, and delay drops drastically as accuracy requirements increase, which negatively impacts both the design-time and running costs of the system. 
Fig.~\ref{fig:initial_design_space} shows the complete design space of 8-bit TOSAM, DSM, and DRUM multipliers. 
The figure highlights that as the accuracy requirements increase, the cost of the optimal design in terms of area, power, and delay increases drastically. 
\textit{Thus, more sophisticated approximations are required that can offer improved accuracy-efficiency trade-offs.}

\subsection{Our Novel Contributions and Concept Overview}
\begin{figure}
    \centering
    \includegraphics[width=1\linewidth]{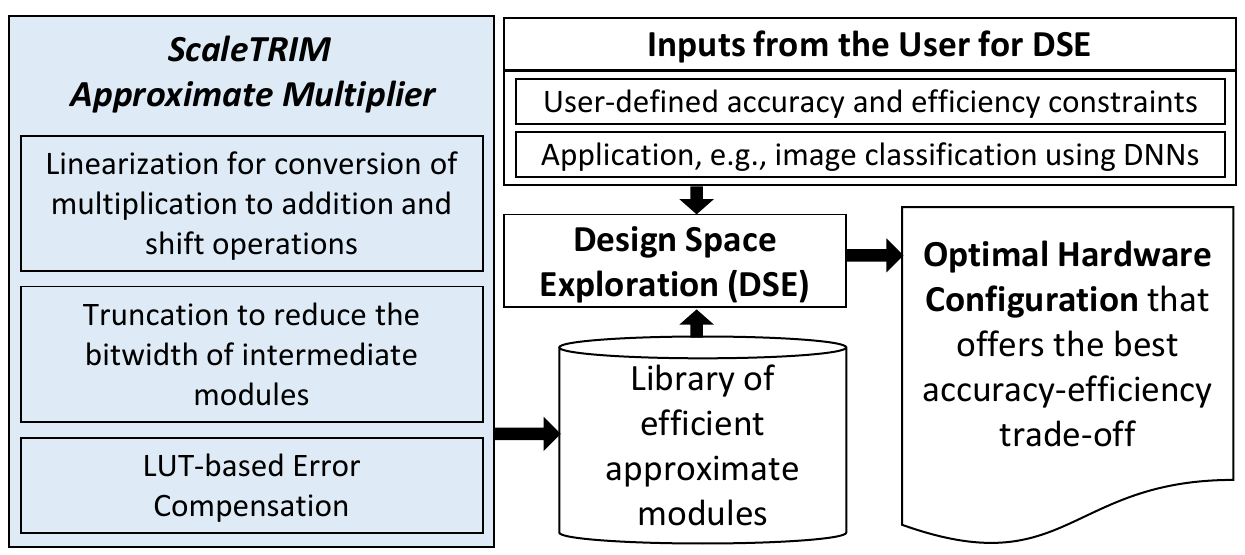}
    \caption{An illustration highlighting the significance of our work from an overall system perspective. scaleTRIM (highlighted in blue) provides highly efficient approximate multiplier configurations that can offer improved accuracy-efficiency trade-offs for error-resilient applications, e.g., image classification using DNNs.}
    \label{fig:novel_contributions}
    \vspace{-15pt}
\end{figure}

Towards expanding the design space of approximate multipliers,
in this paper, we propose a novel scalable approximate multiplier, scaleTRIM, that approximates multiplication using linear functions (i.e., linearization). 
We show that multiplication can be completely replaced by low-cost addition and shift operations by exploiting linearization. 
Moreover, our proposed design utilizes a LUT-based compensation unit to achieve low-cost error reduction, which helps further expand the design space of approximate multipliers and achieve Pareto-optimal configurations.
To improve energy efficiency, input operands are truncated to $h$-bits based on their leading-one positions. 
Then, a curve-fitting method is employed to map the product terms to a linear function, dramatically reducing the area and power/energy consumption of multiplication units. 
Moreover, to further reduce approximation error, we introduce a piecewise error-correction term computed by partitioning the input space into $M$ segments and averaging the errors in each segment. 
Thus, our proposed design supports various configurations by varying the $h$ and $M$ parameters to exploit the accuracy-efficiency trade-off. Fig.~\ref{fig:novel_contributions} illustrates the significance of the contributions from a full-system perspective. In summary, 

\begin{itemize}
    \item We propose a scalable, unsigned approximate multiplier that replaces standard multiplication with low-cost addition and bit-wise shifts through linearization and low-cost error compensation.
    \item We propose a piecewise error-compensation method that partitions the input space into $M$ segments to mitigate approximation errors introduced by truncation and linearization.
    \item We highlight the effectiveness of the proposed design for expanding the design space of approximate multipliers and offering Pareto-optimal configurations, which can be useful for a diverse set of error-resilient applications. 
    \item We investigate the application of our proposed multiplier design for image classification using DNN. Our results demonstrate that the use of scaleTRIM for DNN-based image classification applications results in an improved accuracy-efficiency trade-off compared to state-of-the-art approximate multipliers. 
\end{itemize}

\input{Sections/Related_Works}
\begin{figure*}
    \centering
    \includegraphics[width=1\linewidth]{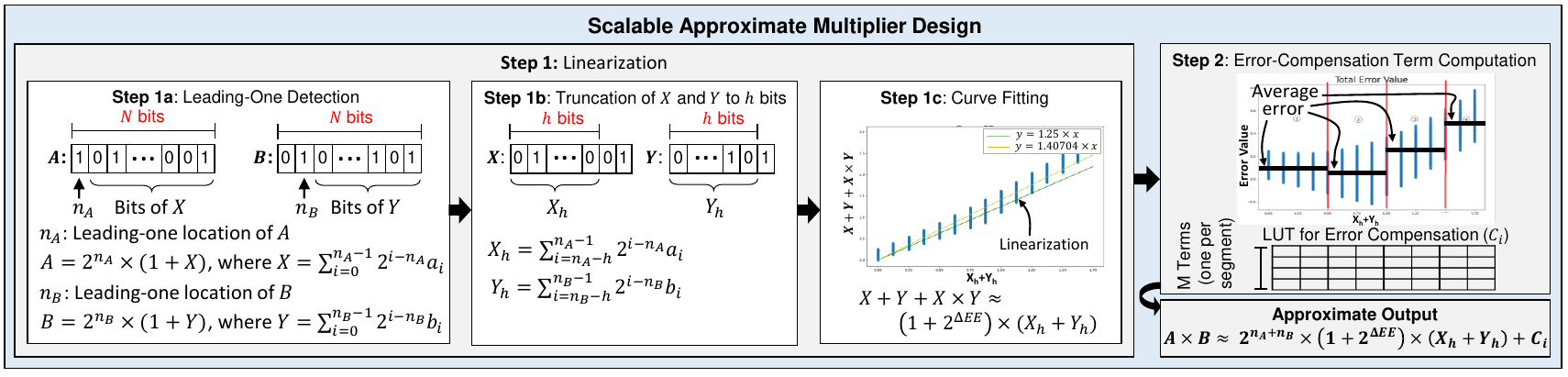}
    \caption{Overview of our novel $scaleTRIM(h,M)$ design methodology. The design is based on two major step: (1) Linearization, and (2) Error Compensation using a low-cost LUT unit. The linearization step is further composed of three sub-steps: (a) Leading-One Detection, (b) Truncation of input operands, and (c) Curve Fitting.\vspace{-10pt}}
    \label{Fig:Design_Methodology_Overview}
\end{figure*}

\section{ScaleTRIM: Proposed Scalable Truncation-based Integer Approximate Multiplier}\label{Sec:proposed}
In this section, we present our approximate multiplier design, scaleTRIM. The overall design methodology for generating scaleTRIM is summarized in Fig.~\ref{Fig:Design_Methodology_Overview}. It is based on the truncation of input operands and the linearization of the multiplication operation. By using linearization along with truncation, the multiplication operation is replaced with low-cost addition and bit-wise shift operations. Moreover, the following are our key ideas:
\begin{itemize}
    \item Introducing an LUT-based error compensation method to further reduce the approximation error.
    \item Dividing the complete approximation error function space into $M$ segments and defining a piecewise constant error-compensation term by computing the average of errors in each segment.
\end{itemize}
In principle, to improve energy efficiency, after finding the leading-one positions of input operands, the operands are truncated to $h$ bits. Then, we use curve fitting based on the sum of the truncated inputs to fit the product to a linear function, a.k.a. \textit{linearization}. This method reduces multiplication to bit-wise shift and addition operations. The flow of the complete linearization process is illustrated as Step~1 in Fig.~\ref{Fig:Design_Methodology_Overview}, and the details are presented in Section ~\ref{subsection:linearization}. The proposed error compensation method is illustrated as Step~2 in Fig.~\ref{Fig:Design_Methodology_Overview} and described in detail in ~\ref{subsection:error_reduction}. ScaleTRIM supports various degrees of truncation (controlled using $h$) and error compensation (controlled using $M$) to exploit the accuracy-efficiency trade-off.
The configurability of scaleTRIM is described in detail in Section~\ref{subsection:error_configurability}, and the hardware design is presented in Section~\ref{subsection:Hardware_design}.

\subsection{Linearization}
\label{subsection:linearization}
In this subsection, we present our method for approximating multiplication by addition and bit-wise shift operations by using a linearization step. We first present a mathematical formulation of the proposed method. Consider two $N$-bit unsigned integers $A$ and $B$, as below:
\begin{equation}
    A=\sum_{i=0}^{N-1}2^{i}a_{i}  \textrm{ , } B=\sum_{i=0}^{N-1}2^{i}b_{i} \quad a_{i}, b_{i}\in\{0,1\}
    \label{binary_representation}
\end{equation}

Assume the leading-one bit positions of $A$ and $B$ are $n_{A}$ and $n_{B}$, respectively. By factoring the $2^{n_{A}}$ and $2^{n_{B}}$ from the binary representation of input operands shown in Eq.~\ref{binary_representation}, we can rewrite the equation as: 

\begin{equation} \label{binary_representation_factoring}
\begin{split}
& A=2^{n_{A}}\times(1+\sum_{i=0}^{n_A-1}2^{i-n_{A}}a_{i})=2^{n_{A}}\times(1+X)\\
& B=2^{n_{B}}\times(1+\sum_{i=0}^{n_B-1}2^{i-n_{B}}b_{i})=2^{n_{B}}\times(1+Y)\\
&\qquad a_{i}=\{0,1\},b_{i}=\{0,1\},0 \leq X <  1,0 \leq  Y <  1
\end{split}
\end{equation}

Thus, the result of $A\times B$ can be calculated using on Eq.~\ref{multiply result}.
\begin{equation}
\begin{split}
    M_{ACC}&=A\times B=2^{n_{A}}\times(1+X)\times2^{n_{B}}\times(1+Y)\\
    &=2^{n_{A}+n_{B}}\times(1+X+Y+X\times Y)
    \label{multiply result}
\end{split}
\end{equation}

where $M_{ACC}$ represents the result of accurate multiplication. 

\begin{figure}
 \subfloat[]{
    \centering
    \includegraphics[width=0.45\columnwidth]{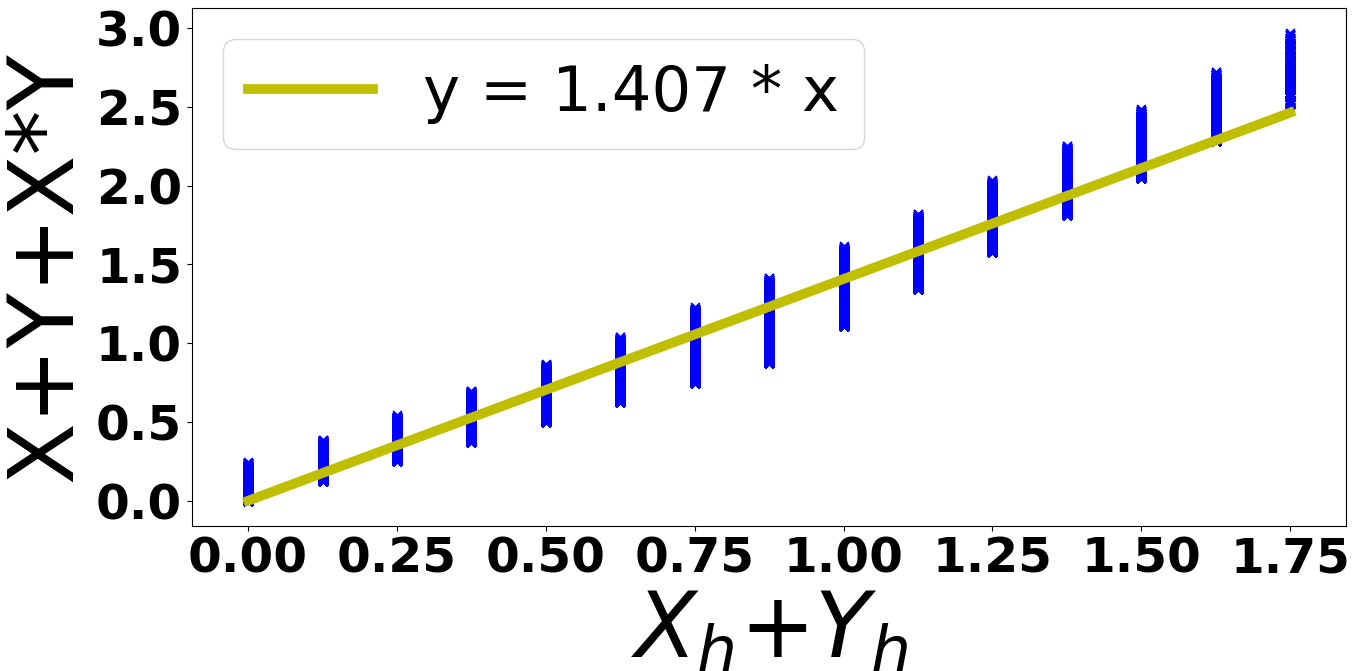}
    \label{curve_fit}
    }
     \subfloat[]{
     \centering
    \includegraphics[width=0.45\columnwidth]{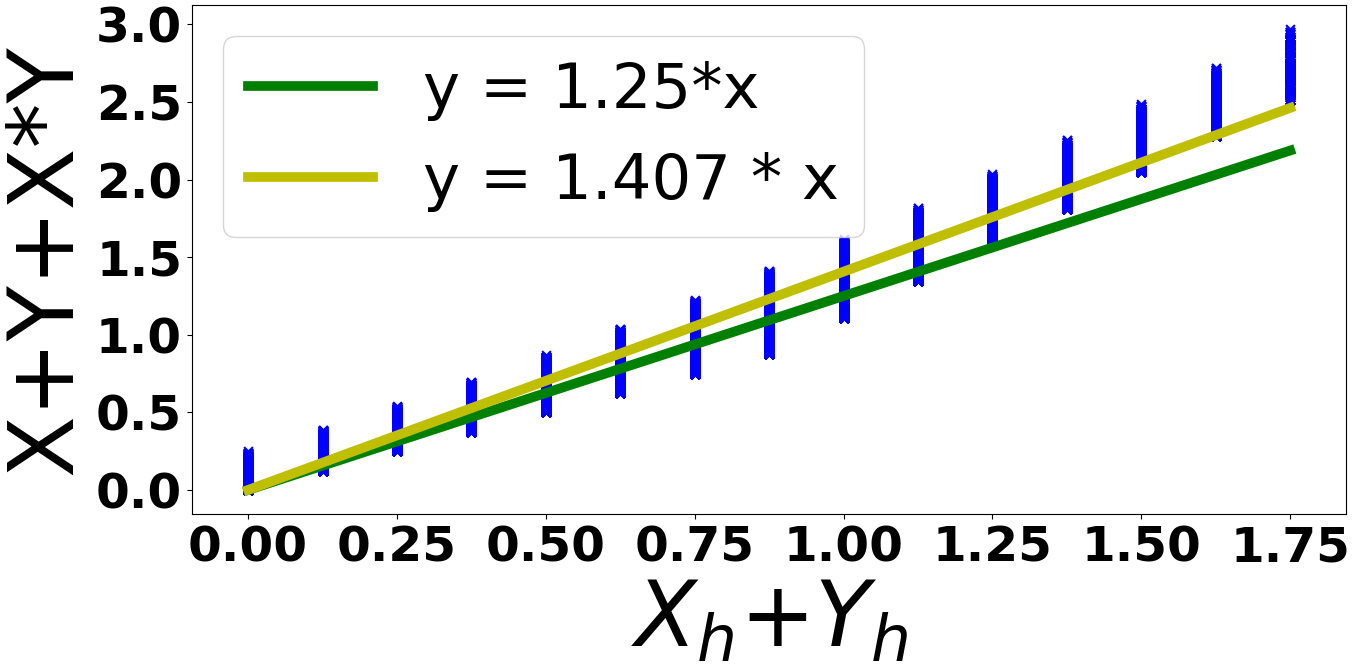}
    \label{rounded_curve_fit}
    }
    \caption{An Example of (a) curve fitting method considering $h=3$. (b) Rounding the $\alpha$ value down to the nearest $1+2^{\Delta EE}$ for efficient hardware implementation.}
    \vspace{-10pt}
\end{figure}

To approximate the multiplication term using only addition and bit-shift operations, we first analyzed the exact expression $X+Y+X \times Y$ against the truncated sum $X_h +Y_h$, where $X_h$ and $Y_h$ are obtained after leading-one detection and truncation to $h$ bits. We generated the full set (or a large representative subset) of operand pairs, computed both $X+Y+X \times Y$ and $X_h +Y_h$, and plotted these values to observe their correlation. We then applied a linear curve fit with zero intercept to this plot to determine the optimal scaling factor $\alpha$ that minimizes the mean approximation error (see Fig.~\ref{curve_fit}).

As the energy consumption and execution time both are heavily dominated by the $(X+Y+X\times Y)$ term in Eq.~\ref{multiply result}, we propose to truncate $X$ and $Y$ to $h$ bits each and approximate the term using Eq.~\ref{Eq:Approximate_Value}. 

\begin{equation}
    X+Y+X\times Y \sim \alpha\times(X_h+Y_h)
    \label{Eq:Approximate_Value}
\end{equation}

Here, $X_h$ and $Y_h$ represent the truncated versions of $X$ and $Y$. To estimate accurate terms, i.e., $(X+Y+X\times Y)$, with approximate terms, i.e., $(X_h+Y_h)$, shown in Eq.~\ref{Eq:Approximate_Value}, we used a standard linear curve fitting approach. This fitting process is performed entirely offline at design time and does not require a full-precision multiplier in the hardware implementation. Once the optimal $\alpha$ is determined, it is quantized to the form $1+2^{\Delta EE}$ by rounding $\alpha -1$ down to the nearest power of two. This quantization enables scaling to be implemented using only a single bit-shift and one addition (see Fig.~\ref{rounded_curve_fit}), avoiding expensive multiplication units. The final shift value $\Delta EE$ is stored as a design constant.

For better illustration, Fig.~\ref{curve_fit} shows an example of the estimation method based on the curve-fitting approach. 
For example, we assume $h=3$ and estimate the term $(X+Y+X\times Y)$ using the above equation to minimize the average error. The value of $\alpha$ comes out to be $1.407$. Thus, according to Eq.~\ref{Eq:Approximate_Value}, $(X+Y+X\times Y) = 1.407\times(X_h+Y_h)$. To further simplify the hardware implementation, we simplify Eq.~\ref{Eq:Approximate_Value} to Eq.~\ref{approximate value shifter}, reducing the overall multiplication to addition and left- or right-bit-wise shift operations. 
Note that according to our experiments, the range of $\alpha$ is between $1$ and $2$.
\begin{equation}
\begin{split}
    \alpha\times(X_h+Y_h)& \sim (1+2^{\Delta EE})\times(X_h+Y_h)\\
    &=(X_h+Y_h)+2^{\Delta EE}\times(X_h+Y_h)
    \label{approximate value shifter}
\end{split}
\end{equation} 
Here, we assume that $\Delta EE$ is an integer. To find $\Delta EE$, we round $\alpha-1$ down to the nearest power of 2, i.e., $(2^{\Delta EE})$. Thus, the scaling factor $\alpha$ is implemented as ($1+2^{\Delta EE}$).
For example, the curve-fitting plot (Fig.~\ref{curve_fit}) yields $\alpha \approx 1.407$ when minimizing mean error between $(X+Y+X\times Y)$ and  $X_h+Y_h$. This value is not directly implemented in hardware; instead, $\alpha -1=0.407$ is rounded down to the nearest power of two, giving $2^{\Delta EE} = 2^{-2} = 0.25$ ($\Delta EE=-2$). As shown in Fig.~\ref{rounded_curve_fit}, $\Delta EE$ comes out to be $-2$, and the estimated line is also shown in Fig.~\ref{rounded_curve_fit}. 
By using a linearization step, which includes $\alpha$-selection, and quantization is done once at design time, ensuring that the deployed multiplier has no large multipliers and only uses low-cost logic. Hence, the equation of the proposed approximate multiplier with $h$ bit truncation simplifies ($M_{AppP}(h)$) to Eq.~\ref{approximate multiplication}. Note that both $n_A$ and $n_B$ must be greater than $h$. For the lower values, input truncation is not possible. Consequently, the bits adjacent to the LODs are utilized for $X_h$ and $Y_h$.
\begin{equation}
\begin{split}
    &M_{AppP}(h)= 2^{n_{A}+n_{B}}\times(1+(X_h+Y_h)+\\& \quad\quad\quad\quad\quad\quad\quad\quad 2^{\Delta EE}\times(X_h+Y_h)) \approx A\times B
    \label{approximate multiplication}
    \end{split}
\end{equation} 
Note that the proposed approximation scheme may result in significant errors in some computations, depending on the input values. 
Thus, to reduce approximation error, we present an LUT-based error-compensation technique in the following subsection. 

\subsection{Error Compensation}
\label{subsection:error_reduction}
\begin{figure}[ht]
    \centering
    \includegraphics[width=0.6\columnwidth]{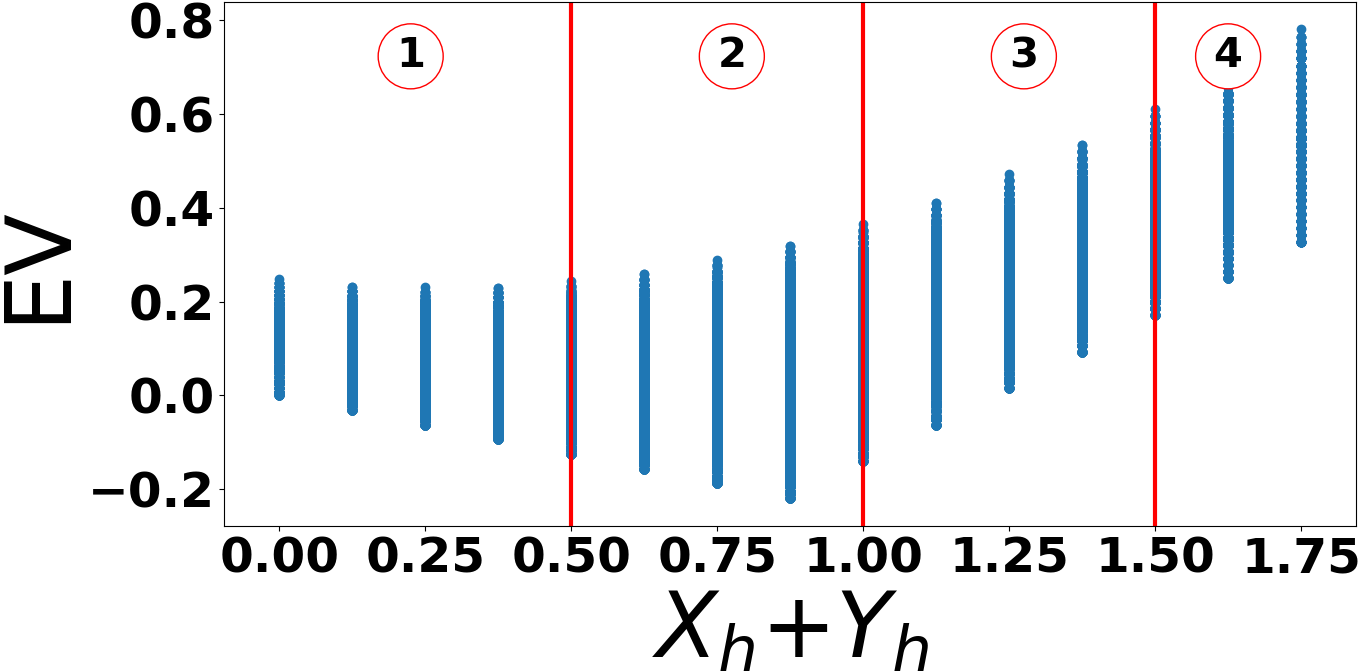}
    \caption{An Example illustrating approximate errors (i.e., Error Values (EVs)) at different $X_h+Y_h$ for the proposed approximate multiplier with $h=3$. EVs fall in different ranges for different intervals of $X_h+Y_h$, and thus, the whole space is divided into multiple segments (in this case four segments) to achieve better error compensation.}
    \label{Error Value}
    \vspace{-10pt}
\end{figure}

The Error Values (EVs) quantify the residual approximation error introduced by the truncation and linearization steps. After performing leading-one detection (LOD) and truncating the operands to obtain $X_h$ and $Y_h$, the exact normalized multiplication term $X+Y+X\times Y$ is approximated using the linearized expression based on $X_h+Y_h$ and the scaling factor ($1+2^{\Delta EE}$). The EV for each input pair is defined as the difference between the exact normalized value and its corresponding linearized approximation. These EVs are computed offline over a large set of input combinations and analyzed as a function of $X_h+Y_h$, as illustrated in Fig.~\ref{Error Value}. This definition directly corresponds to the approximation in Eq.~\ref{approximate multiplication}, where the nonlinear term is replaced by the shift-add form.

From Eq.~\ref{approximate multiplication}, it can be noticed that the accuracy of the proposed approximate multiplier depends on the level of approximation, i.e., $h$, and varies for different input combinations. The Error Values (EVs) for $h=3$ case are plotted against $X_h+Y_h$ in Fig.~\ref{Error Value}. 


\begin{figure}[ht]
    \centering
    \includegraphics[width=0.7\columnwidth]{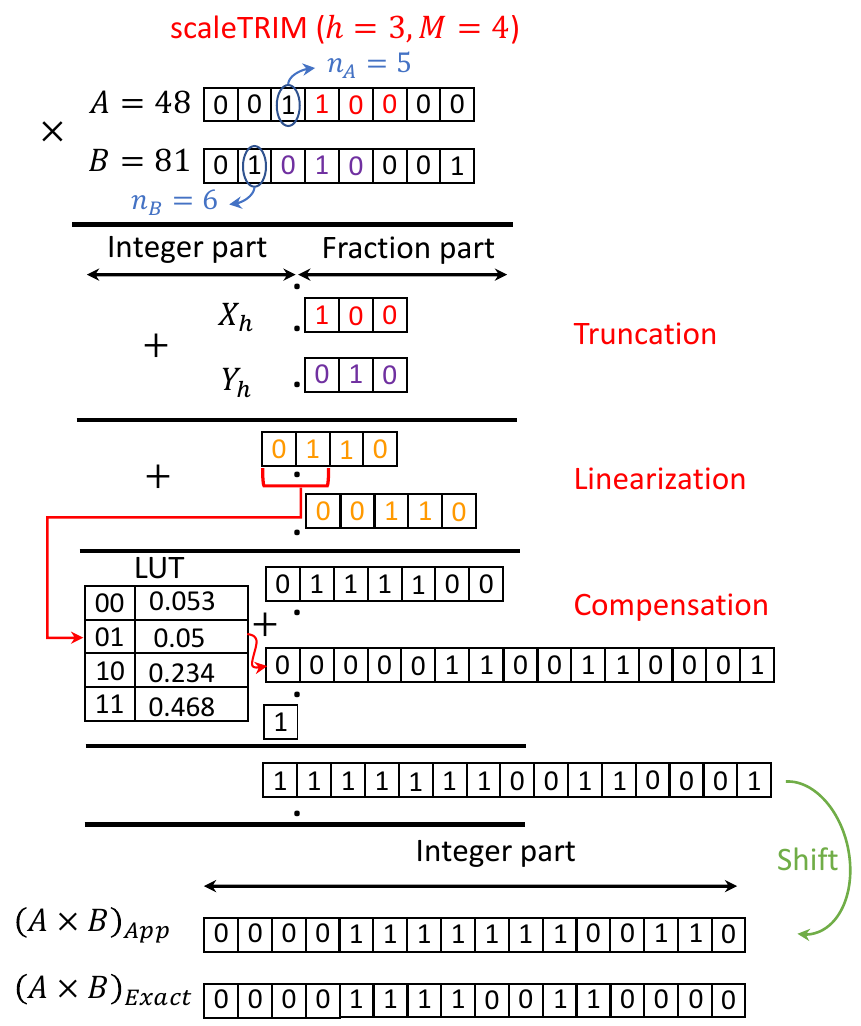}
    \caption{An Example of 8-bit $scaleTRIM(3,4)$ with inputs $A = 48$ and $B = 81$. The approximated multiplication ($(A\times B)_{App}$)  is 4070, while the exact multiplication result ($(A\times B)_{Exact}$)   is 3888. This results in an absolute error of 182.}
    \label{Example of multiplication}
    \vspace{-10pt}
\end{figure}

It can be observed from Fig.~\ref{Error Value} that the EVs fall in different ranges for different intervals of $X_h+Y_h$.
Thus, the whole space can be divided into multiple segments based on $X_h+Y_h$. 
In our case, we divide the complete space into $M$ segments, and then, for each segment, compute a constant error compensation term using the average of errors in the segment. 
These terms can then be combined to form a piece-wise constant $C_i$, which is then added to Eq.~\ref{approximate multiplication} to achieve better accuracy results. 
To store the piece-wise constant, we need to store $M$ values, i.e., one value for each segment of $X_h+Y_h$. As an example, the constant compensation values for different values of $h$ and $M$ in different ranges of $X_h + Y_h$ are illustrated in Table~\ref{tab:error_analysis} in the Appendix. 
We propose adding a piecewise-constant error-compensation term before scaling the approximate multiplier's final output.
The number of elements in the compensation LUT is equal to 
$M$, where each element stores the piecewise constant compensation value corresponding to one segment of $X_h + Y_h$. The LUT is indexed using $\lceil \log_2(M) \rceil$ bits derived from the most significant bits of $X_h + Y_h$.
Each compensation value is represented using 16 bits in our design, a choice that balances storage overhead with the precision needed for effective error correction; however, this bit-width can be configured for other design constraints. Access to the LUT is performed using a simple multiplexer, where the select lines are driven by the most significant bits of the truncated sum $X_h + Y_h$ : the two MSBs for $M=4$ and the three MSBs for $M=8$. This method enables efficient retrieval without the need for large memory blocks or complex addressing logic.
Mathematically, the approximate multiplier function, including the piecewise constant error-compensation term, can be written as follows. 
\begin{equation}
\begin{split}
    M_{AppP}(h)&=A\times B \sim 2^{n_{A}+n_{B}}\times(1+(X_h+Y_h)\\
    &+2^{\Delta EE}\times(X_h+Y_h)+C_i)
    \label{error-compensate approximate multiplication}
\end{split}
\end{equation} 
The $M$ values of $C_i$ are calculated offline before building the multiplier and stored in a LUT.
Given that $h$ and $M$ values can be configured at design time to achieve accuracy-efficiency trade-off, we define our proposed approximate multiplier (scaleTRIM) as $scaleTRIM(h,M)$, where $h$ is the bit-width of $X_h$ and $Y_h$, and $M$ is the number of segments of $X_h+Y_h$ for error compensation. As an example, the steps of multiplying $A$ by $B$ for the case of $h = 3$ and $M = 4$ ($scaleTRIM(3,4)$) are illustrated in Fig.~\ref{Example of multiplication}.

The hardware design of the proposed approximate multiplier is presented in Section~\ref{subsection:Hardware_design}. 

\subsection{Accuracy Configurability}
\label{subsection:error_configurability}

The accuracy–efficiency trade-off in scaleTRIM is controlled at design time through two parameters of $h$ and $M$. The parameter $h$ directly impacts both hardware efficiency and approximation accuracy. A smaller $h$ reduces the size of the intermediate units, leading to lower area, power, and delay, at the cost of increased approximation error. Conversely, a larger $h$ improves accuracy while incurring higher hardware overhead.

Furthermore, increasing $M$ enables finer segmentation of the $X_h+Y_h$ space and more accurate error compensation; however, it also increases the size of the compensation LUT and associated hardware cost.
Thus, by selecting appropriate values of $h$ and $M$, scaleTRIM enables systematic exploration of the accuracy–efficiency design space. In practice, the optimal configuration can be identified through offline design space exploration to satisfy application-specific constraints on accuracy, power, area, or latency.

\subsection{scaleTRIM Hardware Implementation}
\label{subsection:Hardware_design}
\begin{figure}
    \centering
    \includegraphics[scale=0.43]{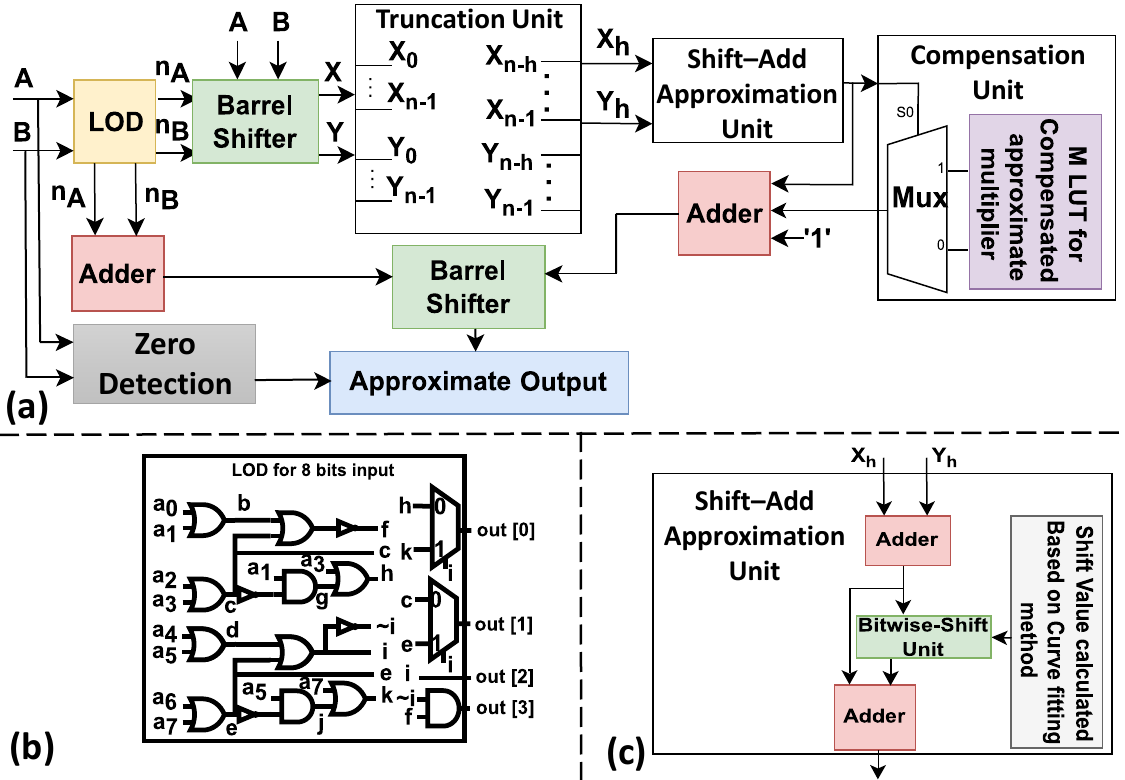}
    \caption{The hardware design of scaleTRIM. (a) is an overview of different hardware blocks of scaleTRIM. (b) is a gate level of LOD block and (c) is an overview of the Shift-Add approximation Unit of scaleTRIM.}
    \label{tot block diagram}
    \vspace{-10pt}
\end{figure}

The block diagram of the proposed scaleTRIM hardware architecture is illustrated in Fig.~\ref{tot block diagram}a. The design consists of five main components: the Zero Detection unit, the LOD, the Truncation unit, the Shift–Add approximation unit, and the Compensation unit.
First, a Zero Detection unit checks whether any input operand is zero and directly forces the output to zero in such cases, avoiding unnecessary computation. For non-zero inputs, the LOD block determines the position of the most significant ‘1’ bit in both operands, as shown in Fig.~\ref{tot block diagram}b. 

The LOD and Shift-Add blocks can be implemented in two ways: (1) using logic gates, as described in~\cite{6704691}, or (2) using lookup tables (LUTs), as presented in~\cite{8626488}. In scaleTRIM, these blocks are implemented using standard logic gates to improve area and power efficiency. Moreover, a small LUT is preserved in the architecture to store fixed pre-computed constants required for the error-compensation stage~\cite{9116315,9086744}. Since LUT-based implementations may introduce area and power overhead, a logic gate-based realization is adopted for the main computational blocks.

The output of the LOD block is $n_A$ and $n_B$. 
These values, together with inputs, are fed into barrel shifters to generate $X$ and $Y$, which are then truncated using the truncation unit to generate $X_h$ and $Y_h$. Note that if $n_A$ or $n_B$ is smaller than $h$, we concatenate zeros to the right of the truncated number to ensure a fixed bit-width ($h$) for the truncated values. For this purpose, we implemented a required multiplexer within the truncation unit to ensure a fixed bit-width for $X_h$ and $Y_h$.


In the next step, $X_h$ and $Y_h$ are fed to the Shift-Add approximation unit (Fig.~\ref{tot block diagram}c). 
First, $X_h$ and $Y_h$ are added to get $X_h + Y_h$. 
The sum is then shifted based on $\Delta EE$ term computed at the design time and added to $X_h + Y_h$ to achieve $(X_h+Y_h)+2^{\Delta EE}\times(X_h+Y_h)$. 
In parallel, the compensation unit produces the error compensation term using $X_h + Y_h$, which is then added to the output of the Shift-Add approximation unit to generate $1+(X_h+Y_h)+2^{\Delta EE}\times(X_h+Y_h)+C_i$. 
Note that the compensation terms are calculated offline and stored in an $M$-sized LUT, using read-only hardwired constants without the use of memory. 
In the last step, the result is bit-wise shifted based on $n_A + n_B$ to obtain the approximate product. 

\textbf{Handling Signed Numbers:} In this paper, we proposed and evaluated the scaleTRIM multiplier as an unsigned integer multiplier. However, extending an unsigned multiplier to support signed numbers is straightforward. For details, the readers are referred to~\cite{7372600,8532287}.

\begin{figure*}[t]
\centering
    \subfloat[Power vs. MRED]{
    \includegraphics[width=0.45\textwidth]{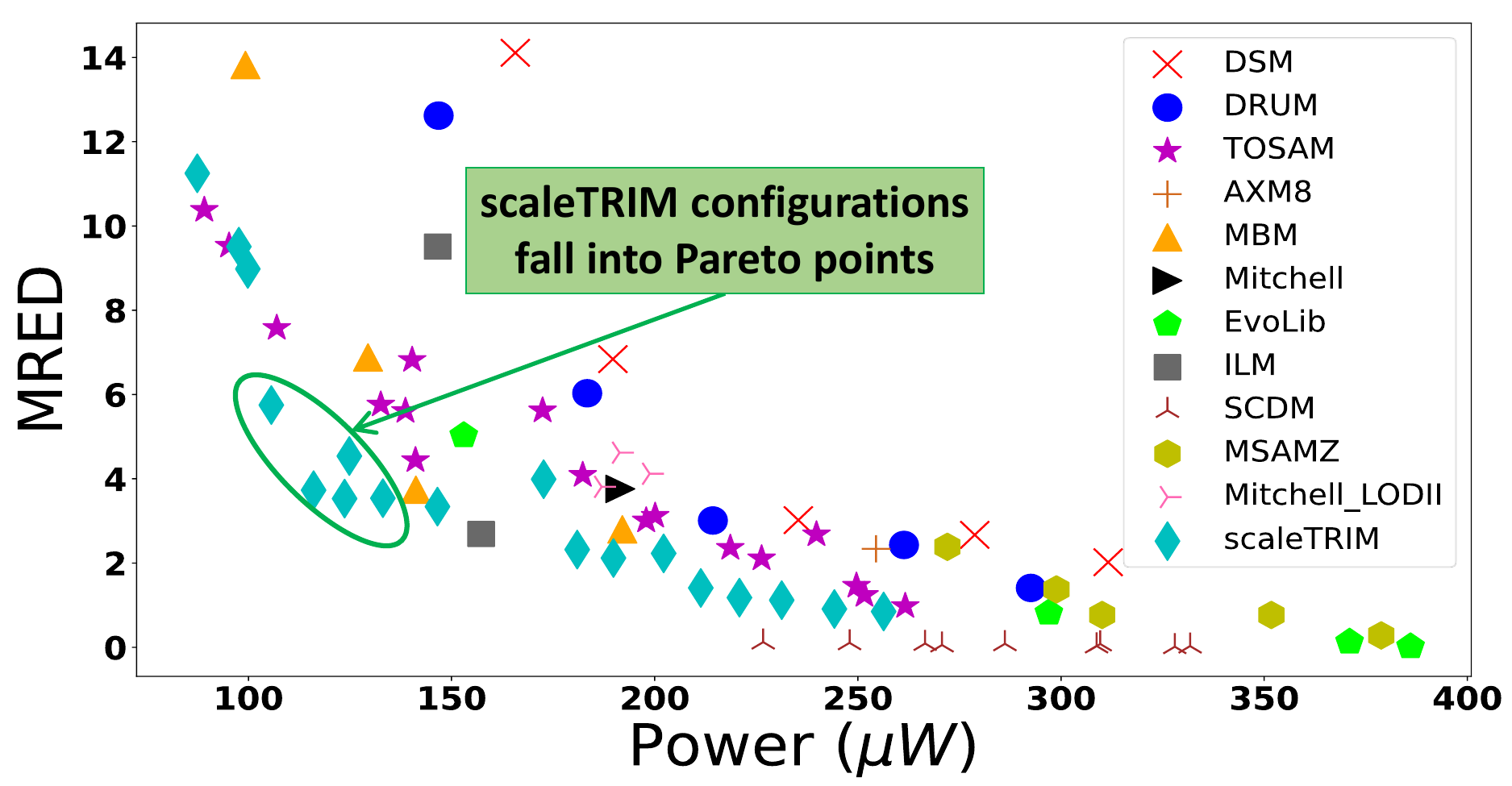}
    \label{Power_8bit_DS}
    }
    \subfloat[Area vs. MRED]{
    \includegraphics[width=0.45\textwidth]{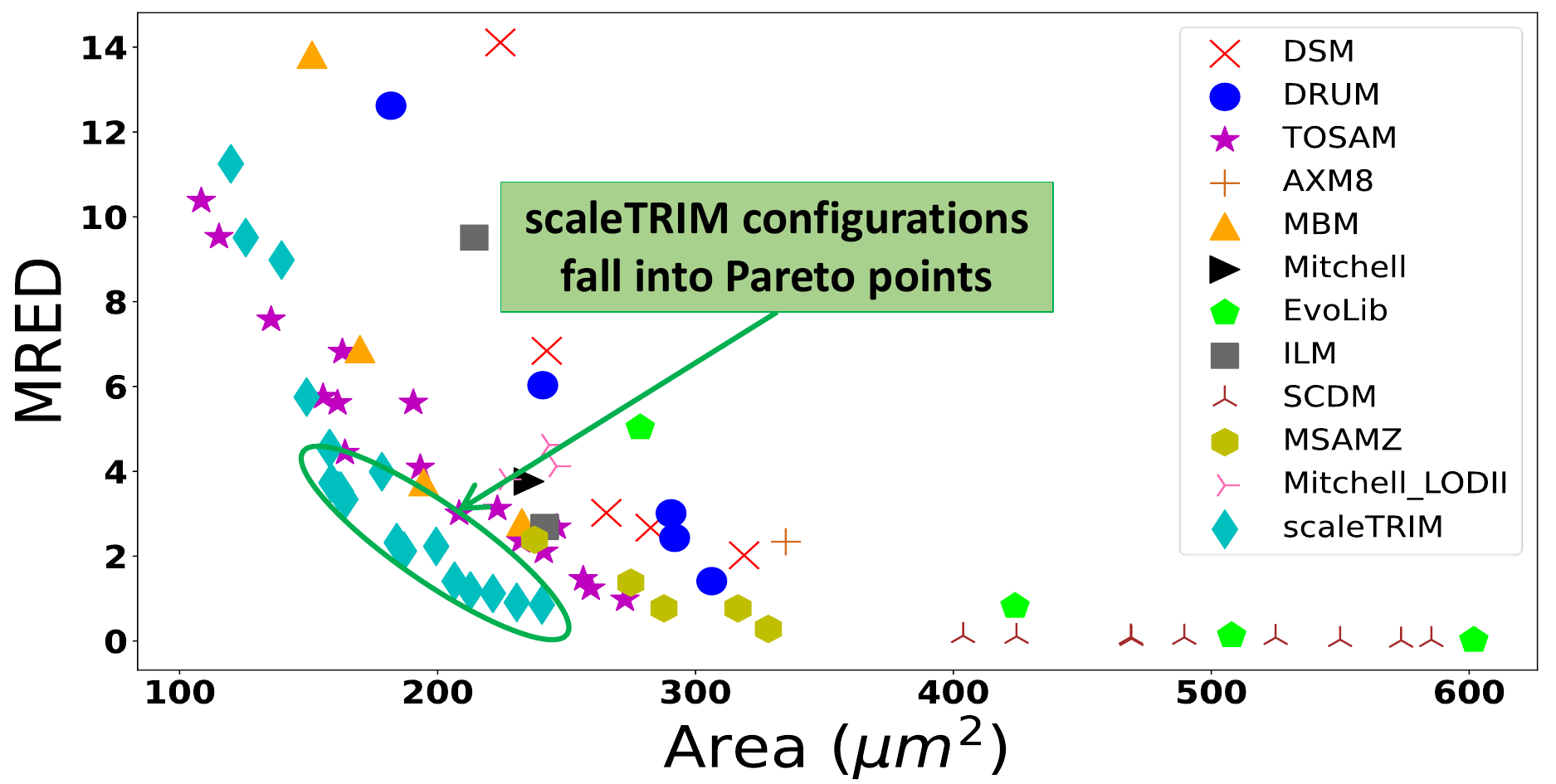}
    }\\
    \subfloat[Delay vs. MRED]{
    \includegraphics[width=0.45\linewidth]{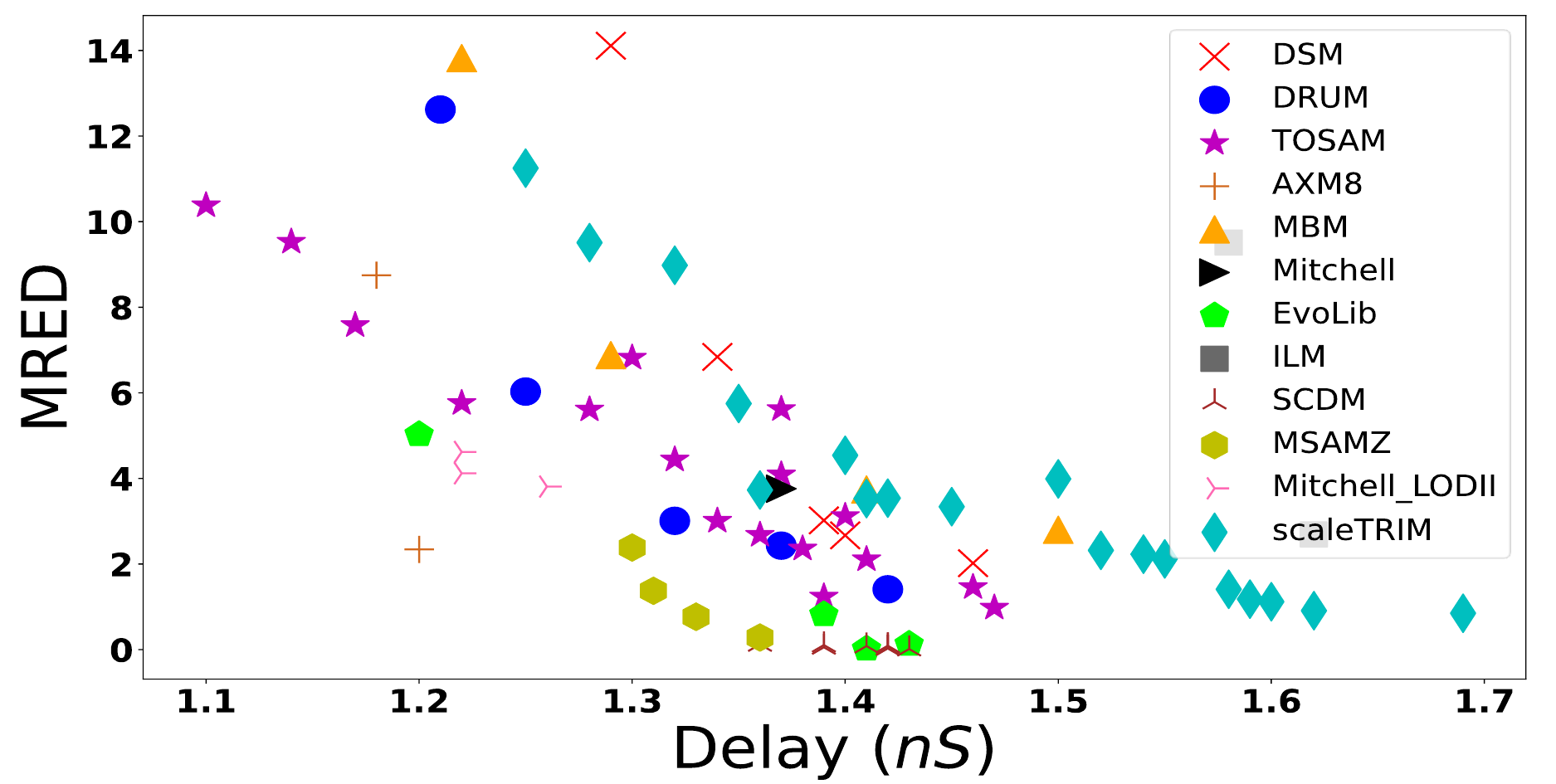}
    \label{Delay_8bit_DS}
    }
    \subfloat[PDP vs. MRED]{
    \includegraphics[width=0.45\linewidth]{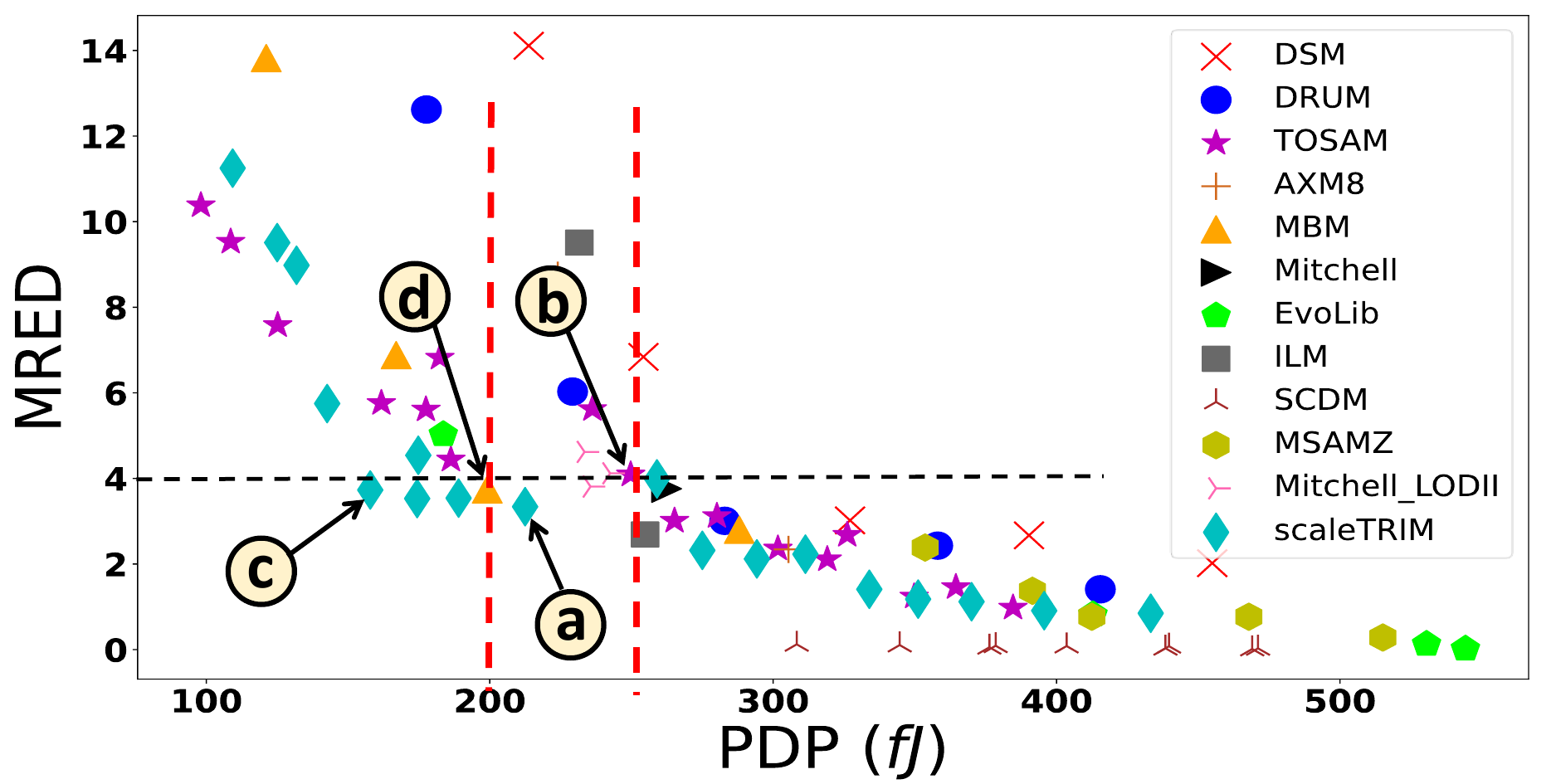}
    \label{PDP_8bit_DS}
    }
     \caption{Design space of comparison the 8-bit scaleTRIM with the state-of-the-art in terms of (a) Power vs. MRED, (b) Area vs. MRED, (c) Delay vs. MRED, and (d) PDP vs. MRED. The exact values of the figure are presented in Table~\ref{tab:multiplier_comparison} in the Appendix.}
    \label{DS_8bit}
    \vspace{-15pt}
\end{figure*}

\begin{figure*}[t]
\centering
    \subfloat[Power vs. MRED]{
    \includegraphics[width=0.45\textwidth]{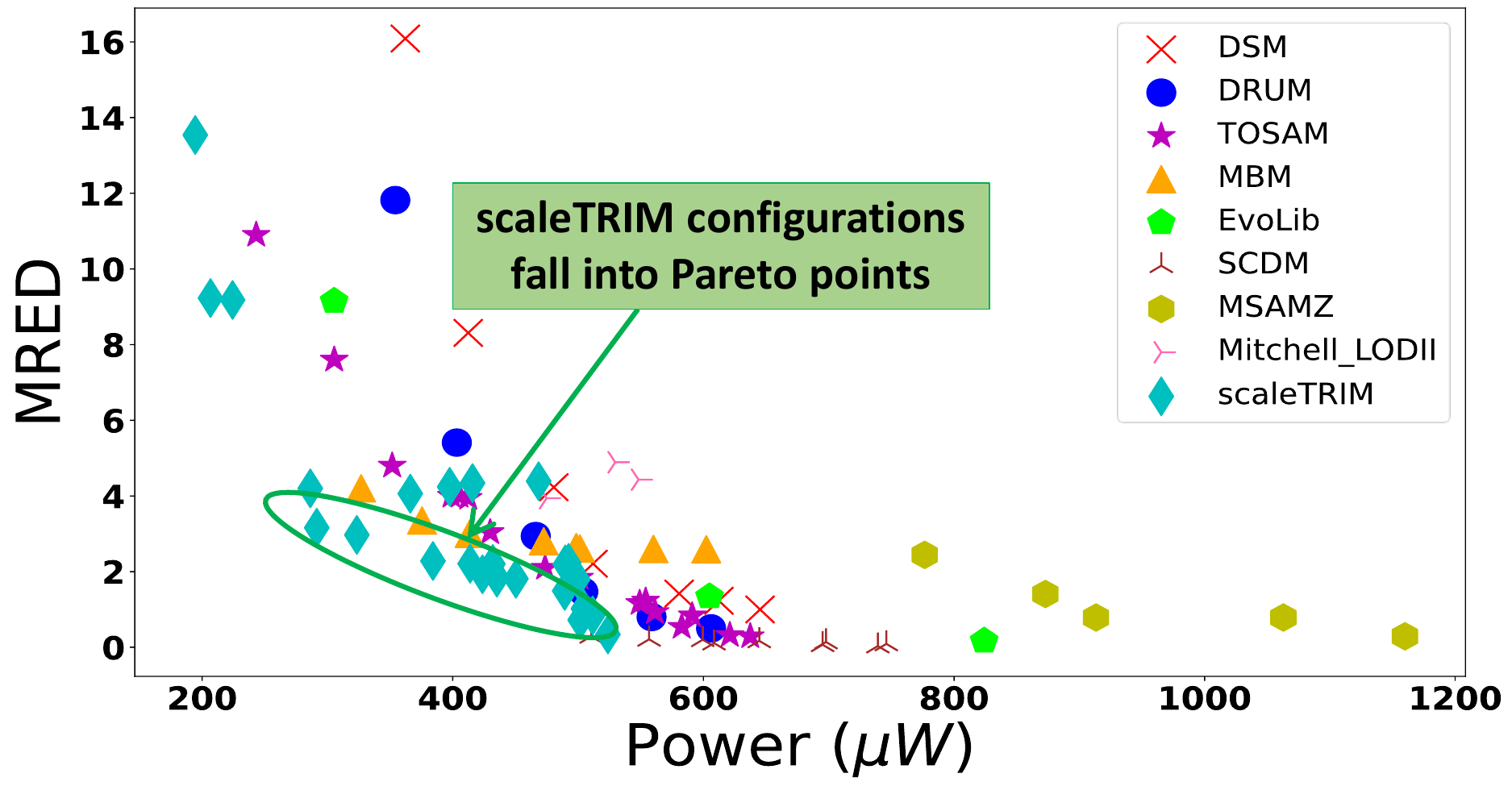}
    \label{Power_16bit_DS}
    }
    \subfloat[Area vs. MRED]{
    \includegraphics[width=0.45\textwidth]{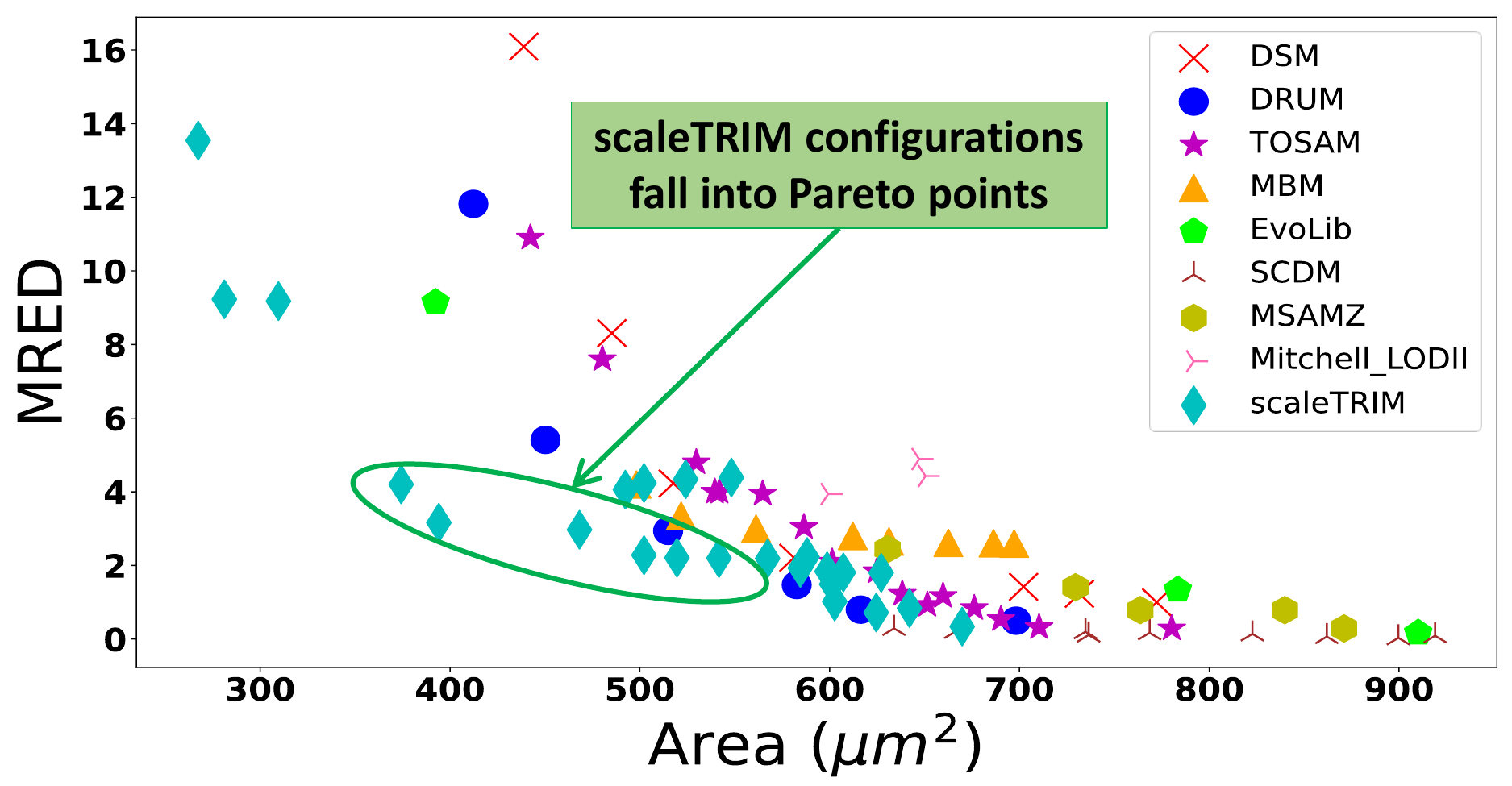}
    }\\
    \subfloat[Delay vs. MRED]{
    \includegraphics[width=0.45\textwidth]{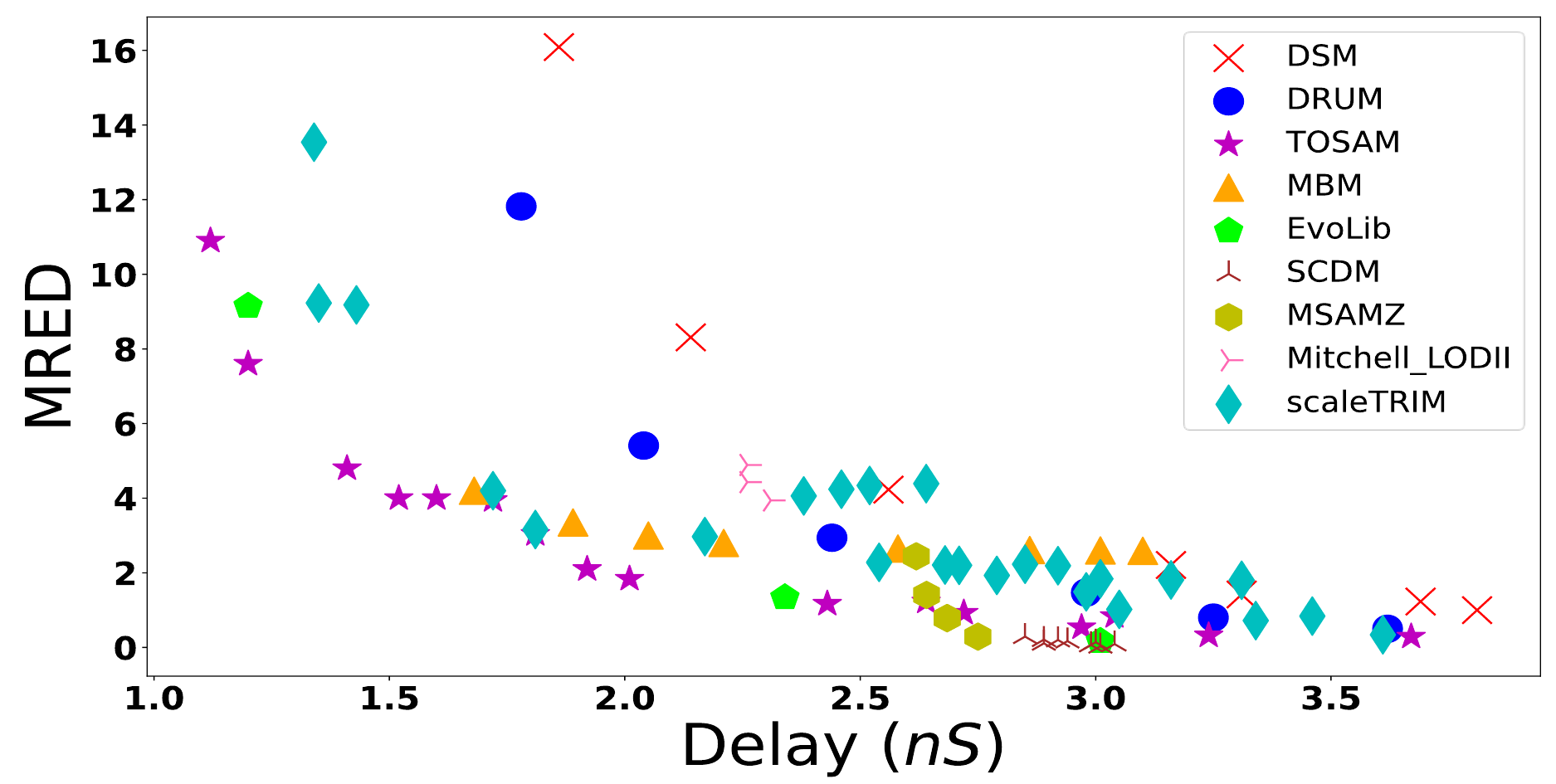}
    \label{Delay_16bit_DS}
    }
    \subfloat[PDP vs. MRED]{
    \includegraphics[width=0.45\textwidth]{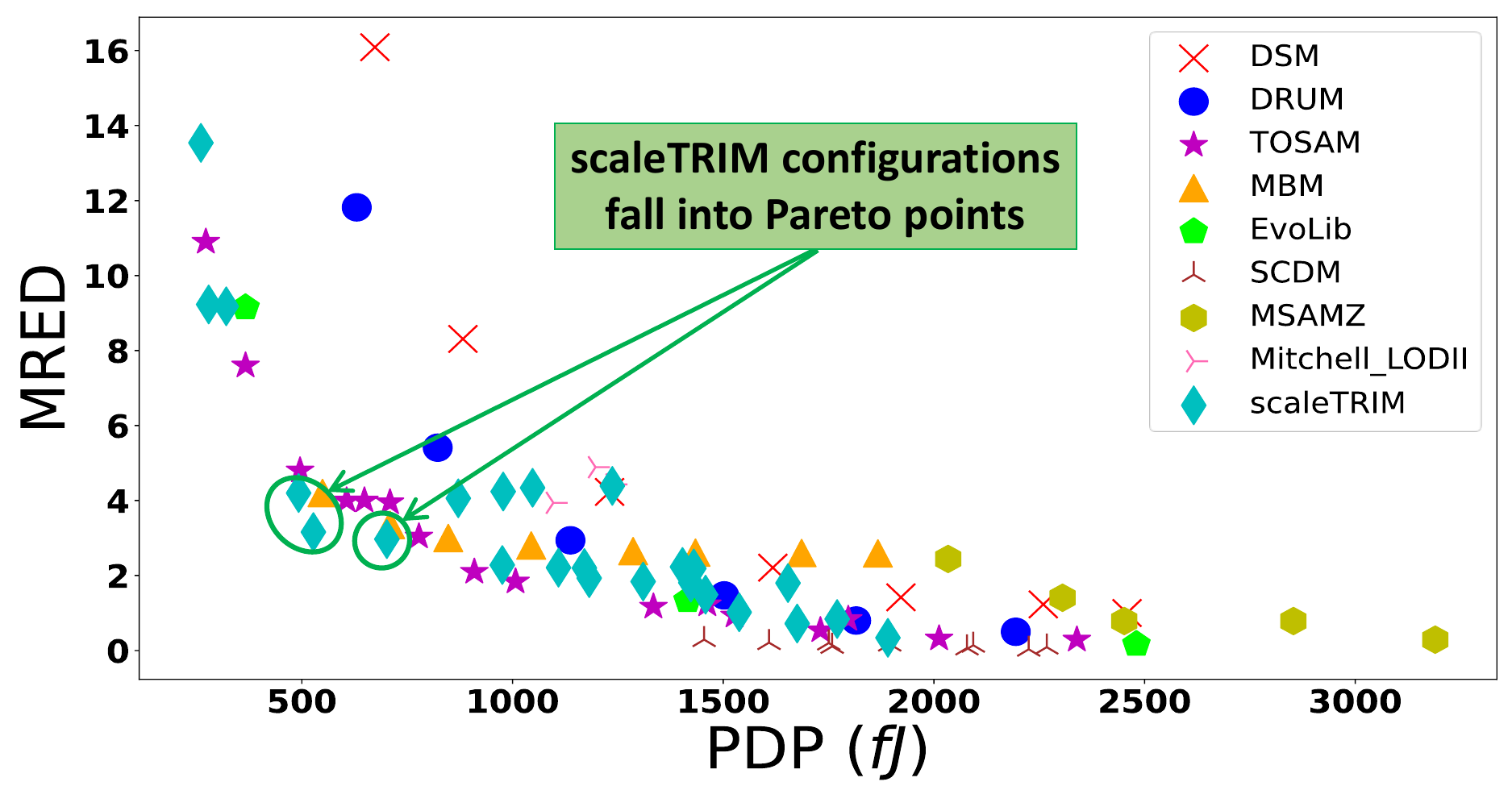}
    \label{PDP_16bit_DS}
    }
     \caption{Design space of comparison the 16-bit scaleTRIM with the state-of-the-art in terms of (a) Power vs. MRED, (b) Area vs. MRED, (c) Delay vs. MRED, and (d) PDP vs. MRED.}
    \label{DS_16bit}
    \vspace{-10pt}
\end{figure*}
\section{Results}
\label{sec:result}
In this section, we evaluate the proposed approximate multiplier (scaleTRIM) in terms of accuracy and efficiency, i.e.,  area, delay, power, and PDP. To facilitate this evaluation, we develop the software and hardware implementation of scaleTRIM. We compare the proposed approximate multiplier with several state-of-the-art approximate multipliers such as DRUM~\cite{7372600}, DSM~\cite{6858039}, Mitchell multiplier~\cite{5219391}, some Pareto-optimal (non-dominated) configurations from the EvoApproxLib benchmark library (EvoLib)~\cite{7926993} which fall into optimal points of Mean Relative Error (MRE) vs. Power plot in EvoLib, MBM~\cite{8493590}, AXM8~\cite{deepsita2023energy}, an Improved Logarithmic Multiplier for Energy-Efficient Neural Computing (ILM)~\cite{908674423}, Fast and low-power leading-one detectors for energy-efficient logarithmic computing (Mitchell\_LODII)~\cite{ansari2021fast}, SCDM~\cite{shakibhamedan2024ace}, MSAMZ~\cite{huang2024energy}, and TOSAM~\cite{8626488}.

\subsection{Accuracy Comparison}
\label{Sec:accuracy}
For error behavior analysis, we developed the scaleTRIM behavioral simulation model with PYTHON. The error metric, such as Mean Absolute Relative Error Distance (MRED), is used to report the accuracy of scaleTRIM.
MRED calculates the average absolute relative error between the approximate and exact results, ensuring that both overestimations and underestimations are equally considered, without being canceled out due to opposite signs. Thus, MRED offers a more accurate and application-relevant measure of approximation quality.
The Absolute Relative Error Distance (ARED) is obtained by Eq.~\ref{ARE}. The mean of ARED across $N$ input samples is known as MRED. Note that the MRED is reported as a percentage in the results.
\begin{equation}
    ARED_i=\lvert\frac{M_{App,i}-M_{Acc,i}}{ M_{Acc,i}}\rvert
    \label{ARE}
\end{equation}
Here $M_{Acc,i}$ and $M_{App,i}$ represent the exact and approximate values of the multiplier output of $i^{th}$ input pair, respectively. $ARED_i$ denotes ARED of $i^{th}$ input pair.

The accuracy of 8-bit scaleTRIM with different configurations is compared with the 8-bit state-of-the-art approximate multipliers in terms of MRED. Approximate multipliers such as DSM, DRUM, some optimal points of MRE vs. Power of EvoApprox library (EvoLib), MBM, ILM, Mitchell\_LODII, SCDM, MSAMZ, and TOSAM are implemented in different configurations to report their respective accuracies. The respective papers describe these configurations in detail. All accuracy values in Figs.~\ref{DS_8bit} and \ref{DS_16bit} are in percentage. Our evaluations show that the MRED of most scaleTRIM configurations offer better results than state-of-the-art approximate multipliers.
For example, with MRED constraint = $4\%$ and PDP constraint in the range of $200 fJ$ to $250 fJ$, the best configuration would be $scaleTRIM(4,8)$ (see pointer a in Fig.~\ref{PDP_8bit_DS}) with $MRED=3.34$ and $PDP=212.47 fJ$ which improves the MRED about $15.23\%$ compared to $TOSAM(1,5)$ with $MRED=4.06$ and $PDP=249.7 fJ$ (see pointer b in Fig.\ref{PDP_8bit_DS}).

\subsection{Hardware Comparison}
\label{Sec:hardware}
We compare the efficiency of scaleTRIM and the other state-of-the-art multipliers using Synopsys Design Compiler in a $freepdk-45nm$ Nangate technology. The metrics, such as area, delay, power, and PDP, are considered to compare the efficiency of scaleTRIM with other approximate multipliers. Synthesis is performed using the “compile\_ultra” command and targeting performance optimization. Post-synthesis timing simulations are performed using Modelsim to obtain precise switching activity. We simulate the examined multipliers for 100,000 random inputs to obtain accurate switching activity estimation. Then, Synopsys PrimeTime is used to calculate the power consumption. 

In Fig.~\ref{DS_8bit}, we can observe that 8-bit scaleTRIM achieves lower area,  power, and PDP in most configurations with the MRED constraint (e.g., MRED $\leq 4\%$). 
For example, with MRED constraint $\leq 4\%$ and PDP constraint = $200 fJ$, the best configuration would be $scaleTRIM(3,4)$ (see pointer c in Fig.~\ref{PDP_8bit_DS}) with $MRED=3.73$ and $PDP=153.75 fJ$ which improves the PDP about $22.8\%$ compared to $MBM-2$ with $MRED=3.74$ and $PDP=199.12 fJ$ (see pointer d in Fig.\ref{PDP_8bit_DS}). However, by noting the Figs.~\ref{Delay_8bit_DS} and \ref{Delay_16bit_DS}, the delay of scaleTRIM compared to other state-of-the-art approximate multipliers, such as TOSAM, can not offer the best configurations, since the TOSAM used LUT-based LODs to find the position of the leading-one, which used more area and power compared to the scaleTRIM. 
Additionally, some optimal 8-bit configurations in EVoLib exhibit superior delay performance when compared to our proposed approximate multiplier configurations, achieving an improvement of approximately $14.5\%$, on average. Nevertheless, 8-bit scaleTRIM configurations outperform in terms of power and area efficiency. Moreover, as observed in the PDP vs. MRED plot (Fig.~\ref{PDP_8bit_DS}), the scaleTRIM configurations fall into optimal points compared with other state-of-the-art approximate multipliers.

\begin{figure*}[t]
\centering
    \subfloat[Power vs. MED]{
    \includegraphics[width=0.45\textwidth]{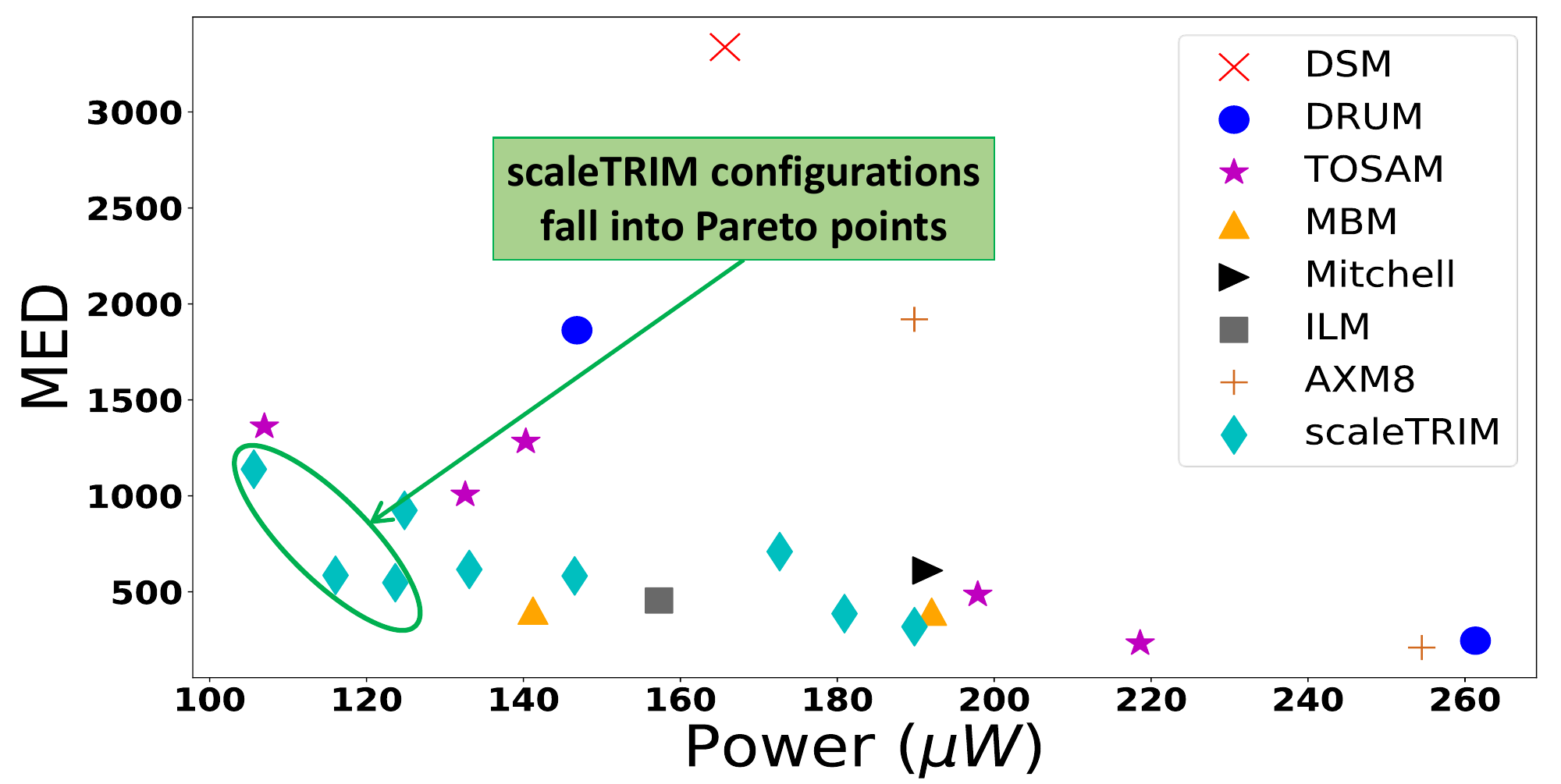}
    }
    \subfloat[Area vs. MED]{
    \includegraphics[width=0.45\textwidth]{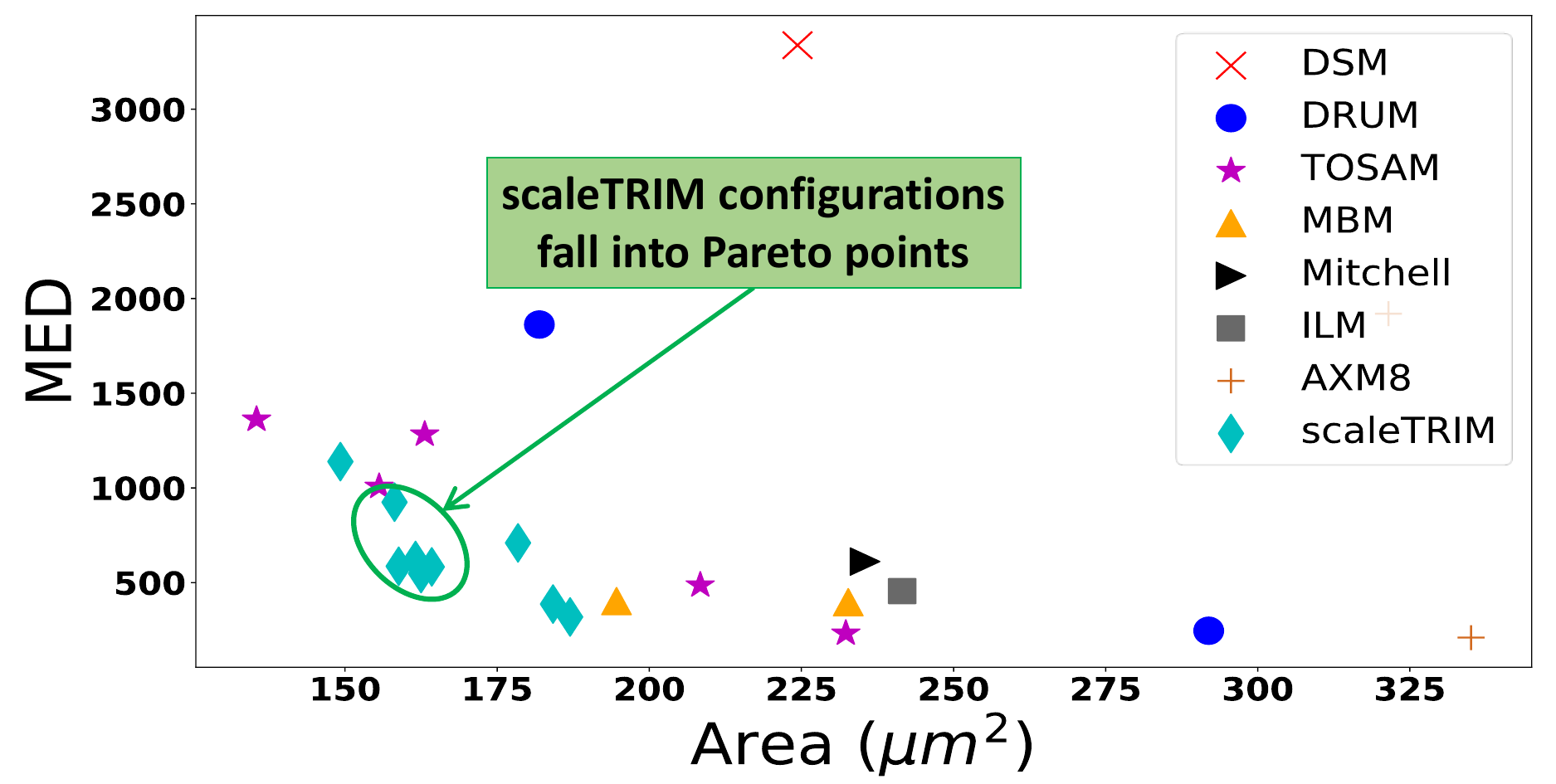}
    }\\
    \subfloat[Delay vs. MED]{
    \includegraphics[width=0.45\textwidth]{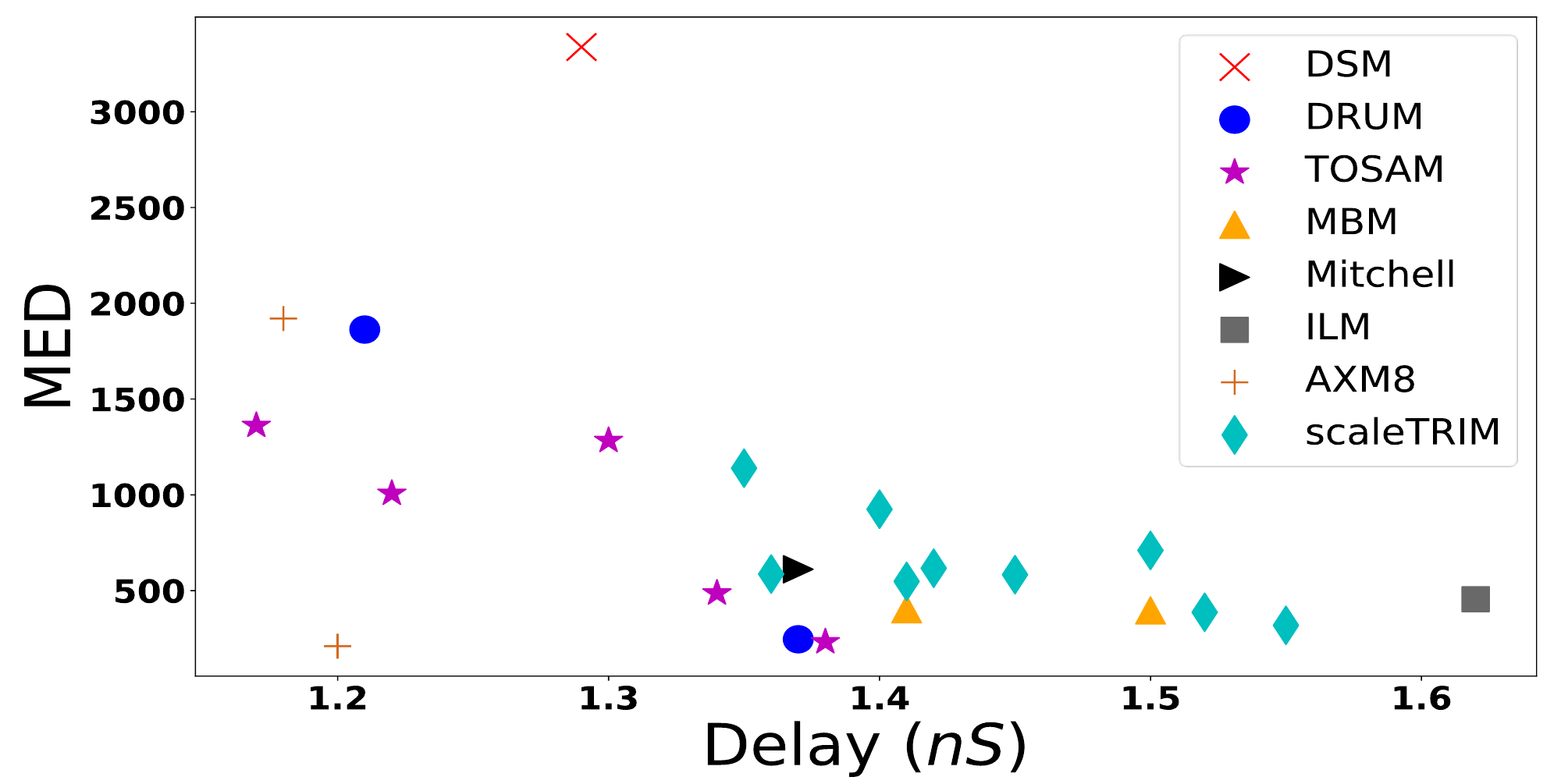}
    \label{MEDDelay_8bit_DS}
    }
    \subfloat[PDP vs. MED]{
    \includegraphics[width=0.45\textwidth]{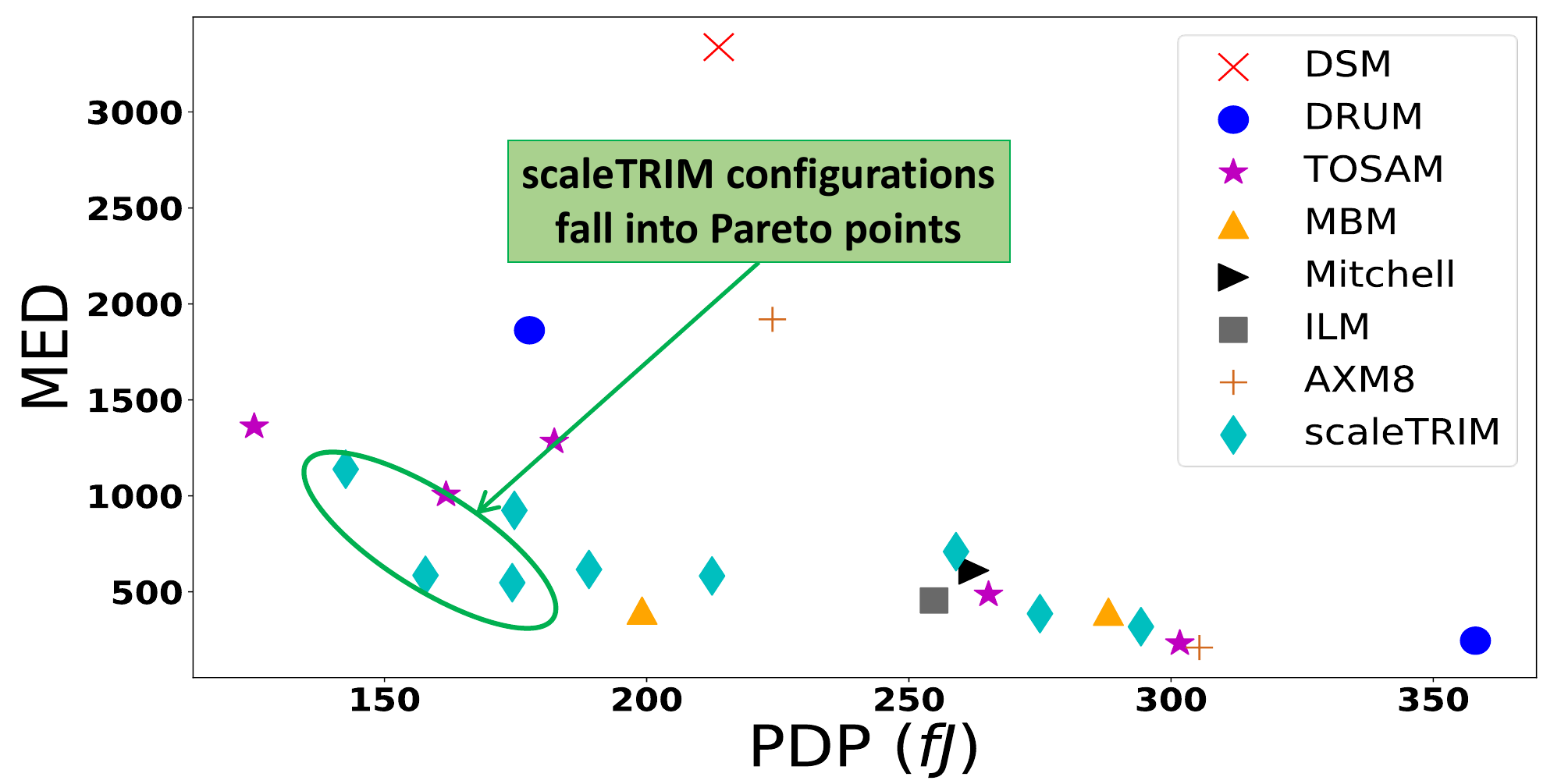}
    \label{MEDPDP_8bit_DS}
    }
     \caption{Design space of comparison of the 8-bit scaleTRIM with the state-of-the-art in terms of (a) Power vs. MED, (b) Area vs. MED, (c) Delay vs. MED, and (d) PDP vs. MED. The exact values of the figure are presented in Table~\ref{tab:detailed_metrics} in the Appendix.}
    \label{DS_MED_8bit}
    \vspace{-10pt}
\end{figure*}

\begin{figure*}[t]
\centering
    \subfloat[Power vs. Max Error]{
    \includegraphics[width=0.45\textwidth]{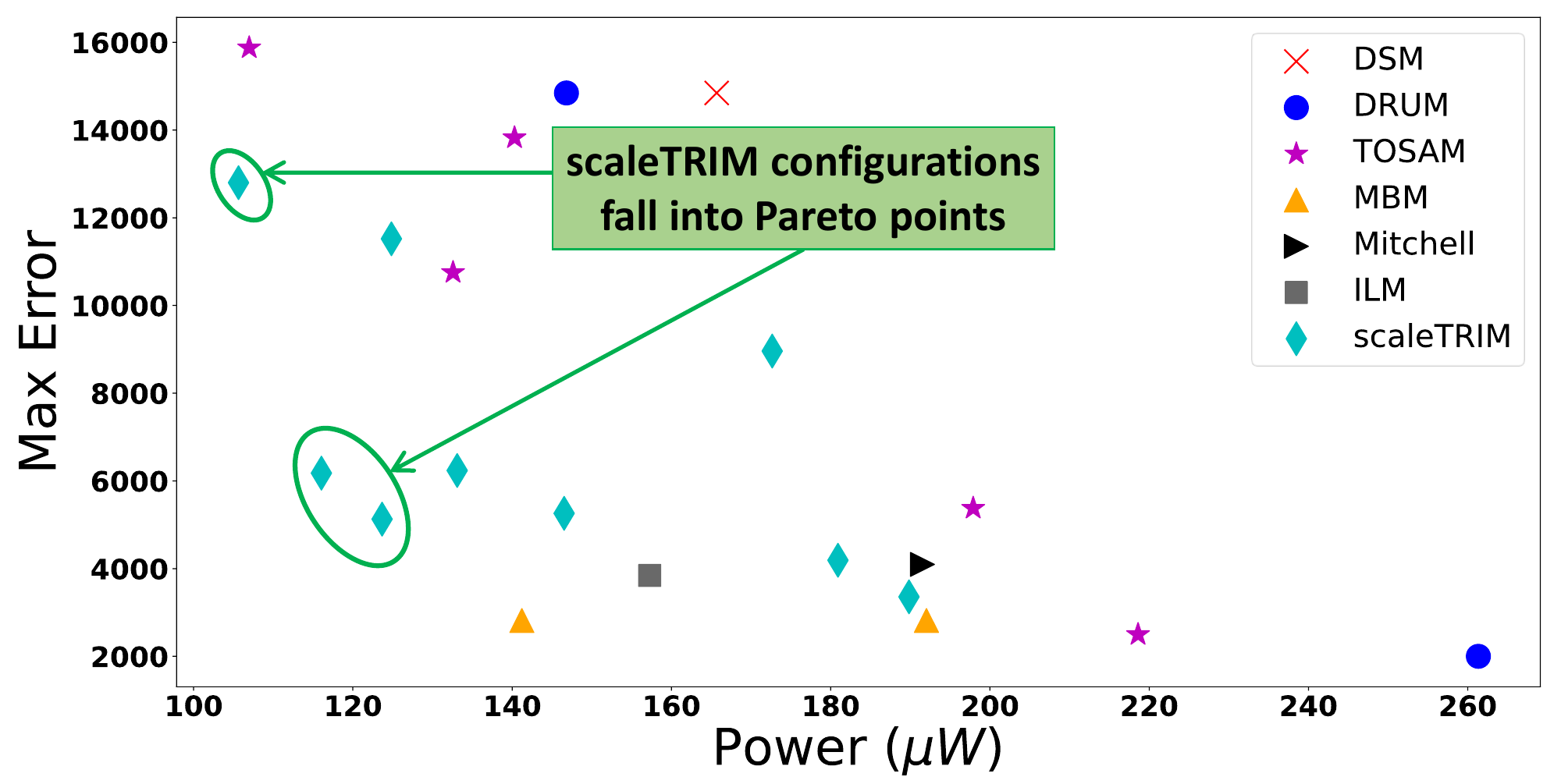}
    }
    \subfloat[Area vs. Max Error]{
    \includegraphics[width=0.45\textwidth]{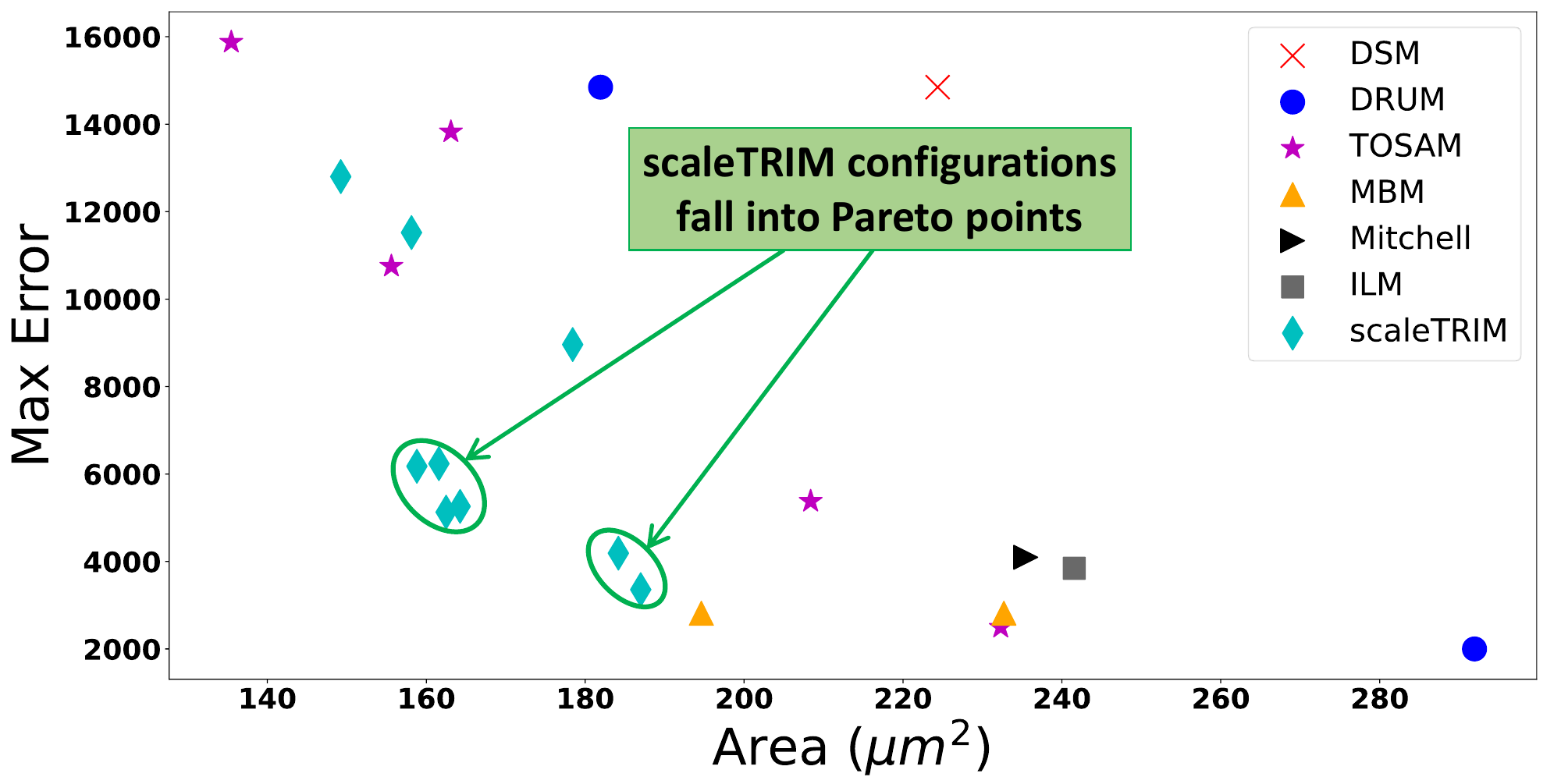}
    }\\
    \subfloat[Delay vs. Max Error]{
    \includegraphics[width=0.45\textwidth]{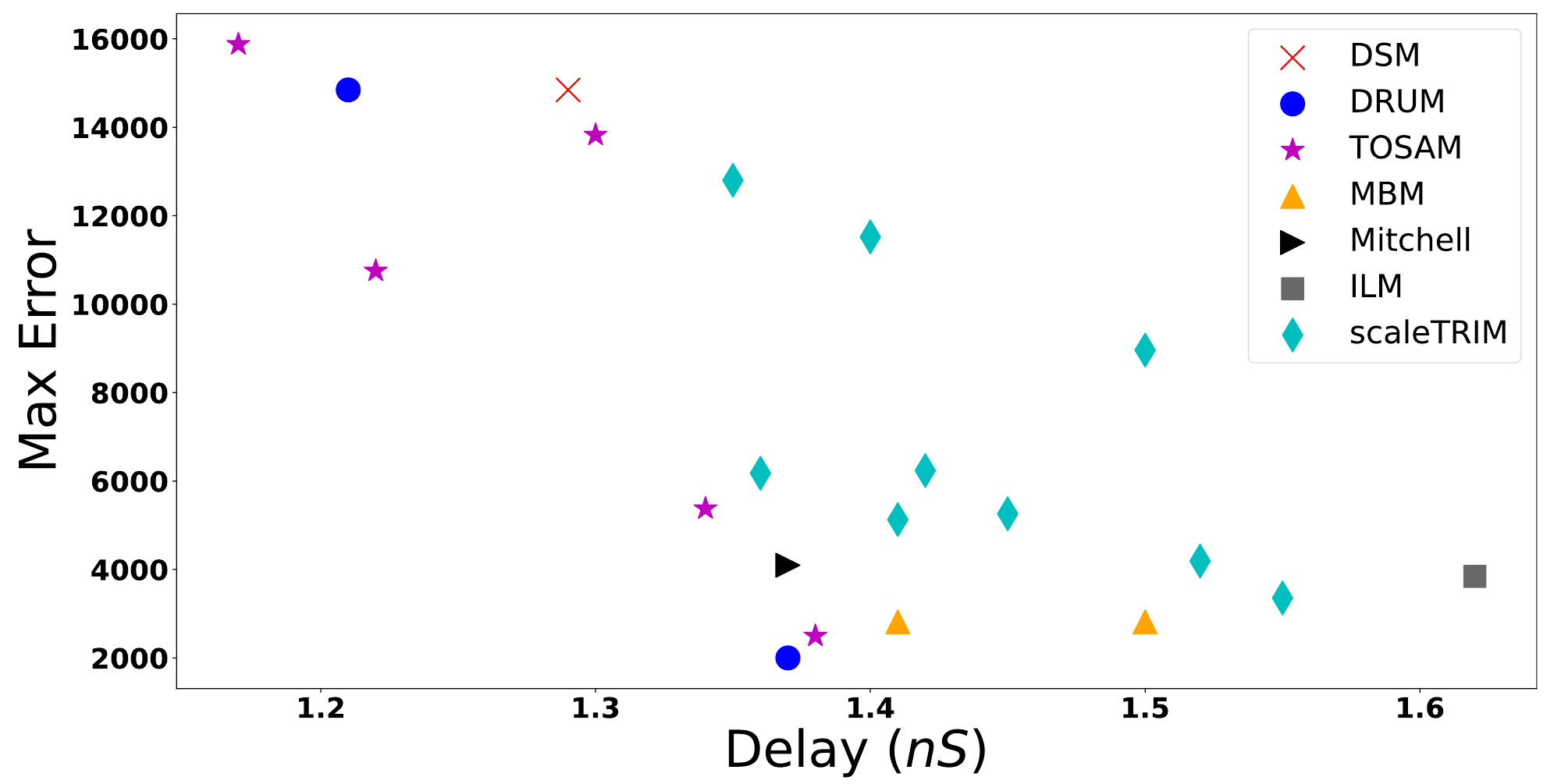}
    \label{MAXDelay_16bit_DS}
    }
    \subfloat[PDP vs. Max Error]{
    \includegraphics[width=0.45\textwidth]{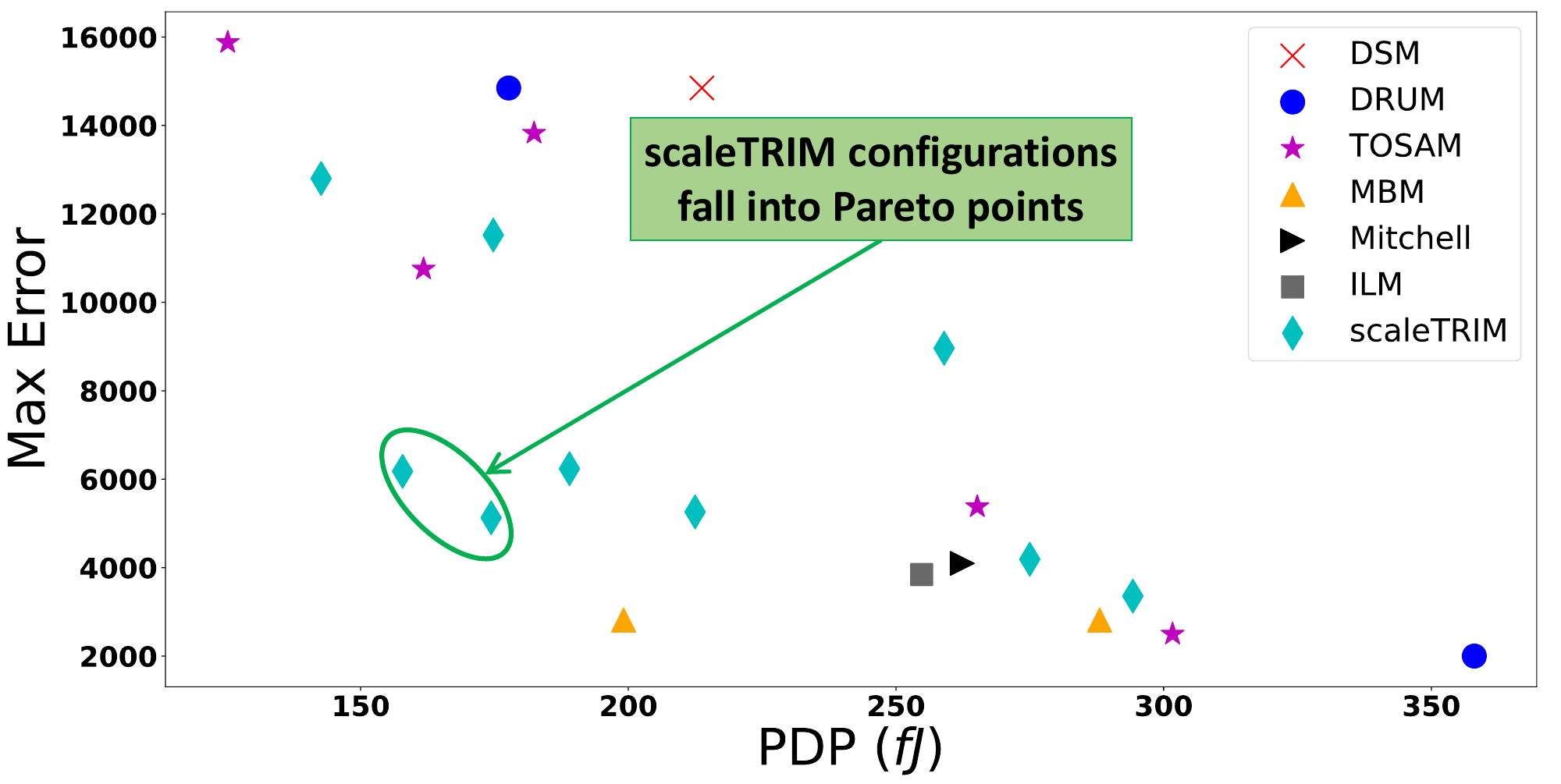}
    }
     \caption{Design space of comparison of the 8-bit scaleTRIM with the state-of-the-art in terms of (a) Power vs. Max Error, (b) Area vs. Max Error, (c) Delay vs. Max Error, and (d) PDP vs. Max Error. The exact values of the figure are presented in Table~\ref{tab:detailed_metrics} in the Appendix.}
    \label{Max_error_DS}
    \vspace{-10pt}
\end{figure*}

\begin{figure*}[t]
\centering
    \subfloat[Power vs. Std]{
    \includegraphics[width=0.45\textwidth]{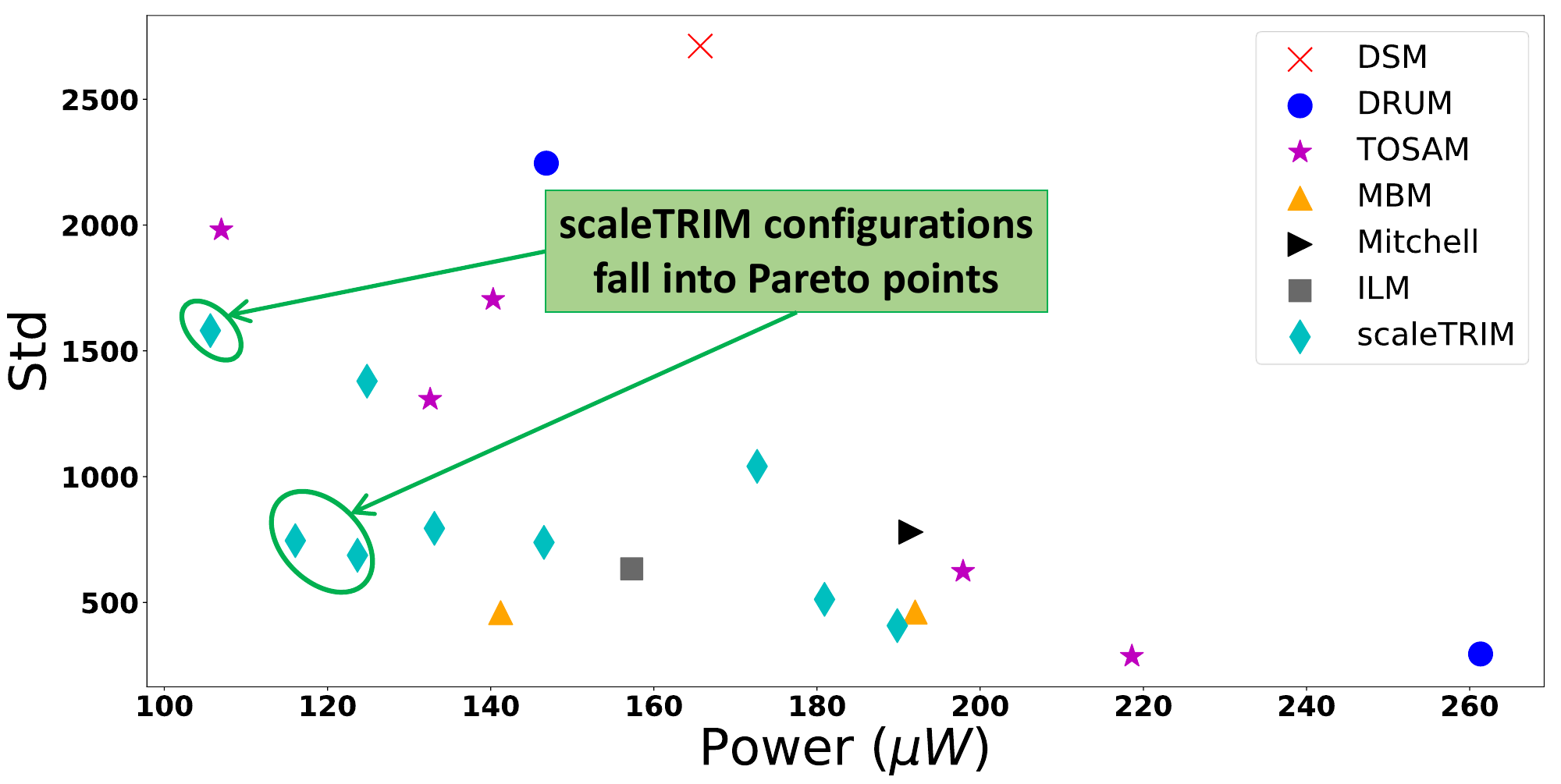}
    }
    \subfloat[Area vs. Std]{
    \includegraphics[width=0.45\textwidth]{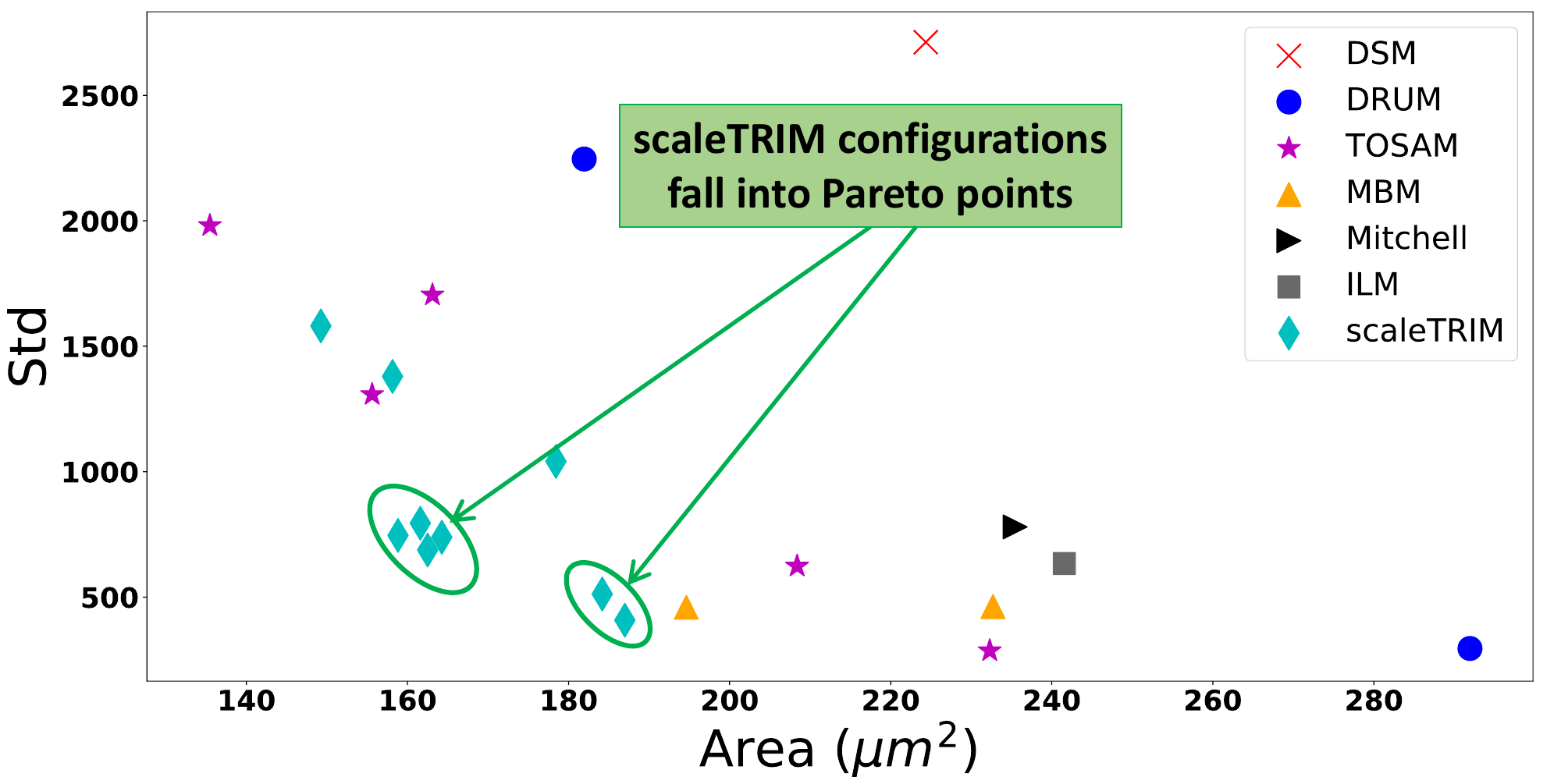}
    }\\
    \subfloat[Delay vs. Std]{
    \includegraphics[width=0.45\textwidth]{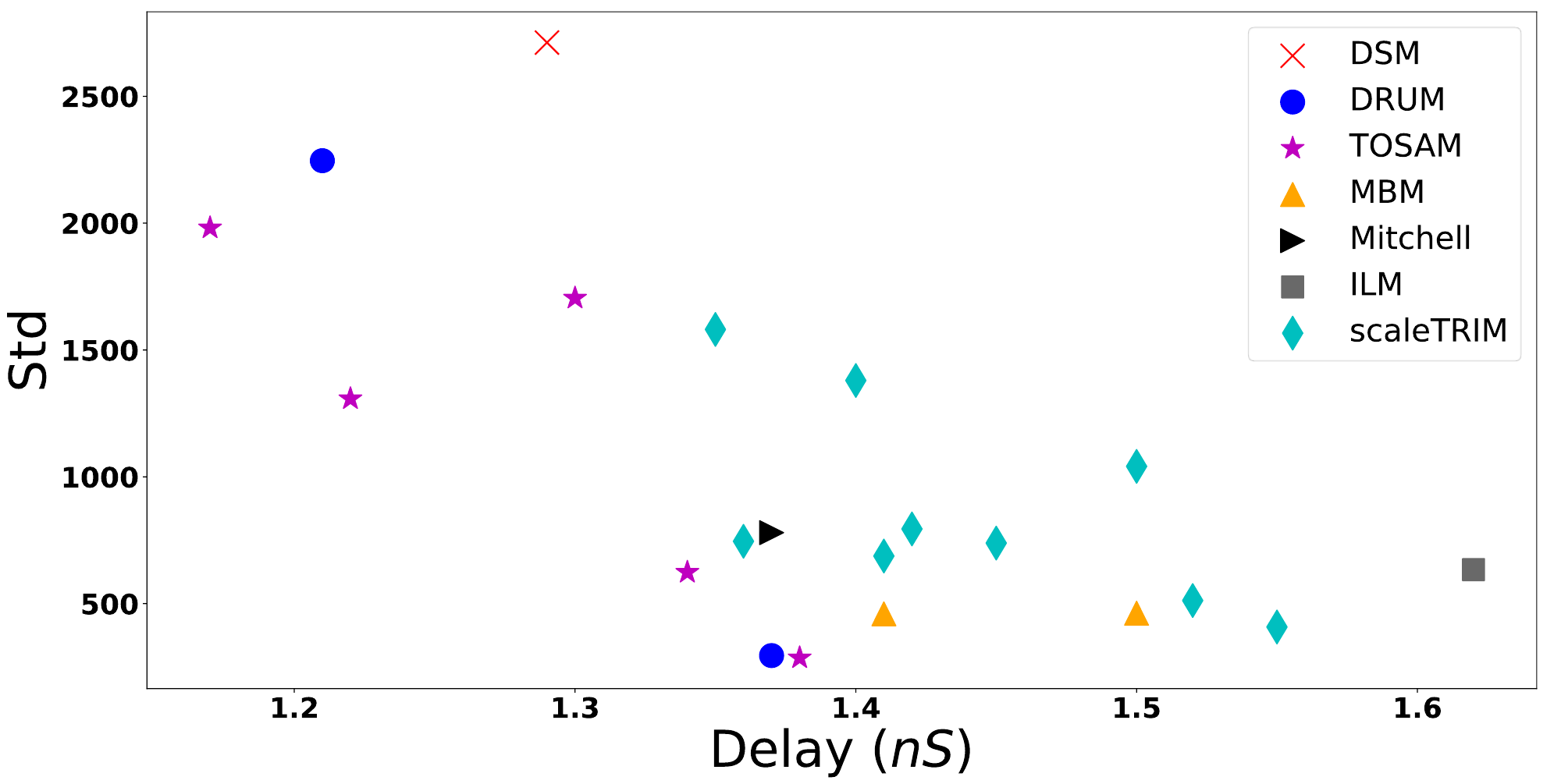}
    \label{Std_delay}
    }
    \subfloat[PDP vs. Std]{
    \includegraphics[width=0.45\textwidth]{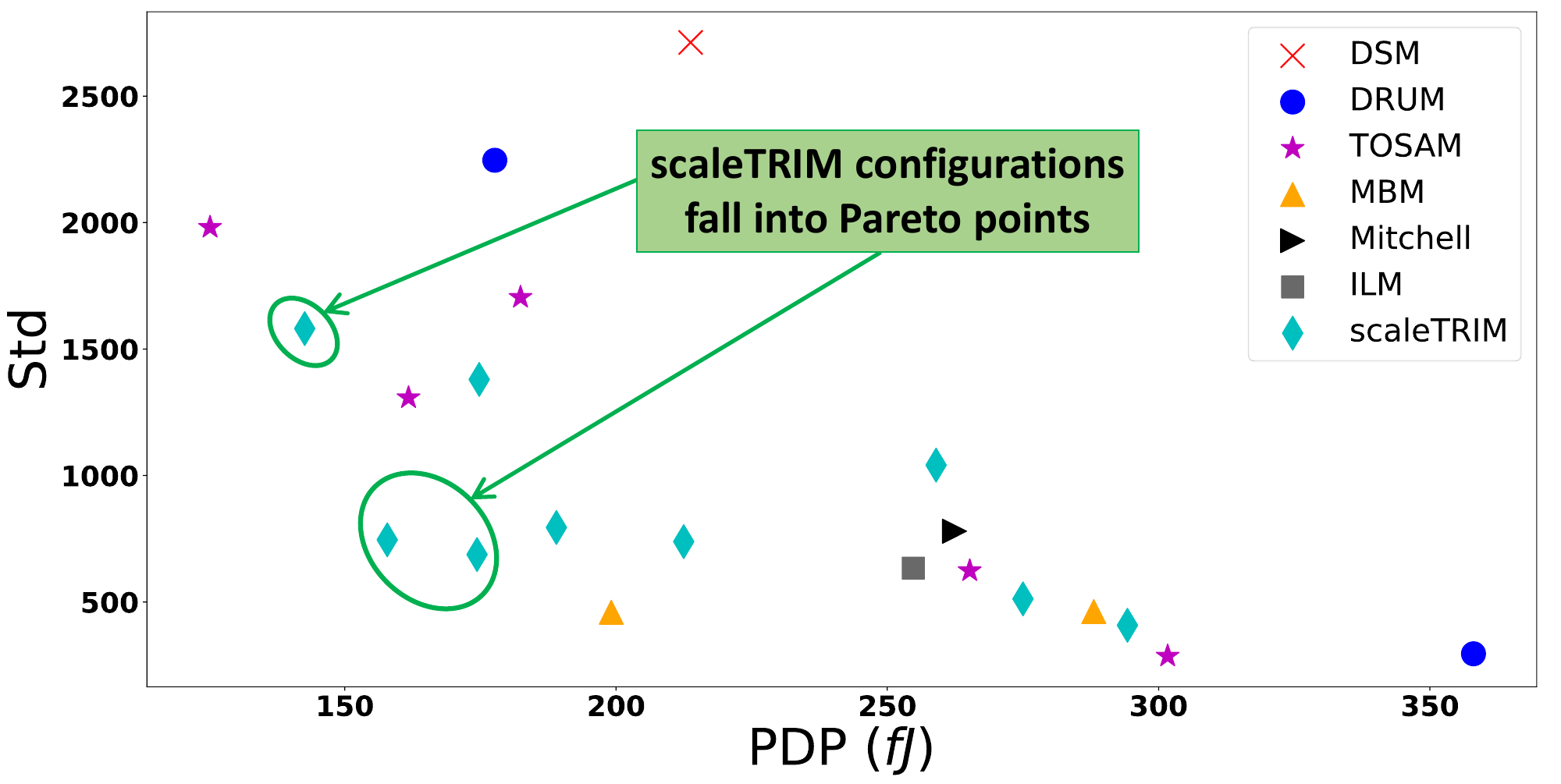}
    }
     \caption{Design space of comparison of the 8-bit scaleTRIM with the state-of-the-art in terms of (a) Power vs. Std, (b) Area vs. Std, (c) Delay vs. Std, and (d) PDP vs. Std. The exact values of the figure are presented in Table~\ref{tab:detailed_metrics} in the Appendix.}
    \label{Std_error_DS}
    \vspace{-10pt}
\end{figure*}

Moreover, we compare some configurations of scaleTRIM with other state-of-the-art approximate multipliers in terms of additional error metrics such as mean error distance (MED), error distance peak (Max-Error), and standard deviation of error distance (Std). Figs.~\ref{DS_MED_8bit}, \ref{Max_error_DS}, and \ref{Std_error_DS} show the comparison of scaleTRIM with other state-of-the-art approximate multipliers in terms of accuracy and performance design space. From Figs.~\ref{DS_MED_8bit}, \ref{Max_error_DS}, and \ref{Std_error_DS}, we can conclude that the scaleTRIM configurations will still fall on the Pareto front points in the design space of accuracy and efficiency.

\color{black}
\subsection{Design-space Comparison}

To further illustrate the flexibility of scaleTRIM, we analyze its design space through a comprehensive evaluation that considers both accuracy and hardware metrics. Unlike sections \ref{Sec:accuracy} and \ref{Sec:hardware}, which individually compare accuracy and hardware performance against prior works, this subsection focuses on the global trade-off trends enabled by the configurable parameters $h$ and $M$.

Figs.~\ref{DS_8bit} and~\ref{DS_16bit} present the complete design space for 8-bit and 16-bit implementations, respectively, showing the relationship between MRED and power, area, delay, and PDP. As observed, scaleTRIM configurations consistently fall into the Pareto frontier of the accuracy–efficiency space. This behavior confirms that the joint tuning of truncation width ($h$) and the number of compensation segments ($M$) enables systematic navigation of trade-offs without requiring structural redesign.  

For 8-bit operands, moderate values of $h$ (e.g., $h=3$ to 4) combined with higher segmentation ($M=4$ or 8) provide balanced operating points with competitive MRED and significantly reduced PDP. Similarly, in the 16-bit case, expanding the admissible range of $h$ allows finer tuning of accuracy, while increasing $M$ improves error-correction granularity at the cost of a modest increase in LUT size. In terms of delay, scaleTRIM generally provides competitive delay performance; however, in latency-critical scenarios, certain baseline designs, such as TOSAM, may be preferable, as their LUT-based LOD implementation can yield lower delay. 

It is important to emphasize that $M$ represents the number of compensation segments used to partition the $X_h+Y_h$ space for piecewise constant error correction.
Increasing $M$ refines the approximation by enabling more localized compensation, thereby reducing average error, while introducing proportional storage overhead. Overall, the results demonstrate that scaleTRIM offers a structured, scalable mechanism for exploring the accuracy–efficiency trade-off across different operand sizes.

Table~\ref{tab:pareto-configs} summarizes the top Pareto-optimal configurations for scaleTRIM and baseline designs.
The configurations reported in Table 2 are selected from the non-dominated design points identified in the complete evaluated design space. For the 8-bit case, we focused on the region defined by $MRED \leq 4\%$ and $200 fJ \leq PDP \leq 250 fJ$ (see Fig.~\ref{PDP_8bit_DS}), and selected the Pareto-front configuration within this region that provides the most favorable trade-off. For the 16-bit case, a representative Pareto-optimal configuration is selected from the PDP vs. MRED design space (Fig.~\ref{PDP_16bit_DS}), and baseline designs with comparable MRED are used for metric comparison. These configurations represent the best trade-offs identified in our design space exploration and can serve as starting points for mapping to application requirements.

\begin{table}[ht]
\centering
\caption{Pareto-optimal configurations for scaleTRIM and SOTA designs. ST = $scaleTRIM(h,M)$.}
\scriptsize
\resizebox{\columnwidth}{!}{%
\begin{tabular}{llccccc}
\toprule
\textbf{Bit-} & \textbf{Config.} & \textbf{MRED} & \textbf{Power} & \textbf{Area} & \textbf{Delay} & \textbf{PDP} \\
\textbf{width} &                 & \textbf{(\%)}  & \textbf{($\mu$W)} & \textbf{($\mu$m$^2$)} & \textbf{(ns)} & \textbf{(fJ)} \\
\midrule
8-bit  & ST(4,8)     & 3.34 & 146.53 & 162.26 & 1.45 & 212.47 \\
8-bit  & TOSAM(1,5)  & 4.06 & 182.28 & 193.32 & 1.37 & 249.72 \\
8-bit  & MBM-2       & 3.74 & 141.22 & 194.62 & 1.41 & 199.12 \\
16-bit & ST(5,8)     & 2.97 & 323.42 & 468.21 & 2.17 & 701.82 \\
16-bit & TOSAM(1,6)	 & 3.04	& 429.83 &	586.47 & 1.81 & 777.99\\
16-bit & DRUM(5)     & 2.94 & 466.2 & 514.9 & 2.44 & 1137.52 \\
\bottomrule
\end{tabular}%
}
\label{tab:pareto-configs}
\end{table}

It is important to note that although extending this investigation to 32-bit operands would provide additional insight, the preprocessing required to compute the piecewise compensation values ($M$) demands significant time and memory resources, making such an evaluation impractical within our current scope.

\subsection{Comparison of Linearization, Logarithmic Approximation, and Piecewise Linearization}

In this section, scaleTRIM is evaluated alongside two widely studied approximation methods: the Mitchell multiplier~\cite{5219391}, representing logarithmic approaches, and piecewise linearization~\cite{8806911}. The evaluation examines both error distribution characteristics and hardware performance metrics, offering insights into the relative strengths, limitations, and trade-offs of each method.

The Mitchell algorithm employs a logarithmic formulation to approximate multiplication, as expressed in Eq.~\ref{eq::Mitchell}:
\begin{equation}
    {\log}_2\left(M_{\mathrm{APP}}\right) = n_A + n_B + X + Y
    \label{eq::Mitchell}
\end{equation}

The approximate product $M_{\mathrm{APP}}$ is then obtained by computing the antilogarithm of Eq.~\ref{eq::Mitchell} according to Eq.~\ref{eq:antilog}:
\begin{equation}
M_{\mathrm{APP}} = 
\begin{cases} 
2^{n_A+n_B}(1 + X+Y), & X+Y< 1, \\
2^{n_A+n_B+1}(X+Y), & X+Y \geq 1.
\end{cases}
\label{eq:antilog}
\end{equation}

Piecewise linearization approximate method divides the input space into $S$ segments, fitting a separate linear model for each segment as computed by Eq.~\ref{eq:segment}.
\begin{equation}
    A \times B \approx \alpha_s(X_h+Y_h)+\beta_s, \quad s\in \{1, \dots, S\}.
    \label{eq:segment}
\end{equation}
Although this approach improves local fitting accuracy within each segment, it also increases hardware complexity in proportion to $S$ due to the additional storage and selection logic required.

\begin{figure*}[ht]
    \centering
    \includegraphics[scale=0.5]{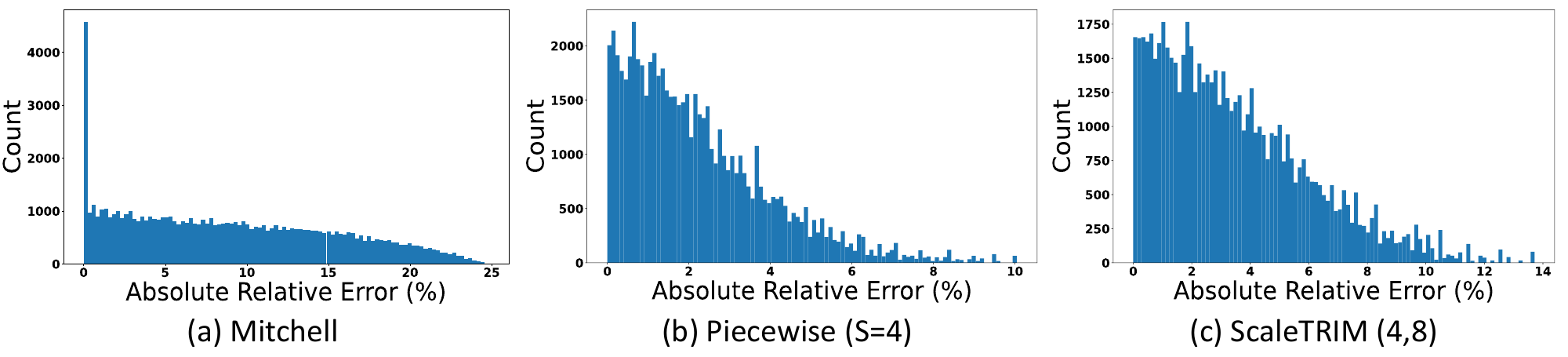}
    \caption{Error distribution histograms (absolute relative error) for 8-bit operands: (a) Mitchell, (b) Piecewise linearization with $S=4$ segments, and (c) $scaleTRIM(4,8)$. The vertical axis (Count) represents the number of operand pairs where the absolute relative error falls within each bin.}
    \label{histogram}
\end{figure*}

To evaluate error behavior, we examined representative configurations of each method for 8-bit operands. 
Fig.~\ref{histogram} presents the error histograms for the three methods. As observed, the Mitchell approximation exhibits a noticeably wider error distribution with heavier tails. In contrast, both the 4-segment piecewise linearization and scaleTRIM concentrate the majority of errors within the lower ARED range. While the piecewise approach shows a slightly tighter upper bound in Fig.~\ref{histogram}, the distribution of scaleTRIM remains similarly compact and avoids the pronounced tail behavior observed in the logarithmic approximation. 
Table~\ref{tab:method_comparison} reports the error statistics and hardware metrics for these configurations, synthesized using the same $45\,nm$ technology and toolflow described in Section~\ref{sec:result}.


\begin{table*}[ht]
\centering
\caption{Error statistics and hardware metrics for three approximation methods over the full 8-bit operand space (excluding zero).}
\scriptsize
\begin{tabular}{lccccccccccc}
\toprule
\textbf{Method} & \textbf{Mean \%} & \textbf{Median \%} & \textbf{95th \%} & \textbf{99th \%} & \textbf{Max \%} & \textbf{MRED (\%)} & \textbf{Area ($\mu$m$^2$)} & \textbf{Power ($\mu$W)} & \textbf{Delay (ns)} & \textbf{PDP (fJ)} \\
\midrule
$scaleTRIM(4,8)$     & 2.36 & 1.96 & 5.97 & 8.32 & 10.95 & 3.34 & 162.26 & 146.53 & 1.45 & 212.47 \\
Mitchell           & 8.91 & 8.17 & 20.34 & 22.87 & 24.80 & 3.76 & 235.45 & 191.52 & 1.37 & 262.38 \\
Piecewise (S=4)    & 2.23 & 1.82 & 5.72 & 7.89 & 10.04 & 3.25 & 210.18 & 172.11 & 1.49 & 256.44 \\
\bottomrule
\end{tabular}
\label{tab:method_comparison}
\end{table*}

The results indicate that scaleTRIM delivers a favorable accuracy–efficiency trade-off. Compared with the Mitchell logarithmic approximation, scaleTRIM yields a substantially tighter error distribution and a large reduction in peak error (Mitchell max error = 24.80\% vs. scaleTRIM max error = 10.95\%). Compared with a 4-segment piecewise linearization, scaleTRIM achieves a comparable average error (MRED 3.34\% vs. 3.25\%) while requiring substantially less area (about 22.8\% reduction in area for the configuration reported in Table
~\ref{tab:method_comparison}), underscoring the design’s improved hardware efficiency for similar accuracy.


\subsection{scaleTRIM for CNNs}

An evaluation is conducted to investigate the impact of scaleTRIM on the performance of the employed CNNs at a higher abstraction level (architectural level). We examined the effect of these multipliers on key performance metrics, including classification accuracy. The performance metrics of scaleTRIM configurations are reported in Fig.~\ref{DS_8bit}. Moreover, to report classification accuracy for the CNNs, we generated behavioral descriptions of these multipliers in Python and integrated them into the CNNs used.

In this study, we took advantage of the flexibility provided during the training phase by using several pre-trained convolutional neural network (CNN) models available in the PyTorch library and the Adapt framework. The selected architectures, including LeNet-5, VGG19, ResNet-18, ResNet-50, and SqueezNet, are widely recognized for their strong performance across diverse computer vision tasks. These models are pre-trained on benchmark datasets such as CIFAR-10 and ImageNet.
To enable efficient inference, we applied post-training quantization (PTQ), converting all model parameters and activations from 32-bit floating-point (float32) to 8-bit integer (int8) representation. To evaluate the effect of approximate arithmetic, we integrated our proposed approximate multipliers into these quantized models by replacing all exact multiplications, without applying any additional fine-tuning.

The scaleTRIM approximate multiplier is designed to be compatible with DNN accelerators, particularly within MAC units, which form the core computational blocks of most neural network architectures. To evaluate the impact of approximate arithmetic, we integrated behavioral models of the examined multipliers into the inference pipeline using the Adapt framework~\cite{9913212}, a PyTorch-based environment for fast emulation of DNN accelerators employing approximate computing units. All exact multiplications in the quantized models are replaced with their approximate counterparts without additional fine-tuning, allowing direct assessment of the effect of arithmetic approximation on inference accuracy.



\begin{figure*}[t]
\centering
    \subfloat[Accuracy vs. PDP, LeNet-5]{
    \includegraphics[width=0.45\textwidth]{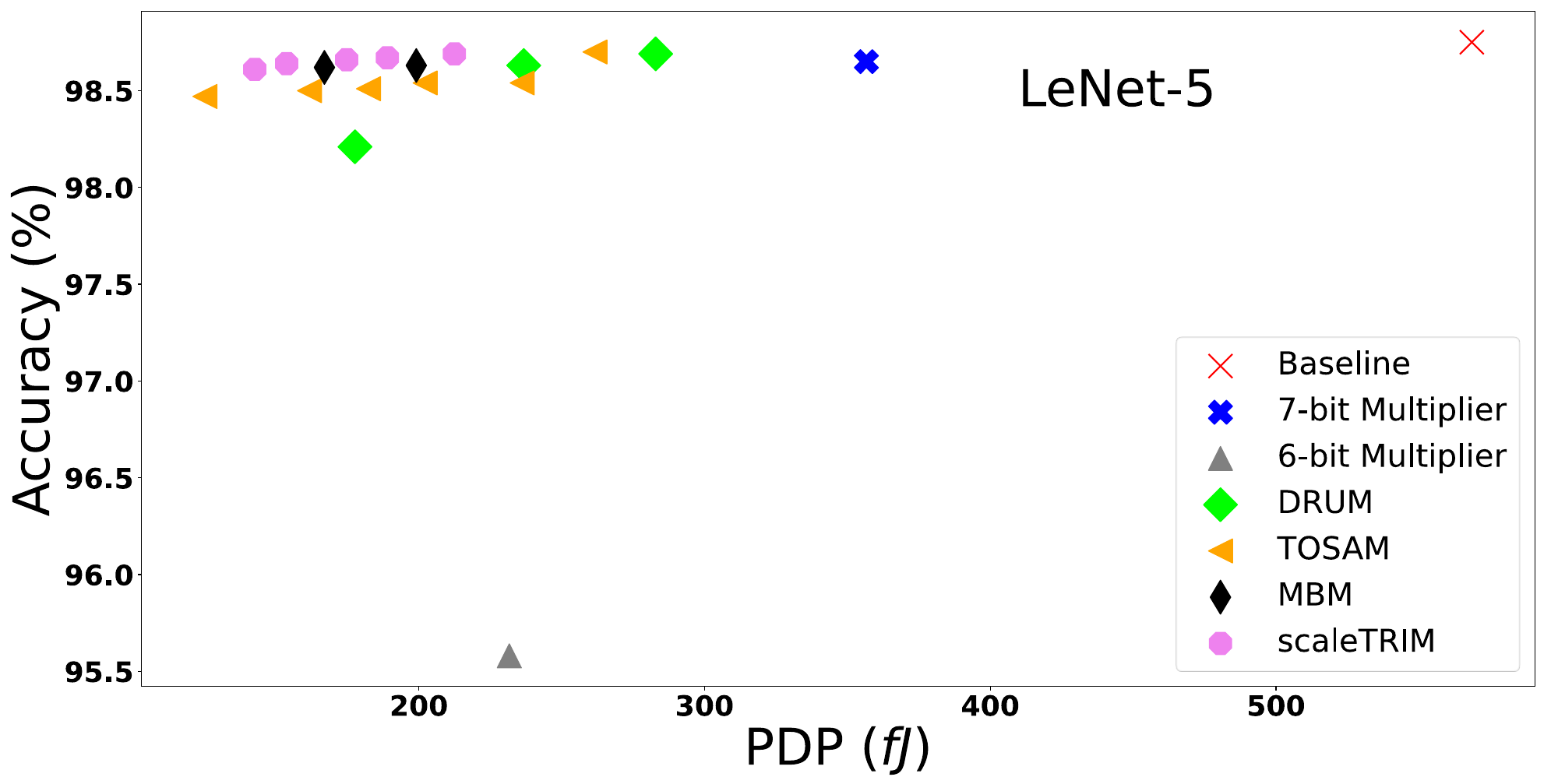}
    }
    \subfloat[Accuracy vs. PDP, VGG19]{
    \includegraphics[width=0.45\textwidth]{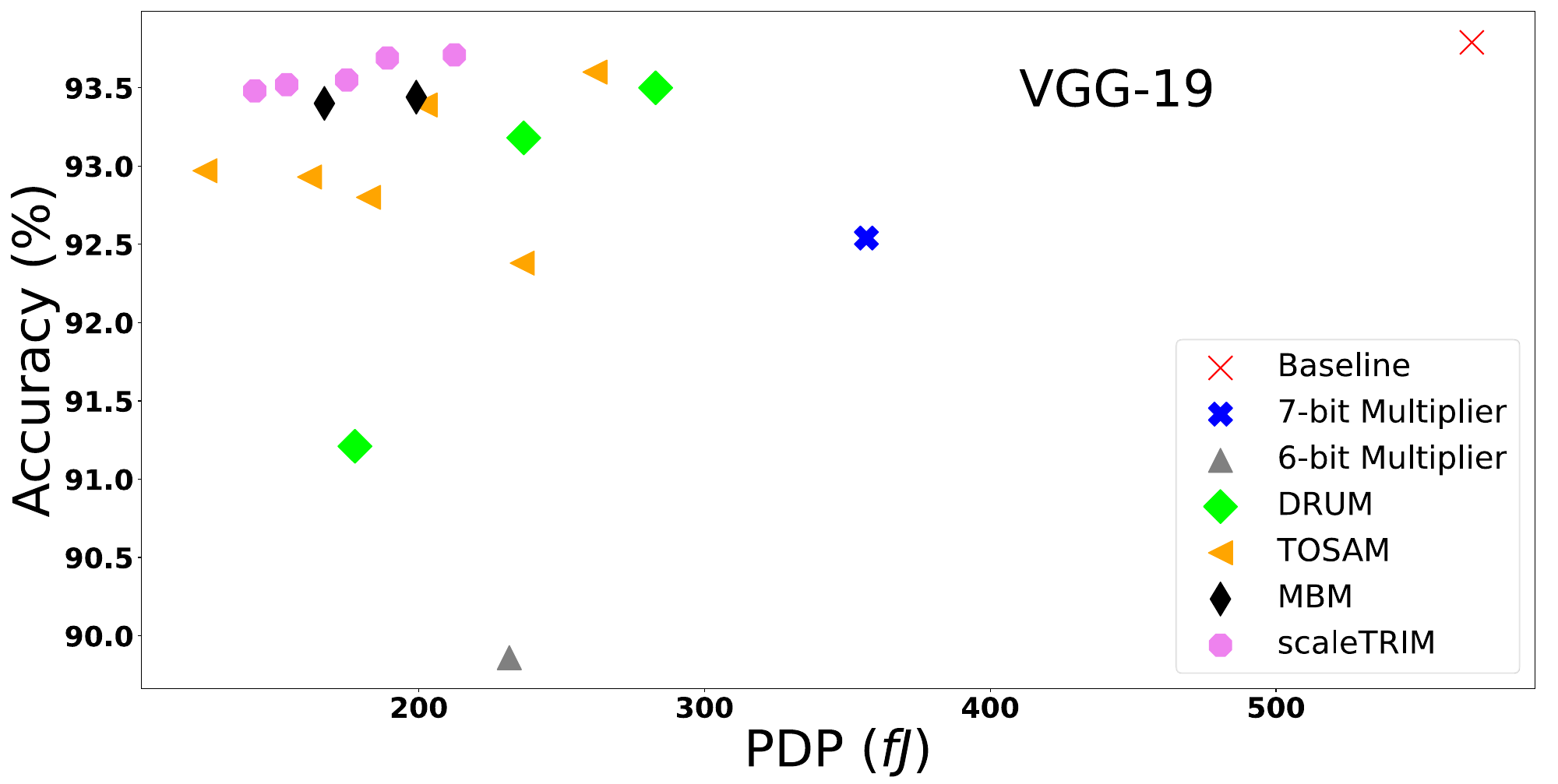}
    }\\
    \subfloat[Accuracy vs. PDP, ResNet-18]{
    \includegraphics[width=0.45\textwidth]{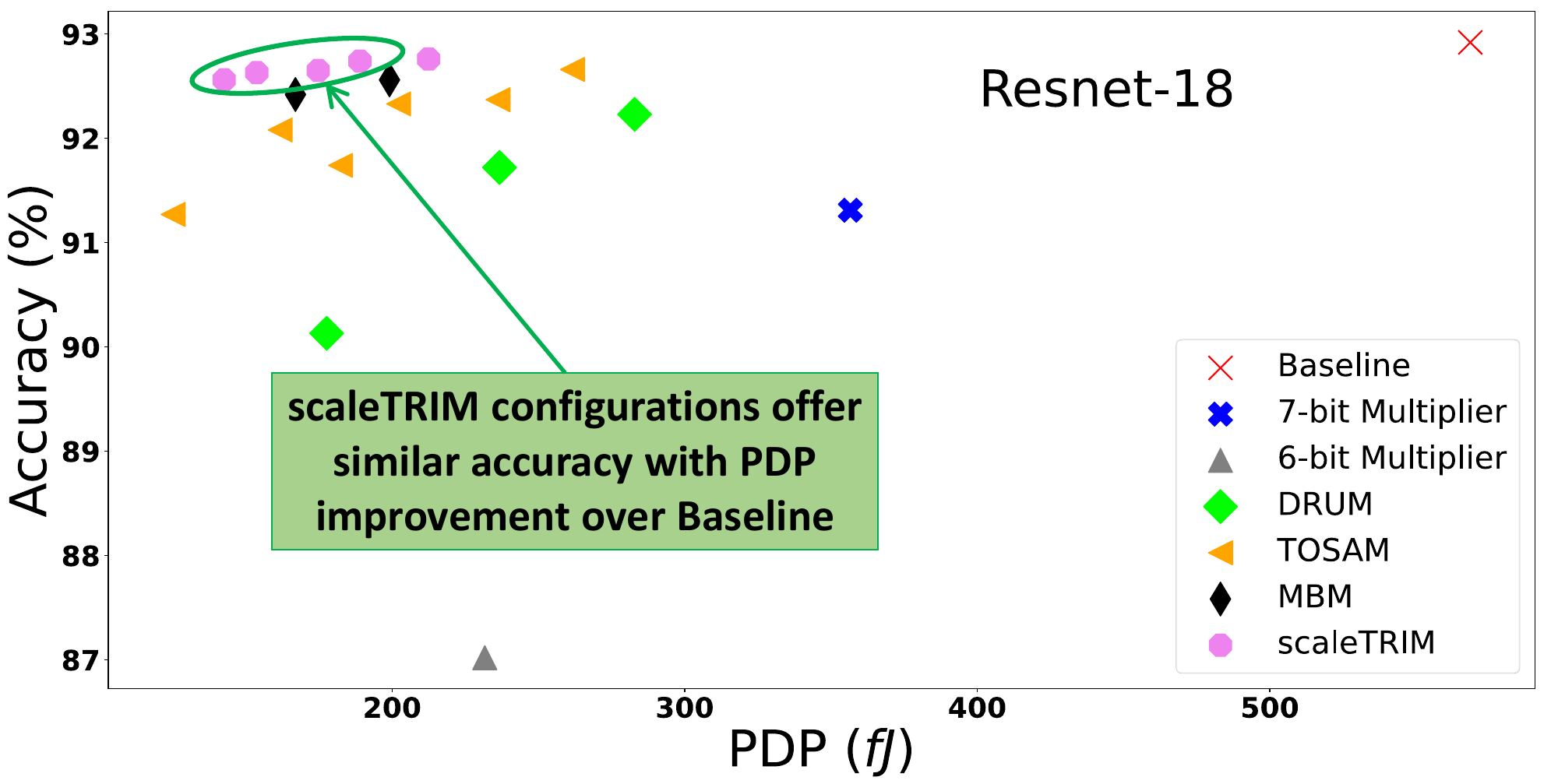}
    \label{Std_delay}
    }
    \subfloat[Accuracy vs. PDP, ResNet-50]{
    \includegraphics[width=0.45\textwidth]{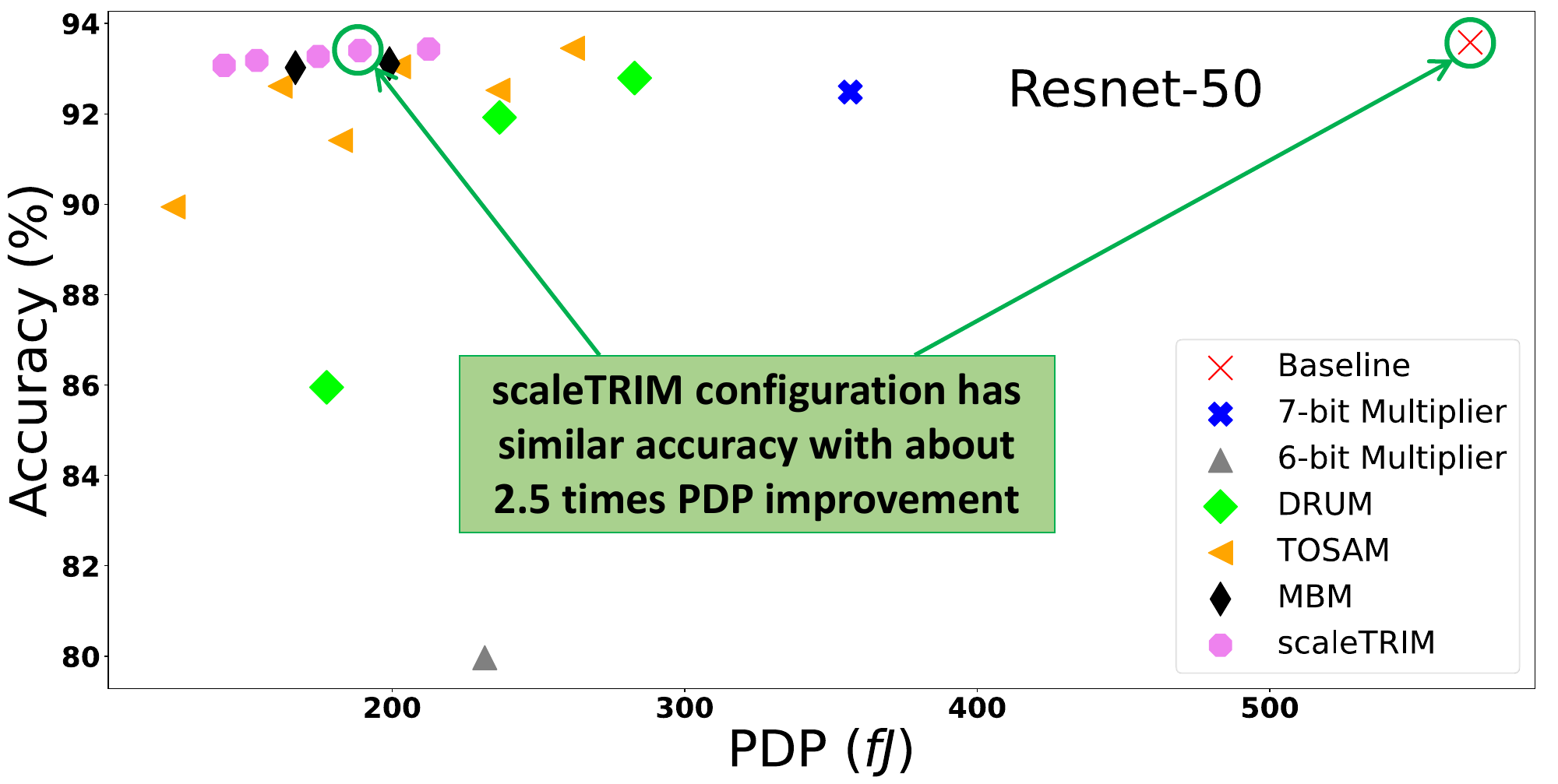}
    }
     \caption{Accuracy vs. PDP comparison of various approximate multipliers for CNN inference. The comparison includes: (a) LeNet-5 with MNIST, (b) VGG19 with CIFAR-10, (c) ResNet-18 with CIFAR-10, and (d) ResNet-50 with CIFAR-10. The evaluation covers multiple configurations of state-of-the-art approximate multipliers, accurate multipliers, and several 8-bit configurations of our scaleTRIM multiplier design.}
    \label{CNN_ACC}
    \vspace{-10pt}
\end{figure*}

We evaluated representative CNN architectures across standard benchmark datasets to analyze the accuracy–efficiency trade-off under approximate multiplication. The performance of scaleTRIM is compared with accurate and state-of-the-art approximate multipliers in terms of classification accuracy and hardware-efficiency metrics. Fig.~\ref{CNN_ACC} presents a comparison of our scaleTRIM with accurate and state-of-the-art approximate multipliers for efficient yet accurate CNN inference across five different CNN models, i.e., LeNet-5 using MNIST dataset, VGG-19, ResNet-18, and ResNet-50, using CIFAR-10 datasets, and SqueezeNet using ImageNet dataset. The evaluation indicates that a CNN with scaleTRIM configurations, such as $scaleTRIM(4,4)$, which has approximately 2.5 times lower PDP than the accurate multiplier, achieves almost the same accuracy as a CNN with an accurate multiplier.


\begin{figure*}[t]
\centering
    \subfloat[Top-1 Accuracy vs. PDP]{
    \includegraphics[width=0.45\textwidth]{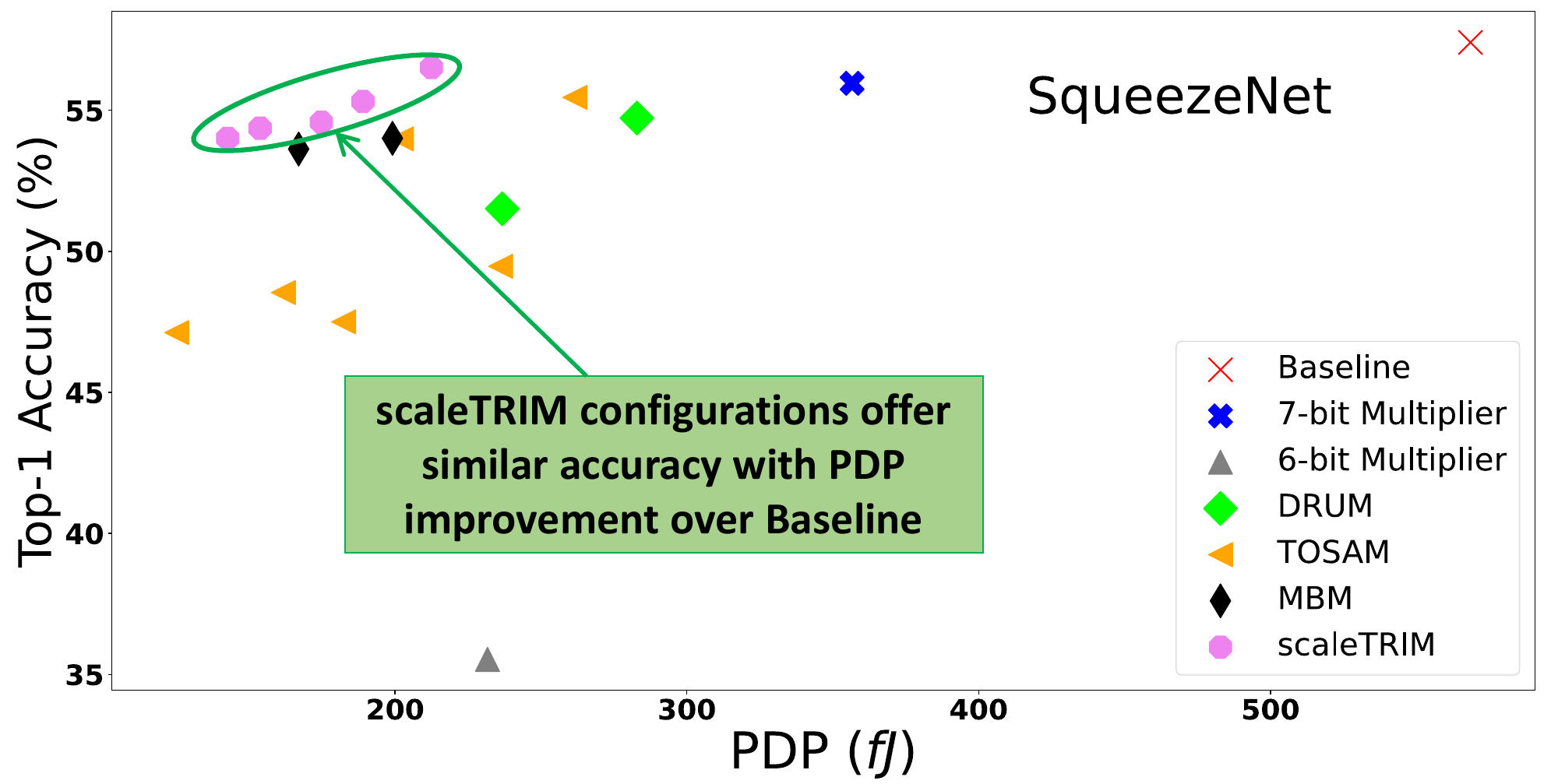}
    }
    \subfloat[Top-5 Accuracy vs. PDP]{
    \includegraphics[width=0.45\textwidth]{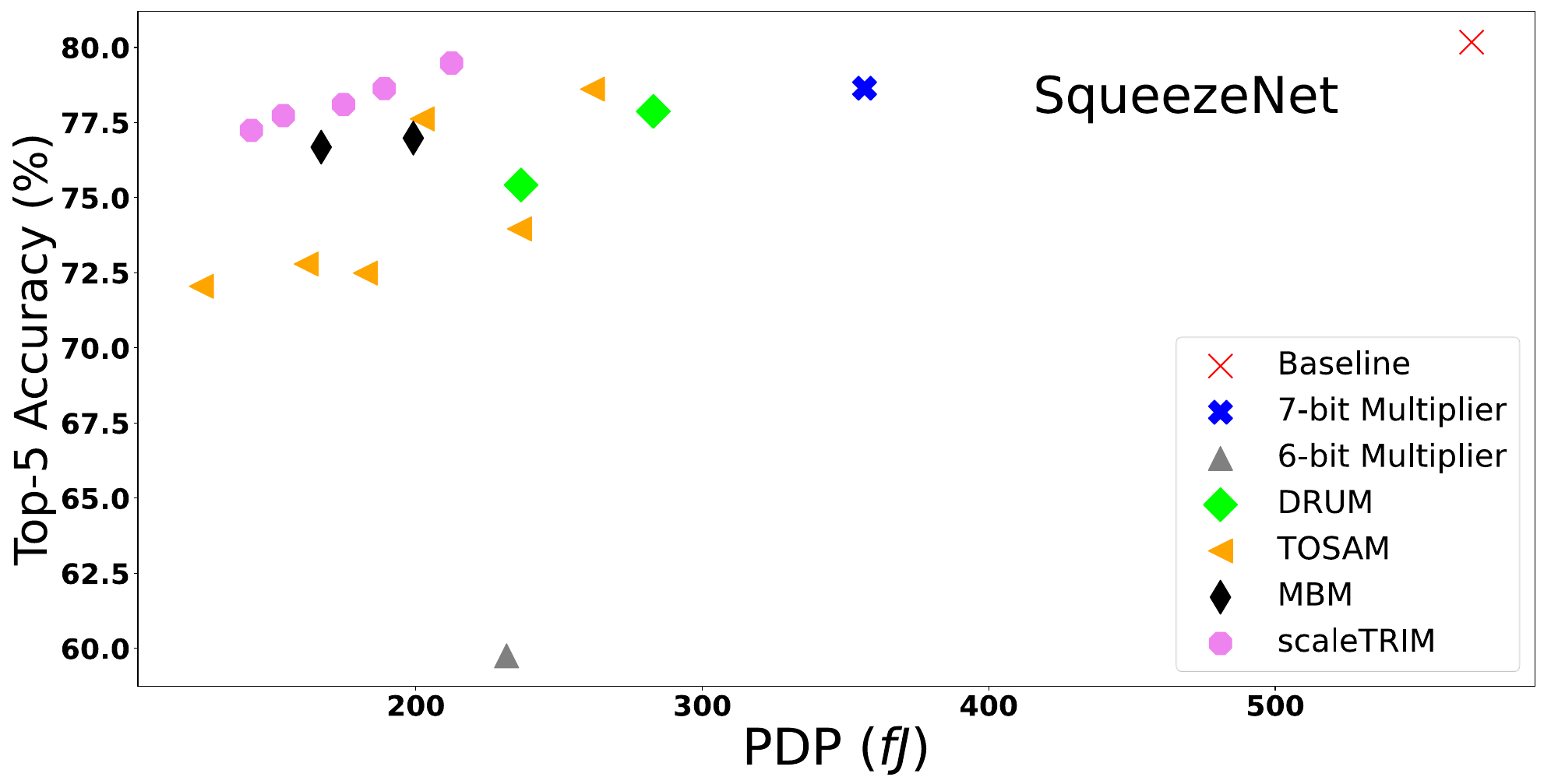}
    }
     \caption{Comparison of (a) Top-1 accuracy and (b) Top-5 accuracy of SqueezeNet architecture used for ImageNet dataset by implementing an accurate, some configurations of state-of-the-art works, and some configurations of 8-bit scaleTRIM multiplier. The exact values of the figure are presented in Table~\ref{tab:accuracy_pdp} in the Appendix.}
    \label{CNN_ACC_ImageNet}
    \vspace{-10pt}
\end{figure*}

Similarly, Fig.~\ref{CNN_ACC_ImageNet} compares the top-1 and top-5 accuracy values of SqueezeNet inference when implemented using scaleTRIM and state-of-the-art approximate multipliers in comparison to accurate multipliers for the ImageNet dataset. Notably, $scaleTRIM (4,8)$ achieves an approximate 62\% improvement in PDP, while the top-5 accuracy decreases only by 0.69\%. Therefore, from Figs.~\ref{CNN_ACC} and \ref{CNN_ACC_ImageNet}, we can observe that scaleTRIM offers a better accuracy-efficiency trade-off in comparison to state-of-the-art approximate multipliers such as DRUM, TOSAM, and MBM across different CNNs and datasets.

\section{Future Work}

While the proposed scaleTRIM multiplier demonstrates a scalable and energy-efficient architecture through design-time configurability of its truncation width ($h$) and segmentation level ($M$), several promising directions remain open for future research. First, future work will explore workload-aware adaptive mechanisms that dynamically adjust these parameters based on runtime characteristics. One potential approach involves investigating a two-stage adaptive architecture consisting of: (1) a lightweight runtime monitoring unit that profiles operand statistics and identifies workload variations, and (2) a reconfiguration controller that selects or updates pre-optimized configurations stored in memory. Realizing such adaptability presents several open research challenges. The monitoring and control logic must be designed to remain energy- and area-efficient, ensuring that the additional hardware overhead does not offset the energy savings gained through approximation. Furthermore, issues related to reconfiguration latency, stability, and responsiveness must be thoroughly examined to prevent potential performance fluctuations during dynamic workload transitions. Finally, a systematic exploration of design trade-offs, including monitoring granularity, reconfiguration frequency, and memory requirements, will be needed to assess the practicality of such an adaptive framework.

Second, another promising research avenue involves shared and reconfigurable LUT-based error compensation. In the current implementation, each scaleTRIM instance employs a dedicated LUT to store pre-computed compensation values. Future designs could explore a centralized or shared LUT architecture, where multiple scaleTRIM units access common compensation data through lightweight indexing or workload-scheduling mechanisms. Such an approach could substantially reduce area and memory overhead while enhancing scalability, resource utilization, and adaptability across parallel processing systems. These enhancements would further expand the versatility, efficiency, and practical integration of scaleTRIM into next-generation low-power and adaptive computing systems.

\section{Conclusion}
In this paper, we proposed scaleTRIM, a scalable integer approximate multiplier based on linearization and error compensation. In scaleTRIM, the input operands are first truncated to $h$ bits based on their leading one-bit position, and then a curve-fitting method is used to fit the product term to a linear function. 
To compensate for approximation errors introduced by truncation and linearization, we proposed an LUT-based error-reduction technique. The results show that compared to state-of-the-art approximate multipliers, scaleTRIM configurations lie on the Pareto front in the design space. Moreover, we demonstrated that using scaleTRIM in DNN-based image classification results in a better accuracy-efficiency trade-off than many state-of-the-art designs. While our evaluation covers 8-bit and 16-bit configurations, extending the design space exploration to 32-bit operands remains future work. The preprocessing required to generate compensation values ($M$) for 32-bit inputs incurs substantial computational and memory costs, making such an evaluation impractical.
Moreover, another limitation of the proposed design is its relatively higher delay due to the added logic for error compensation. As part of future work, we plan to explore architectural optimizations to reduce delay while preserving accuracy, and to extend the design space exploration toward applications with more stringent timing requirements, such as real-time embedded systems or edge AI devices.
\section*{Acknowledgment}
This research is partly supported by the ASPIRE AARE Grant (S1561) on ”Towards Extreme Energy Efficiency through Cross-Layer Approximate Computing".

\color{black}
\printbibliography
\section*{Appendix}

Table~\ref{tab:multiplier_comparison} shows the exact values presented in Fig.~\ref{DS_8bit}.

\begin{table*}[ht]
\centering
\caption{MRED and Performance comparison of  8-bit scaleTRIM configurations with state-of-the-art approximate multipliers}
\label{tab:multiplier_comparison}
\begin{tabular}{lrrrrr}
\toprule
\textbf{8-bit Multiplier} & \textbf{MRED} & \textbf{Delay (nS)} & \textbf{Area (µm²)} & \textbf{Power (µW)} & \textbf{PDP (fJ)} \\
\midrule
MBM=1  & 2.80  & 1.50 & 232.70 & 192.03 & 288.045 \\
MBM=2  & 3.74  & 1.41 & 194.62 & 141.22 & 199.1202 \\
MBM=3  & 6.88  & 1.29 & 169.92 & 129.43 & 166.9647 \\
MBM=4  & 13.82 & 1.22 & 151.34 & 99.28  & 121.1216 \\
MBM=5  & 26.57 & 1.15 & 129.56 & 89.31  & 102.7065 \\
Mitchell & 3.76  & 1.37 & 235.45 & 191.52 & 262.3824 \\
DSM(3) & 14.11 & 1.29 & 224.36 & 165.69 & 213.7401 \\
DSM(4) & 6.84  & 1.34 & 242.33 & 189.71 & 254.2114 \\
DSM(5) & 3.02  & 1.39 & 265.45 & 235.34 & 327.1226 \\
DSM(6) & 2.67  & 1.40 & 282.62 & 278.76 & 390.264 \\
DSM(7) & 2.02  & 1.46 & 318.86 & 311.59 & 454.9214 \\
DRUM(3) & 12.62 & 1.21 & 181.94 & 146.82 & 177.6522 \\
DRUM(4) & 6.03  & 1.25 & 240.78 & 183.38 & 229.225 \\
DRUM(5) & 3.01  & 1.32 & 290.54 & 214.31 & 282.8892 \\
DRUM(6) & 2.43  & 1.37 & 291.93 & 261.34 & 358.0358 \\
DRUM(7) & 1.41  & 1.42 & 306.31 & 292.56 & 415.4352 \\
TOSAM(0,2) & 10.38 & 1.10 & 108.39 & 89.15 & 98.065 \\
TOSAM(1,2) & 9.53  & 1.14 & 115.26 & 95.24 & 108.5736 \\
TOSAM(0,3) & 7.58  & 1.17 & 135.46 & 106.98 & 125.1666 \\
TOSAM(1,3) & 5.76  & 1.22 & 155.61 & 132.58 & 161.7476 \\
TOSAM(2,3) & 5.61  & 1.28 & 161.23 & 138.65 & 177.472 \\
TOSAM(0,4) & 6.82  & 1.30 & 163.10 & 140.30 & 182.39 \\
TOSAM(1,4) & 4.44  & 1.32 & 164.12 & 141.12 & 186.2784 \\
TOSAM(2,4) & 3.01  & 1.34 & 208.38 & 197.90 & 265.186 \\
TOSAM(3,4) & 2.68  & 1.36 & 246.24 & 239.80 & 326.128 \\
TOSAM(0,5) & 5.62  & 1.37 & 190.62 & 172.40 & 236.188 \\
TOSAM(1,5) & 4.09  & 1.37 & 193.32 & 182.28 & 249.7236 \\
TOSAM(2,5) & 2.36  & 1.38 & 232.30 & 218.60 & 301.668 \\
TOSAM(3,5) & 1.24  & 1.39 & 259.41 & 251.61 & 349.7379 \\
TOSAM(0,6) & 3.12  & 1.40 & 223.20 & 200.10 & 280.14 \\
TOSAM(2,6) & 2.11  & 1.41 & 241.20 & 226.30 & 319.083 \\
TOSAM(2,7) & 1.46  & 1.46 & 256.47 & 249.64 & 364.4744 \\
TOSAM(3,7) & 0.98  & 1.47 & 272.67 & 261.65 & 384.6255 \\
scaleTRIM (2,0) & 11.25 & 1.25 & 119.86 & 87.42 & 109.275 \\
scaleTRIM (2,4) & 9.51  & 1.28 & 125.64 & 97.65 & 124.992 \\
scaleTRIM (2,8) & 8.98  & 1.32 & 139.54 & 99.86 & 131.8152 \\
scaleTRIM (3,0) & 5.75  & 1.35 & 141.24 & 105.64 & 142.614 \\
scaleTRIM (3,4) & 3.73  & 1.36 & 150.82 & 113.05 & 153.748 \\
scaleTRIM (3,8) & 3.53  & 1.41 & 154.50 & 123.67 & 174.3747 \\
scaleTRIM (4,0) & 4.54  & 1.40 & 156.14 & 124.84 & 174.776 \\
scaleTRIM (4,4) & 3.54  & 1.42 & 160.59 & 133.10 & 189.002 \\
scaleTRIM (4,8) & 3.34  & 1.45 & 162.26 & 146.53 & 212.4685 \\
scaleTRIM (5,0) & 3.99  & 1.50 & 178.43 & 172.66 & 258.99 \\
scaleTRIM (5,4) & 2.32  & 1.52 & 184.18 & 180.92 & 274.9984 \\
scaleTRIM (5,8) & 2.12  & 1.55 & 186.99 & 189.84 & 294.252 \\
scaleTRIM (6,0) & 2.23  & 1.54 & 199.47 & 202.19 & 311.3726 \\
scaleTRIM (6,4) & 1.41  & 1.58 & 206.59 & 211.34 & 333.9172 \\
scaleTRIM (6,8) & 1.18  & 1.59 & 212.74 & 220.84 & 351.1356 \\
scaleTRIM (7,0) & 1.12  & 1.60 & 221.45 & 231.25 & 370.00 \\
scaleTRIM (7,4) & 0.91  & 1.62 & 230.70 & 244.21 & 395.6202 \\
scaleTRIM (7,8) & 0.85  & 1.69 & 240.46 & 256.34 & 433.2146 \\
EVO-lib1 & 0.019 & 1.41 & 601.80 & 386.00 & 544.26 \\
EVO-lib2 & 0.13  & 1.41 & 507.90 & 371.00 & 523.11 \\
EVO-lib3 & 0.82  & 1.39 & 423.90 & 297.00 & 412.83 \\
EVO-lib4 & 5.03  & 1.20 & 278.60 & 153.00 & 183.60 \\
ILM0 & 2.69  & 1.62 & 241.56 & 157.28 & 254.7936 \\
ILM5 & 9.51  & 1.58 & 214.23 & 146.59 & 231.6122 \\
AXM8-4 &	8.7&	1.18&	321.48&	189.82	& 223.9876\\
AXM8-3 &	2.3&	1.2	&335.04&	254.49&	305.388\\
Mitchel\_LODII\_0 & 3.81  & 1.26 & 226.81 & 186.94 & 235.5444 \\
Mitchel\_LODII\_4 & 4.12  & 1.22 & 246.13 & 198.75 & 242.475 \\
Mitchel\_LODII\_8 & 4.62  & 1.22 & 243.26 & 191.34 & 233.4348 \\
\bottomrule
\end{tabular}
\end{table*}

Table~\ref{tab:detailed_metrics} shows the exact values presented in Figs.~\ref{DS_MED_8bit}, \ref{Max_error_DS}, and \ref{Std_error_DS}.

\begin{table*}[ht]
\centering
\caption{MED, Max Erro, Std and Performance comparison of  8-bit scaleTRIM configurations with state-of-the-art approximate multipliers}
\label{tab:detailed_metrics}
\begin{tabular}{lrrrrrrrr}
\toprule
\textbf{Multiplier} & \textbf{Delay (nS)} & \textbf{Area (µm²)} & \textbf{Power (µW)} & \textbf{PDP (fJ)} & \textbf{MRED} & \textbf{Max Error} & \textbf{MED} & \textbf{Std} \\
\midrule
Mitchell & 1.37 & 235.45 & 191.52 & 262.3824 & 3.76 & 4096 & 611.16 & 779.87 \\
DSM(3) & 1.29 & 224.36 & 165.69 & 213.7401 & 14.11 & 14849 & 3337.88 & 2711.92 \\
DRUM(3) & 1.21 & 181.94 & 146.82 & 177.6522 & 12.62 & 14849 & 1862.78 & 2246.22 \\
DRUM(6) & 1.37 & 291.93 & 261.34 & 358.0358 & 2.43 & 2000 & 245.64 & 295.28 \\
MBM-1 & 1.50 & 232.70 & 192.03 & 288.045 & 2.80 & 2816 & 396.47 & 462.18 \\
MBM-2 & 1.41 & 194.62 & 141.22 & 199.1202 & 3.74 & 2816 & 402.22 & 459.51 \\
ILM0 & 1.62 & 241.56 & 157.28 & 254.7936 & 2.69 & 3844 & 455.05 & 633.94 \\

AXM8-4& 1.18 &	321.48 &	189.82 &	223.9876& 8.7&-& 1919.891&-\\
AXM8-3& 1.2&335.04&	254.49&	305.388& 2.3&-&209.427&-\\

TOSAM(0,3) & 1.17 & 135.46 & 106.98 & 125.1666 & 7.58 & 15873 & 1361.74 & 1981.23 \\
TOSAM(1,3) & 1.22 & 155.61 & 132.58 & 161.7476 & 5.76 & 10753 & 1007.15 & 1307.62 \\
TOSAM(0,4) & 1.30 & 163.10 & 140.30 & 182.39 & 6.82 & 13825 & 1283.11 & 1704.46 \\
TOSAM(2,4) & 1.34 & 208.38 & 197.90 & 265.186 & 3.01 & 5377 & 486.43 & 623.64 \\
TOSAM(2,5) & 1.38 & 232.30 & 218.60 & 301.668 & 2.36 & 2497 & 232.12 & 286.30 \\
scaleTRIM (3,0) & 1.35 & 141.24 & 105.64 & 142.614 & 5.75 & 12801 & 1138.86 & 1580.89 \\
scaleTRIM (3,4) & 1.36 & 150.82 & 113.05 & 153.748 & 3.73 & 6177 & 586.15 & 745.78 \\
scaleTRIM (3,8) & 1.41 & 154.50 & 123.67 & 174.3747 & 3.53 & 5128 & 547.78 & 687.67 \\
scaleTRIM (4,0) & 1.40 & 156.14 & 124.84 & 174.776 & 4.54 & 11521 & 924.47 & 1379.74 \\
scaleTRIM (4,4) & 1.42 & 160.59 & 133.10 & 189.002 & 3.54 & 6237 & 616.67 & 794.53 \\
scaleTRIM (4,8) & 1.45 & 162.26 & 146.53 & 212.4685 & 3.34 & 5260 & 582.91 & 738.72 \\
scaleTRIM (5,0) & 1.50 & 178.43 & 172.66 & 258.99 & 3.99 & 8961 & 709.63 & 1041.10 \\
scaleTRIM (5,4) & 1.52 & 184.18 & 180.92 & 274.9984 & 2.32 & 4190 & 386.55 & 512.30 \\
scaleTRIM (5,8) & 1.55 & 186.99 & 189.84 & 294.252 & 2.12 & 3356 & 318.44 & 407.95 \\
\bottomrule
\end{tabular}
\end{table*}

Table~\ref{tab:accuracy_pdp} shows the exact values presented in Fig.~\ref{CNN_ACC_ImageNet}.

\begin{table*}[ht]
\centering
\caption{Comparison of Top-1 accuracy and Top-5 accuracy of SqueezeNet architecture used for ImageNet dataset by implementing an accurate, some configurations of state-of-the-art works, and some configurations of 8-bit scaleTRIM multiplier}
\label{tab:accuracy_pdp}
\begin{tabular}{lrrr}
\toprule
\textbf{Multiplier} & \textbf{Top 5 Accuracy} & \textbf{Top 1 Accuracy} & \textbf{PDP (fJ)} \\
\midrule
8-bit Accurate multiplier & 80.17 & 57.41 & 568.53 \\
7-bit Accurate multiplier & 78.64 & 55.96 & 356.64 \\
scaleTRIM (3,0) & 77.24 & 54.01 & 142.61 \\
scaleTRIM (3,4) & 77.73 & 54.37 & 153.75 \\
scaleTRIM (4,0) & 78.10 & 54.58 & 174.77 \\
scaleTRIM (4,4) & 78.63 & 55.32 & 189.00 \\
scaleTRIM (4,8) & 79.48 & 56.52 & 212.47 \\
DRUM-3 & 35.50 & 16.76 & 177.65 \\
DRUM-4 & 75.42 & 51.51 & 236.73 \\
DRUM-5 & 78.87 & 55.73 & 282.89 \\
TOSAM (0,3) & 72.05 & 47.12 & 125.16 \\
TOSAM (1,3) & 72.79 & 48.54 & 161.75 \\
TOSAM (0,4) & 72.49 & 47.50 & 182.39 \\
TOSAM (2,4) & 77.62 & 53.99 & 202.21 \\
TOSAM (0,5) & 73.96 & 49.47 & 236.19 \\
TOSAM (2,5) & 78.61 & 55.46 & 261.65 \\
MBM-3 & 77.54 & 54.23 & 199.12 \\
MBM-4 & 78.20 & 54.81 & 166.96 \\
\bottomrule
\end{tabular}
\end{table*}
Table~\ref{tab:error_analysis} shows compensation values stored in the LUTs for different values of $h$ and $M$ for 8-bit $scaleTRIM$. 
\begin{table*}[ht]
\centering
\caption{Compensation values stored in the LUTs for different values of $h$ and $M$ in different ranges of $X_h + Y_h$}
\label{tab:error_analysis}
\begin{tabular}{lrrrrrrrr}
\toprule
 & \multicolumn{2}{c}{h=3} & \multicolumn{2}{c}{h=4} & \multicolumn{2}{c}{h=5} & \multicolumn{2}{c}{h=6} \\
\cmidrule(lr){2-3} \cmidrule(lr){4-5} \cmidrule(lr){6-7} \cmidrule(lr){8-9}
 & M=4 & M=8 & M=4 & M=8 & M=4 & M=8 & M=4 & M=8 \\
\midrule
$0 < X_h + Y_h < 0.25$ & \multirow{2}{*}{0.053} & 0.073 & \multirow{2}{*}{-0.015} & 0.008 & \multirow{2}{*}{-0.046} & -0.020 & \multirow{2}{*}{-0.059} & -0.032 \\
$0.25 < X_h + Y_h < 0.5$ &  & 0.039 &  & -0.028 &  & -0.058 &  & -0.070 \\
\cmidrule(lr){1-9}
$0.5 < X_h + Y_h < 0.75$ & \multirow{2}{*}{0.050} & 0.032 & \multirow{2}{*}{-0.035} & -0.042 & \multirow{2}{*}{-0.073} & -0.076 & \multirow{2}{*}{-0.089} & -0.090 \\
$0.75 < X_h + Y_h < 1$ &  & 0.066 &  & -0.030 &  & -0.071 &  & -0.088 \\
\cmidrule(lr){1-9}
$1 < X_h + Y_h < 1.25$ & \multirow{2}{*}{0.234} & 0.182 & \multirow{2}{*}{0.114} & 0.063 & \multirow{2}{*}{0.058} & 0.008 & \multirow{2}{*}{0.035} & -0.016 \\
$1.25 < X_h + Y_h < 1.5$ & & 0.317 & & 0.190 &  & 0.132 &  & 0.106 \\
\cmidrule(lr){1-9}
$1.5 < X_h + Y_h < 1.75$ & \multirow{2}{*}{0.468} & 0.468 & \multirow{2}{*}{0.354} & 0.336 & \multirow{2}{*}{0.301} & 0.274 & \multirow{2}{*}{0.277} & 0.248 \\
$1.75 < X_h + Y_h < 2$ &  & 0.410 &  & 0.467 &  & 0.412 &  & 0.387 \\
\bottomrule
\end{tabular}
\end{table*}


\clearpage
\begin{IEEEbiography}[{\includegraphics[width=1in,height=1.25in,clip,keepaspectratio]{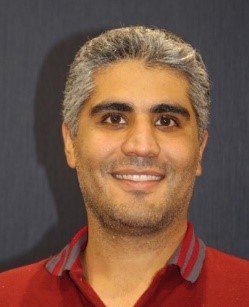}}]{Ebrahim Farahmand}
 received the B.Sc. degree in electrical engineering-communication systems and the M.Sc. degree in electrical engineering-electronics from the Shahid Bahonar University of Kerman (SBUK), Iran, in 2012 and 2016, respectively. He is currently pursuing a Ph.D. degree in the embedded machine intelligence laboratory (EMIL) with the College of Health Solutions at Arizona State University (ASU). His research interests include brain-inspired computing, deep learning, tinyML approximate computing, machine learning accelerator, fault-tolerant design, and network systems.
\end{IEEEbiography}
\begin{IEEEbiography}[{\includegraphics[width=1in,height=1.25in,clip,keepaspectratio]{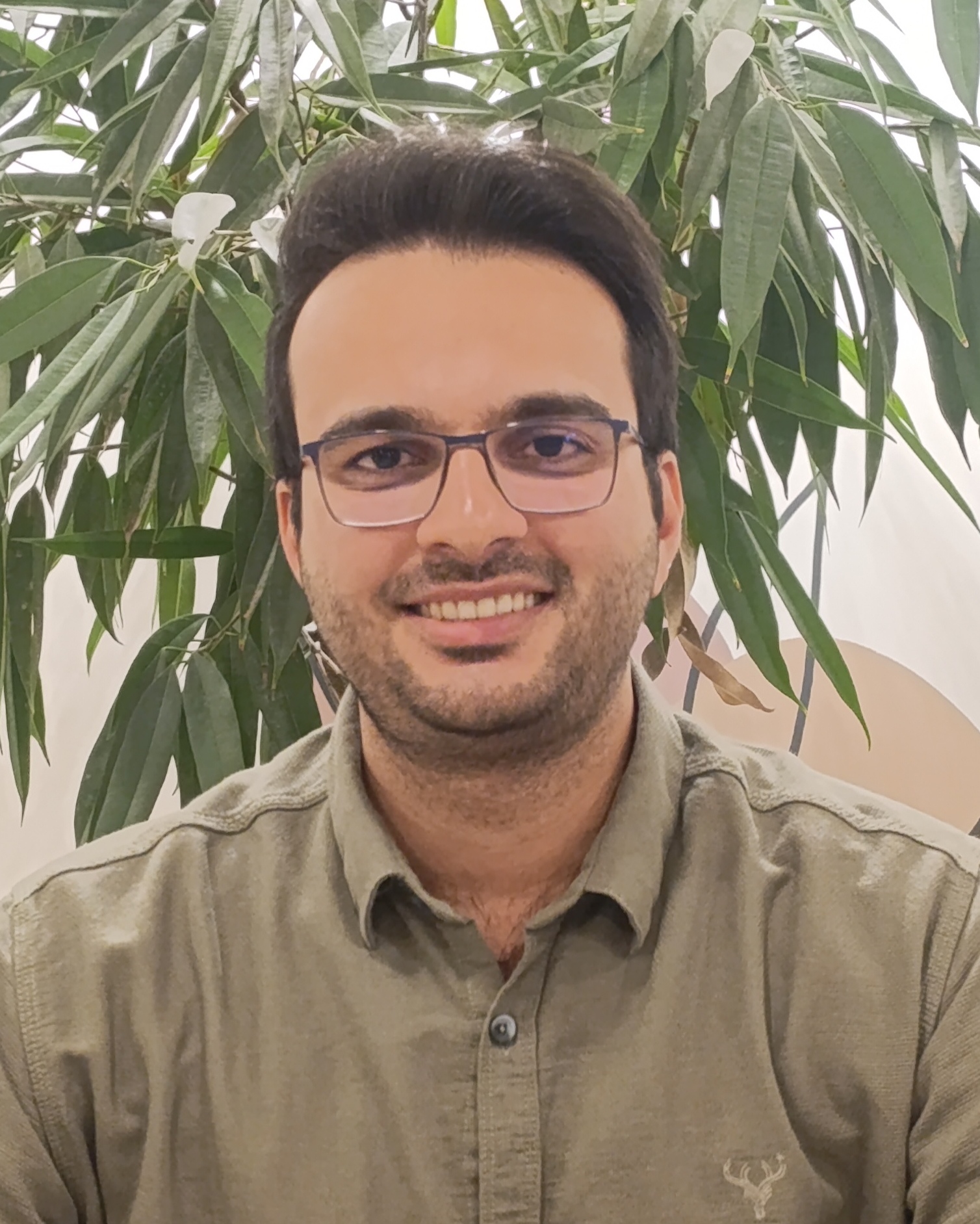}}]{Mohammad Javad Askarizadeh} is a recent graduate with a Master of Science in Electrical Engineering. He obtained his bachelor's degree in the same field from Ferdowsi University of Mashhad, Iran, in 2020. He then pursued his Master's degree at Shahid Bahonar University of Kerman, Iran, graduating in 2023. Currently, his research interests lie in adversarial machine learning.
\end{IEEEbiography}
\begin{IEEEbiography}[{\includegraphics[width=1in,height=1.25in,clip,keepaspectratio]{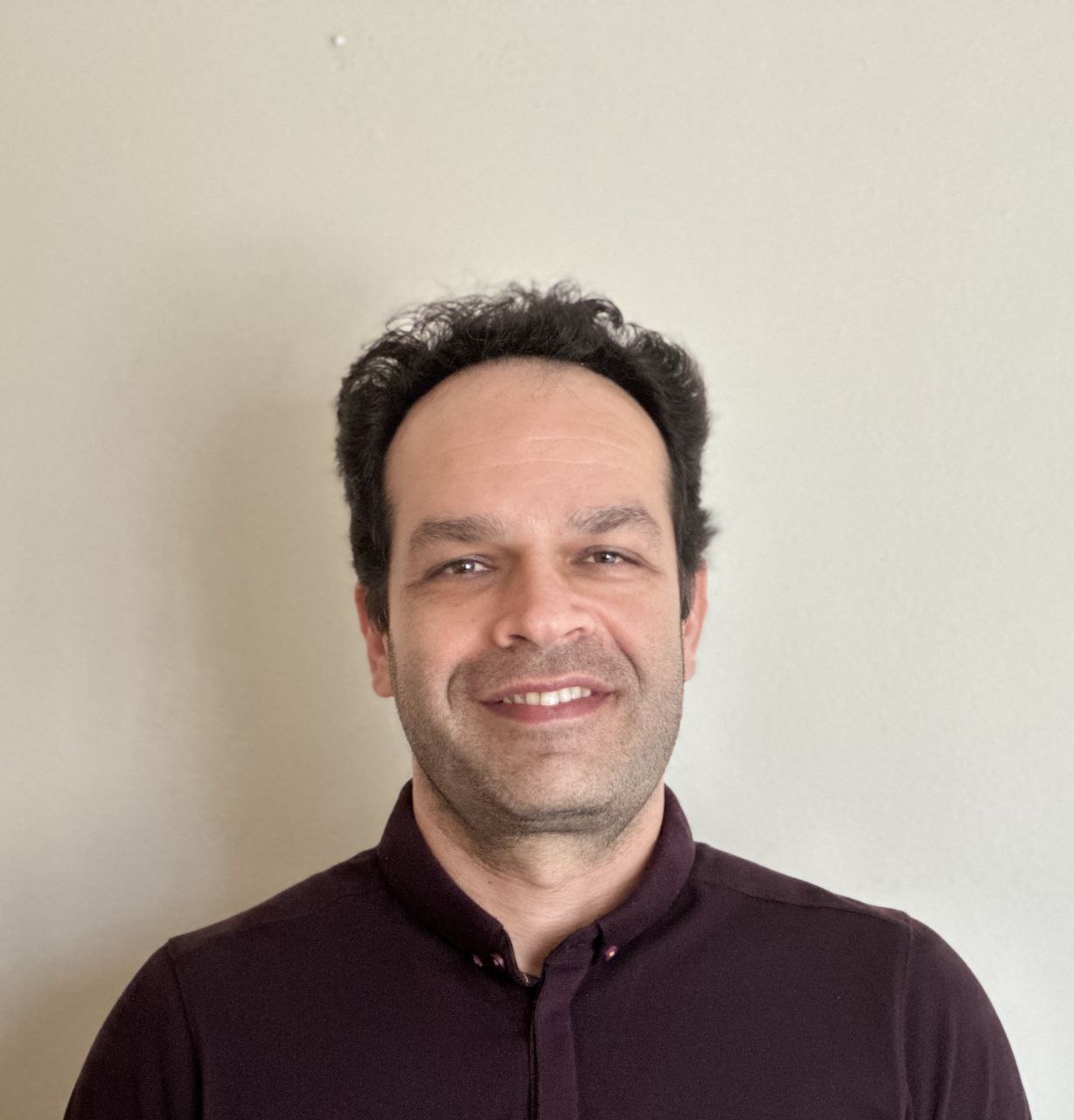}}]{Ali Mahani} received his B.Sc. degree in Electronic Engineering from Shahid Bahonar University of Kerman, Iran, in 2001, followed by M.Sc. and Ph.D. degrees in Electronic Engineering from Iran University of Science and Technology (IUST), Tehran, Iran, in 2003 and 2009, respectively. Since 2009, he has been with the Department of Electrical Engineering at Shahid Bahonar University of Kerman, where he is currently an Associate Professor and the Director of the Reliable and Smart Systems (RSS) Laboratory. Additionally, since June 2022, he has been a Research Scientist with the Department of Electrical Engineering and Computer Science at York University, Toronto, Canada. His research interests include computer architecture, fault-tolerant design, hardware accelerators, and approximate computing.
\end{IEEEbiography}
\begin{IEEEbiography}[{\includegraphics[width=1in,height=1.25in,clip,keepaspectratio]{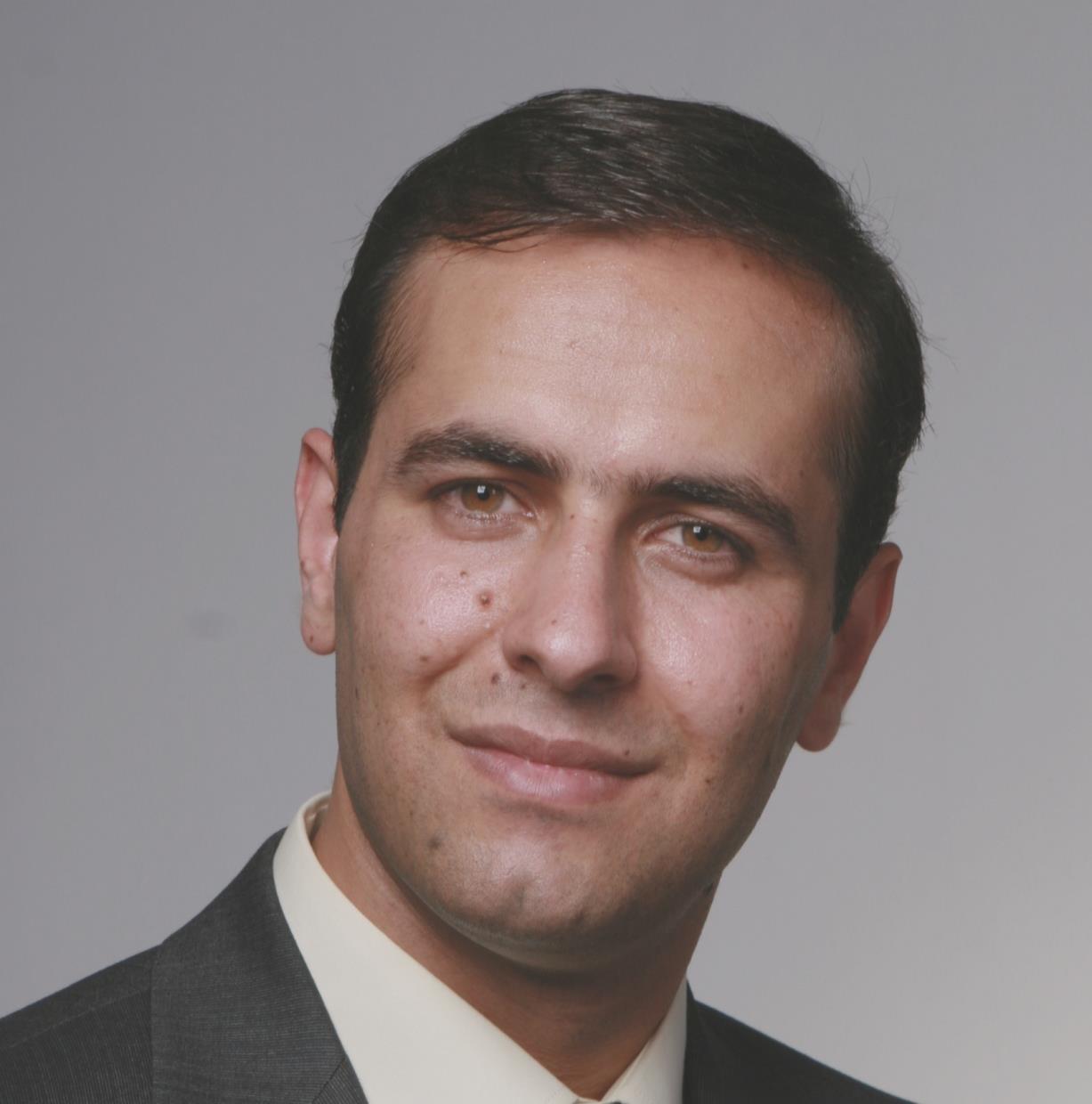}}]{Behnam Ghavami} (Senior Member, IEEE) received his Ph.D. in Computer Engineering from Amirkabir University of Technology, Tehran, Iran, in 2012. He was previously an Associate Professor at Shahid Bahonar University of Kerman. Currently, he is a Visiting Professor at Simon Fraser University's Reconfigurable Computing Lab and a member of The University of British Columbia's SoC Research Group. His research focuses on digital system design automation and VLSI systems.
\end{IEEEbiography}
\begin{IEEEbiography}[{\includegraphics[width=1in,height=1.25in,clip,keepaspectratio]{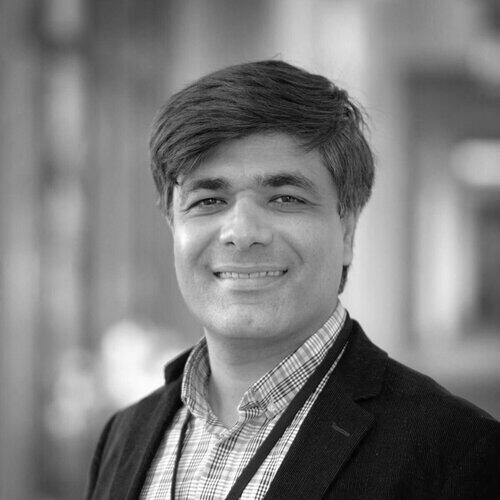}}]{Hassan Ghasemzadeh} (Senior Member, IEEE), received the BSc degree from the Sharif University of Technology, Tehran, Iran, in 1998, the MSc degree from the University of Tehran, Tehran, Iran, in 2001, and the PhD degree from the University of Texas at Dallas, Richardson, TX, in 2010, all in computer engineering. He was on the faculty of Azad University from 2003- 2006, where he served as founding chair of the Computer Science and Engineering Department at the Damavand branch, Tehran, Iran. He spent the academic year 2010- 2011 as a postdoctoral fellow at the West Wireless Health Institute, La Jolla, CA. He was a research manager at the UCLA Wireless Health Institute in 2011-2013. Currently, he is an associate professor of biomedical informatics, the director of the undergraduate biomedical informatics program, and a graduate faculty member of computer science, computer engineering, and biomedical engineering at Arizona State University (ASU). Prior to joining ASU, he was an assistant/associate professor of computer science at Washington State University (WSU 2014-2021). The focus of his research is on algorithm design and system-level optimization of embedded and pervasive systems with applications in healthcare and wellness.
\end{IEEEbiography}

\begin{IEEEbiography}[{\includegraphics[width=1in,height=1.25in,clip,keepaspectratio]{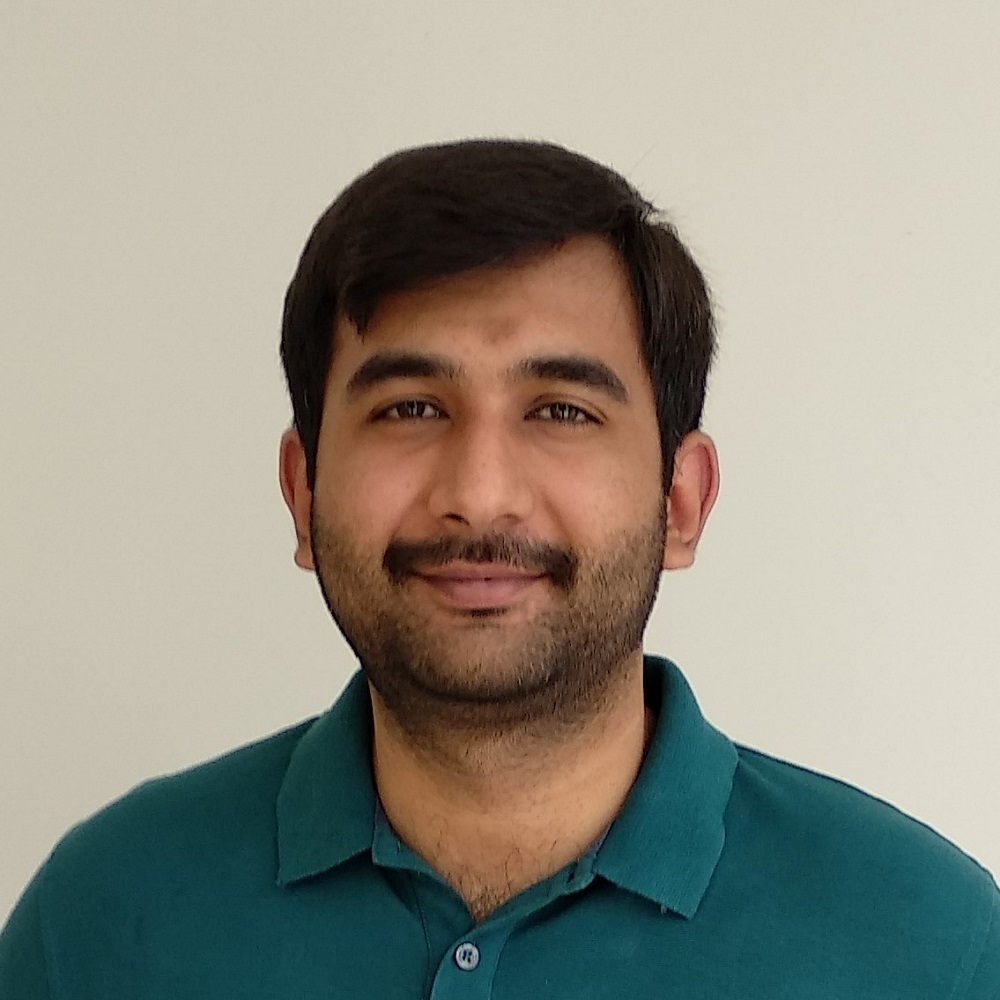}}]{Muhammad Abdullah Hanif} received the B.Sc. degree in electronic engineering from the Ghulam Ishaq Khan Institute of Engineering Sciences and Technology (GIKI), Pakistan, the M.Sc. degree in electrical engineering with a specialization in digital systems and signal processing from the School of Electrical Engineering and Computer Science, National University of Sciences and Technology (NUST), Islamabad, Pakistan, and the Ph.D. degree in computer engineering from the Vienna University of Technology (TU Wien), Austria. He is currently a Postdoctoral Associate with New York University (NYU) Abu Dhabi, United Arab Emirates. His research interests include brain-inspired computing, machine learning, approximate computing, computer architecture, energy-efficient design, robust computing, system-on-chip design, and emerging technologies.
\end{IEEEbiography}
\begin{IEEEbiography}[{\includegraphics[width=1in,height=1.25in,clip,keepaspectratio]{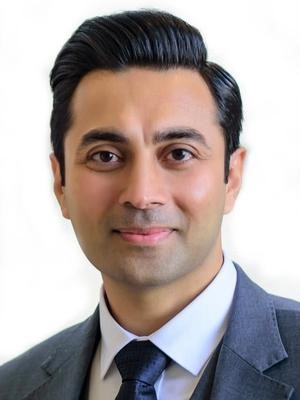}}]{Muhammad Shafique}(M’11 - SM’16) received the Ph.D. degree in computer science from the Karlsruhe Institute of Technology (KIT), Germany, in 2011. In Oct.2016, he joined the Faculty of Informatics at TU Wien, Vienna, Austria as a Full Professor of Computer Architecture and Robust, Energy-Efficient Technologies. Since Sep.2020, Dr. Shafique is with the New York University (NYU), where he is currently a Full Professor and the director of eBRAIN Lab and iCAS Lab at the NYU-Abu Dhabi in UAE, and a Global Network Professor at the Tandon School of Engineering, NYU-New York City in USA. He is also a Co-PI/Investigator in multiple NYUAD Centers on Cybersecurity, Quantum Computing, AI \& Robotics, and Smart Cities. His research interests are in AI \& machine learning hardware and system-level design, brain-inspired computing, EdgeAI, tinyML, machine learning security and privacy, quantum machine learning, cognitive autonomous systems, wearable healthcare, AI for healthcare/medical imaging, energy-efficient systems, robust computing, hardware security, emerging technologies, electronic design automation, FPGAs, MPSoCs, and embedded systems. The researched technologies and tools are deployed in application use cases from IoT, Smart CPS, Healthcare and Robotics domains.

Dr. Shafique has given several Keynotes, Invited Talks, and Tutorials, as well as organized many special sessions at premier venues. He has served as the PC Chair, General Chair, Track Chair, and PC member for several prestigious IEEE/ACM conferences. Dr. Shafique holds one U.S. patent, and has (co-)authored 10 Books, 25+ Book Chapters, 450+ papers in premier journals and conferences, and 200+ archive articles. He received the 2015 ACM/SIGDA Outstanding New Faculty Award, the AI-2000 Chip Technology Most Influential Scholar Awards (2020, 2022, 2023; Honorable Mention 2024, 2025), the ASPIRE AARE Research Excellence Award in 2021, six gold medals, several best paper awards and nominations at prestigious conferences, several HiPEAC paper awards, and multiple competition awards. He is a senior member of the IEEE and IEEE Signal Processing Society (SPS), and a senior member of the ACM, SIGARCH, SIGDA, SIGBED, and HIPEAC.
\end{IEEEbiography}

\EOD

\end{document}

%% file: Sections/Related_Works.tex
\section{Related Work}

A wide range of approximate multiplier designs have been proposed over the years. 
The Dynamic Segment Method (DSM) is proposed in~\cite{6858039} for approximating fixed-point multipliers. 
It takes $m$ bits from each $n$-bit operand from one of the two (or three) fixed bit positions, depending on the leading-one-bit position, to feed into the $m \times m$ multiplier for calculating the approximate product of two input operands. 
This method results in a high Mean Relative Error (MRE). 
To reduce the error of the truncated multiplier, a dynamic range unbiased multiplier (DRUM) is proposed in~\cite{7372600}. DRUM captured the $m$ bits of $n$-bit inputs from the leading-one-bit position and set the Least Significant Bit (LSB) of the $m$ bits to $'1'$. Then, the $m$ bits of inputs are fed into the $m\times m$ multiplier to calculate the approximate product. 

\begin{figure}
    \centering
    \includegraphics[width=1\linewidth]{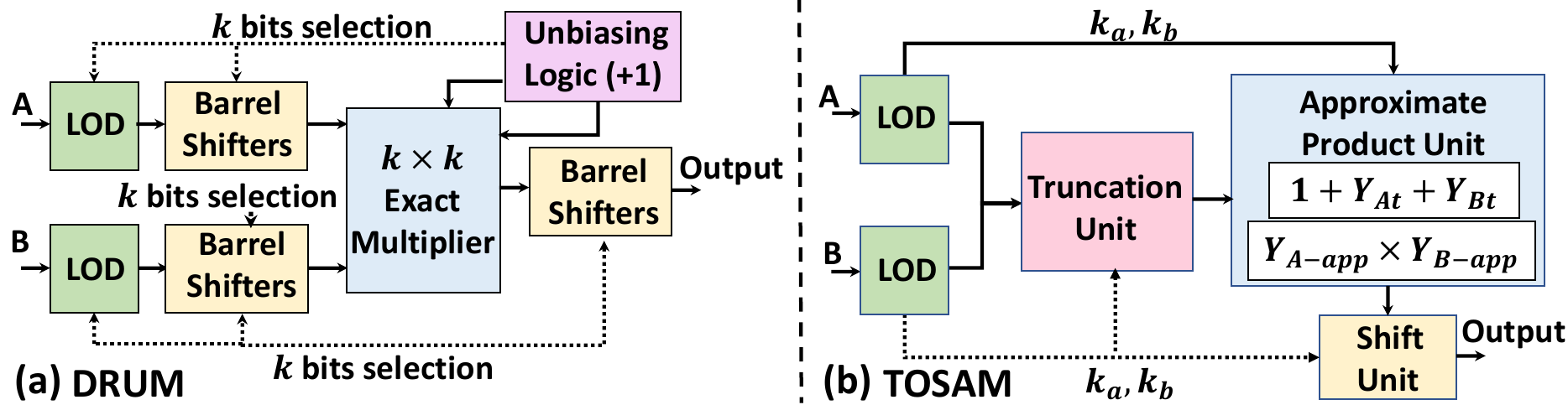}
    \caption{Simplified high-level block diagrams of (a) DRUM and (b) TOSAM approximate multipliers. The diagrams illustrate the main architectural components, including leading-one detection (LOD) and truncation.}
    \label{highlevel_DRUM_TOSAM}
    \vspace{-15pt}
\end{figure}

LETAM~\cite{VAHDAT20171} is another method that truncates the input operands to achieve an improved approximate multiplier design. 
In RoBA~\cite{7517375}, the authors used the rounding of the input operands to the nearest power of 2 to reduce the complexity of multipliers. 
Furthermore, TOSAM~\cite{8626488} provides a scalable approximate multiplier based on truncation and rounding. First, the leading-one positions of the input operands are computed. Then, the next $t$ bits, next to leading-one position, of each operand are fed into a $t$-bit adder. Moreover, to decrease the error of approximation, the $h$ bits, next to the leading-one position, are captured and concatenated with $'1'$ at the LSB location. Then, $(h+1)$ bits of both the operands are fed to an $(h+1)\times(h+1)$ multiplier to compute the product. To facilitate structural comparison with our proposed architecture, simplified high-level block diagrams of DRUM and TOSAM are shown in Fig.~\ref{highlevel_DRUM_TOSAM}.

Another set of works focuses on improving multiplication speed by approximating multiplication using addition operations. 
One prominent approach in this direction is to use linearization to transform the multiplication into a linear function, for example, ApproxLP~\cite{8806911}. 
However, ApproxLP targets the floating-point format and is mainly tested for GPU-based computations.
In \cite{shakibhamedan2024ace} proposed ACE-CNN, a framework that incorporates an approximate multiplier of signed carriage disregard multipliers (SCDM8) designed for energy-efficient convolutional neural network (CNN)-based image classification. The proposed SCDM8 multipliers are based on approximate array architectures that strategically disregard carry propagation in selected partial product units, resulting in simpler circuits and reduced critical path delay. A total of 100 configurations were evaluated, of which 20 were identified as offering acceptable accuracy and integrated into several pre-trained CNN models.

Some approximate multipliers, such as ATCM~\cite{yang2017low}, proposed high-speed multipliers by introducing approximation directly within the tree compressor. Similarly, LSAM and CSAM~\cite{guo2024hardware} and AXMAC~\cite{deepsita2023energy} explored hardware-efficient architectures, with the latter employing a novel recursive multiplication strategy to improve delay efficiency in Multiply-Accumulate (MAC) units, focusing on optimizing general-purpose arithmetic units. A key motivation for this research is image processing, where inherent error tolerance enables significant hardware savings. 

Several studies, including PAM~\cite{anguraj2023design}, the approximation method in~\cite{rashidi2024efficient}, and ADMs~\cite{anusha2020design}, have designed area- and delay-efficient approximate multipliers specifically for this domain, validating their effectiveness through image processing metrics. Beyond ASICs, FPGA-specific optimizations have also been investigated. \cite{toan2020fpga} introduced multi-level approximate multipliers for high-performance FPGA applications, while CORDIC-based~\cite{khurshid2025high} presented high-performance CORDIC-based approximate MAC architectures, demonstrating the adaptation of approximation techniques to reconfigurable hardware platforms.

Another approach to approximating multiplier modules, with a focus on ASIC implementation, is presented by Mitchell~\cite{5219391}, which converts the numbering system into a logarithmic system where multiplication can be achieved through an addition operation. 
Several studies have focused on improving the Mitchell approximate multiplier design~\cite{8493590,9116315,9086744}. In MBM~\cite{8493590}, an error-reduction mechanism is introduced for the Mitchell multiplier to minimize approximation errors. In this approach, the complexity arises from the calculations involving the logarithm and antilogarithm of the operands and outputs, respectively. In this paper, we target a scalable approximate multiplier based on truncation, linearization, and error compensation.

Moreover, EvoApproxLib~\cite{7926993} is a widely adopted open-source library of approximate arithmetic circuits generated using evolutionary optimization techniques. It provides Pareto-optimal configurations of approximate multipliers under various accuracy and hardware efficiency constraints (e.g., power, area, and delay). Due to its comprehensive design space and standardized benchmarking methodology, EvoApproxLib is frequently used as a reference baseline in approximate computing research.

A novel approximate multiplier architecture called the Most Significant One-driven Approximate Multiplier (MSAMZ)~\cite{huang2024energy} incorporates a dynamic weight approximation strategy that selectively approximates the lower-significance bits of the multiplier while preserving precision in the higher-order bits. By partitioning the operand space through an approximation factor (k) and a precision factor (m), MSAMZ facilitates configurable trade-offs between computational accuracy and hardware efficiency. Various approximation schemes are investigated, including one-dominating and zero-dominating strategies, with optional compensation mechanisms to reduce error. These techniques are intended to lower energy consumption by simplifying arithmetic operations to shifts and additions. Despite the promising results at the application level, the proposed multiplier does not consistently achieve superior outcomes in terms of power and area.

LHTAM~\cite{izadi2025lhtam} integrates an efficient bit-shifting mechanism, a custom-designed priority encoder, and a simplified result production unit using MUX-based logic to reduce complexity, power, and delay. Unlike traditional designs that rely on Leading-One Detector (LOD) circuits or lookup tables, LHTAM eliminates these blocks, resulting in significant hardware savings. However, LHTAM is primarily utilized in floating-point multiplication and is limited to specific configurations (e.g., 16-bit, with a fixed number of significant bits z=5). It does not encompass broader design-space exploration or adaptability to varying precision and accuracy trade-offs.

A summary of the related works and their comparison with the contribution of this paper, i.e., scaleTRIM, is presented in Table~\ref{Related_state_of-the-art}. 

\begin{table*}[ht]
  \centering
  \footnotesize
  \caption{A summary of the related works and their comparison with scaleTRIM}
    \label{Related_state_of-the-art}
    \begin{tabularx}{\linewidth}{>{\centering\arraybackslash}p{1.5cm} >{\centering\arraybackslash}p{4.9cm} >{\centering\arraybackslash}p{3.4cm} >{\centering\arraybackslash}p{1.5cm} >{\centering\arraybackslash}p{2cm} >{\centering\arraybackslash}p{2.2cm}} 
    \toprule
    \textbf{Approximate Multiplier} & \textbf{Types of Operations to Realize Multiplication} & \textbf{Type of Error Compensation} & \textbf{Employs Truncation} & \textbf{Logarithmic Approximation} & \textbf{Design time Reconfigurability} \\ 
    \midrule
    DRUM~\cite{7372600} & Leading-one bit detection, Truncation & Concatenate bit '1' to LSB & Yes & No & Yes \\
    DSM~\cite{6858039} & Segment the fixed bits width next to leading-one bit & No & Yes & No & No \\
    LETAM~\cite{VAHDAT20171} & Leading-one bit detection, Truncation & No & Yes & No & Yes \\
    TOSAM~\cite{8626488} & Leading-one bit detection, Truncation & Rounded the value & Yes & No & Yes \\
    ROBA~\cite{7517375} & Rounding and bit-wise shifter & No & No & No & No \\
    ApproxLP~\cite{8806911} & Piece-wise linear-plane curve fitting & No & Yes & No & Yes \\
    Mitchell~\cite{5219391} & Leading-one detection, $\log(1+x) \approx x$ & No & No & Yes & No \\
    MBM~\cite{8493590} & Leading-one detection, $\log(1+x) \approx x$ & Add a fixed value & Yes & Yes & Yes \\
    SCDM8~\cite{shakibhamedan2024ace} & Array-based, partial product carry disregard & Statistical evaluation & No & No & Yes \\
    MSAMZ~\cite{huang2024energy} & Dynamic weight, Shift-and-add ops & One-/Zero-dominating & Yes & No & Yes \\
    LHTAM~\cite{izadi2025lhtam} & Bit-shifting, MUX-based result logic & Simplified logic-based & Yes & No & No \\
    \midrule
    \textit{\textbf{scaleTRIM}} & \textbf{Leading-one detection, Linearization} & \textbf{LUT-based compensation} & \textbf{Yes} & \textbf{No} & \textbf{Yes} \\
    \bottomrule
    \end{tabularx}
\end{table*}